\documentclass[12pt, floatfix]{iopart}
\usepackage[english]{babel}

\pdfoutput=1

\usepackage{graphicx}
\usepackage{latexsym}
\usepackage{bm}
\usepackage[protrusion=true, expansion=true]{microtype}

\usepackage{color}
\definecolor{navyblue}{rgb}{0.0, 0.0, 0.5} 
\usepackage[colorlinks=true, breaklinks=false, linkcolor=navyblue, urlcolor=navyblue, citecolor=navyblue]{hyperref}

\usepackage{iopams}
\usepackage{appendix}
\usepackage{cite}

\newcommand{\ee}{\mathrm{e}}
\newcommand{\ii}{\mathrm{i}}
\newcommand{\D}[1]{\,\mathrm{d}#1}
\newcommand{\Dev}{\xi}
\newcommand{\JumpR}{\chi}

\newcommand{\expes}[1]{\langle #1 \rangle}

\newcommand{\bra}[1]{\mathinner{\langle{#1}|}}
\newcommand{\ket}[1]{\mathinner{|{#1}\rangle}}

\newcommand{\fixme}[1]%
   {\begingroup{\color{blue}\it (FIXME: #1)}\endgroup}

\begin{document}
\review{Quantum systems under frequency modulation}
\author{M~P~Silveri$^{1,2}$, J~A~Tuorila$^{1,3}$, E~V~Thuneberg$^1$, G~S~Paraoanu$^3$}
\address{$^1$Department of Physics, University of Oulu, P.O. Box 3000, FI-90014, Finland\\
$^2$Department of Physics, Yale University, New Haven, CT 06520, USA\\
$^3$Department of Applied Physics, Aalto University School of Science, Aalto University, P.O. Box 15100,
 FI-00076 AALTO, Finland}
\ead{sorin.paraoanu@aalto.fi}
\vspace{10pt}
\begin{indented}
\item[]\today
\end{indented}


\begin{abstract}
We review the physical phenomena that arise when quantum mechanical energy levels are modulated in time. The dynamics resulting from changes in the transition frequency is a problem studied since the early days of quantum mechanics. 
It has been of constant interest both experimentally and theoretically since, with the simple two-state model providing an inexhaustible source of novel concepts.
When the transition frequency of a quantum system is modulated, several phenomena can be observed, such as Landau-Zener-St\"uckelberg-Majorana interference, motional averaging and narrowing, and the formation of dressed states with the presence of sidebands in the spectrum.
Adiabatic changes result in the accumulation of geometric phases, which can be used to create topological states.
In recent years, an exquisite experimental control in the time domain was gained through the parameters entering the Hamiltonian, and high-fidelity readout schemes allowed the state of the system to be monitored non-destructively. 
These developments were made in the field of quantum devices, especially in superconducting qubits, as a well as in atomic physics, in particular in ultracold gases. As a result of these advances, it became possible to demonstrate many of the fundamental effects that arise in a quantum system when its transition frequencies are modulated. 
The purpose of this review is to present some of these developments, from two-state atoms and harmonic oscillators to multilevel and many-particle systems.
\end{abstract}

\begin{indented}
\item {\it Keywords}: Landau-Zener-St\"uckelberg-Majorana interference, quantum control, frequency
modulation, topological phases and transitions, motional averaging and narrowing,
superconducting qubits, artificial gauge fields
\end{indented}

\maketitle

\microtypesetup{protrusion=false}
{\small \tableofcontents}
\microtypesetup{protrusion=true}

\newpage 

\section{Introduction}
\label{sec:introduction}

The year 1932 can be perhaps considered as the year of birth of the two-level model. Almost simultaneously, Lev Landau \cite{Landau32,Landau32i}, Clarence Zener \cite{Zener32}, Ernst St\"uckelberg \cite{Stueckelberg32}, and Ettore Majorana \cite{Majorana32} published the solution for the occupation probabilities of the adiabatic energy eigenstates of a spin-1/2 particle under a nonadiabatic sweep across an avoided level crossing. It was already then realized that a phase difference between the two states will be generated during this process, and if the passage is reversed then the resulting occupation probabilities will be  oscillating functions of the phase collected in between the crossings. This phase is typically referred to as the St\"uckelberg phase. The corresponding population oscillations are conceptually similar to the interference pattern in a Mach-Zehnder interferometer, with the two states playing the role of the interferometer paths travelled by the photons. In many natural phenomena, the basic physics can indeed be described by a model of two crossing energy levels. For example in atomic collisions, the incoming atoms in the ground state make transitions to  higher excited states as they approach each other. In inelastic scattering processes, St\"uckelberg interference has been seen already since the 1960's in several experiments as oscillations in the atomic collisional cross section as a function of the scattering angle \cite{Nikitin72}.

In a parallel theoretical development, the issue of what happens when a system is subjected to noise that shifts its energy levels became a paradigmatic problem in the theory of dephasing. Several theoretical approaches can be followed, depending on the structure of the noise power spectrum. A conventional microscopic description for dephasing in a two-state system is to couple it to a bath of harmonic oscillators with an interaction Hamiltonian which is diagonal in the eigenbasis of the system and linear for the oscillator coordinates~\cite{Leggett87}. When the bath can be assumed to be in equilibrium at a certain temperature, the effect of the environment is encapsulated in the spectral density of the bath and, consequently, the thermal fluctuations in the coordinates of the harmonic oscillators  produce dephasing in the off-diagonal density matrix elements of the two-state system.

Equally important, the transverse driving of a two-state system  is a topic that bears similarities to frequency modulation. The excitation of a quantum system by an oscillating magnetic field was observed for the first time in 1939 in the experiments of Isidor Rabi with molecular beams aiming at measuring nuclear magnetic moments~\cite{Rabi39}. When this type of driving is resonant, the population oscillates between the ground state and the first excited state with the so-called Rabi frequency  which is determined by the strength of the effective coupling between the system and the field. This result was already obtained in a general form  by Majorana~\cite{Majorana32}, as well as in the early theoretical works by Rabi~\cite{Rabi37}. In a further development, stimulated by the need to understand the geometric mapping of quantum states introduced by Majorana~\cite{Bloch45}, Felix Bloch wrote the standard Bloch equations used widely to describe decoherence processes in NMR (nuclear magnetic resonance), and introduced the spherical representation for the state of a spin-1/2 particle~\cite{Bloch46}.

This review aims at presenting the phenomena that arise when the quantum mechanical energy levels are modulated. A large part of the review concentrates on the two-state system, on which different types of modulations are demonstrated. The treatment of other quantum systems is based on some selected examples rather than on a full coverage of recent literature. For standard topics such as Landau-Zener-St\"uckelberg-Majorana interference, spin-boson model, or tunneling, there already exists authoritative reviews --  see {\it e.g.} references~\cite{Shevchenko10,Grifoni98, Leggett87} -- but we will introduce some of the key results in such way that our presentation is self-contained. The theoretical considerations and calculations are rather general, and therefore they are in principle valid for any physical realization of the problem. However, in the most cases we also describe concrete experimental demonstrations of the discussed effects. 

Overall, a major driving force behind much of the recent experimental progress has been the intriguing possibility of using two-state systems as the basic building blocks of quantum computers and quantum simulators~\cite{Feynman82}. In a nutshell, the information in a quantum computer is stored and processed in logical bits formed by quantum two-state systems, i.e. qubits. Accordingly, the working principles of such a device have to be described in terms of quantum laws and, thus, the resulting behaviour relies on the concepts of interference and entanglement. It has been shown that if such a machine were ever to be built, it could speed-up certain computational tasks considerably~\cite{Deutsch92,Grover96,Shor97}. Different experimental approaches~\cite{Ladd10} have made the realisation of this dream plausible. However, improvements in the gate and measurement fidelities as well as increasing the coherence times are still essential. Naturally, since a quantum processor consists of many qubits, this raises also the question of the role of dephasing in a many-body system. Several experimental advances have been made here; this direction has just opened, and more will certainly be done in the following years.

Our examples come mainly from the contexts of superconducting circuits consisting of Josephson qubits and resonators or cavities, as well as degenerate ultracold gases. These technologies have experienced significant progresses recently, approaching and reaching the same level of maturity as the more established ones in the NMR and ion trap communities. Superconducting (Josephson) qubits are electrical circuits which behave as artificial atoms with a Hamiltonian constructed from various combinations of the Josephson energy, the capacitive charging energy, the inductive energy, and the energy associated with current biasing \cite{Makhlin01,You05,Clarke08,Schoelkopf08,Paraoanu14}. The aim is to construct a circuit that behaves quantum-mechanically as a two-state system or, in general, as a nonlinear multilevel system. The energy level separation can be biased to a fixed value but also modulated externally very fast around this value, such that the system can be excited to the higher levels. Similar degree of control is achieved with SQUID (superconducting quantum interference device) arrays and SQUID-terminated resonators, where the magnetic flux can be used to tune rapidly the SQUID inductance and thus the resonance frequency. The second class of systems that we employ extensively are Bose-Einstein condensed (BEC) ultracold atomic gases \cite{Dalfovo99,Leggett01,Bloch08}, which can be trapped in periodic optical lattices, thus realizing standard textbook solid-state models \cite{Bloch05,Bloch005}. In this case, the trapping potential is modulated, which provides a way to externally control the tunneling rate between the lattice sites. In the near future, there are many exciting prospects for novel realizations of the effects discussed in this review in systems such as nanomechanical resonators \cite{Aspelmeyer14} and nitrogen-vacancy (NV) centers in diamond \cite{Doherty13}, where significant experimental progress in fabrication and control has been achieved in the recent years.

In perspective, it is only a recent experimental achievement that the two-state physics was realized in single quantum systems with well-characterized qubits and full control of external parameters. In this sense, our review complements the standard results with relatively recent developments, especially triggered by advances in the experiments.
An important precursor to modern experiments has been the observation  of St\"uckelberg interference, which was made in the early 1990's with thermal Rydberg atoms driven by short microwave pulses~\cite{Baruch92}. In recent experiments on mesoscopic systems~\cite{Oliver05,Sillanpaa06,Wilson07,Izmalkov08,Sun09,LaHaye09,Lanting14}, it became possible to drive the Landau-Zener transitions and simultaneously monitor the time evolution of its quantum state, and also  to create artificial noise for the system, allowing the studies of dephasing under various types of fluctuations. Sometimes this brings a fresh view on classical results as well. For example, let us look at a spin-1/2 system under frequency modulation. By simply changing the basis through a $\pi/2$ rotation around the $y$-axis the frequency modulation will transform into a Rabi-type drive. Although extremely simple, this connection has only recently been studied systematically~\cite{Ashhab07} and observed experimentally using a nitrogen-vacancy spin in diamond~\cite{Zhou14}. Another example is the strongly nonadiabatic regime of frequency modulation, where the Landau-Zener transition formula cannot be applied~\cite{Mullen89,Vitanov96,Garraway97}. In this case, even for sudden modulations, St\"uckelberg interference appears but the formalism for calculating the interference maxima and minima has to be modified accordingly\cite{Silveri15}.

The paper is organized as follows. We start in section \ref{sec:background} with a brief introductory material on the perturbation theory in the interaction picture. This allows us to derive rather general expressions for the transition probability under a perturbation that shifts the energy levels of a multilevel system, and to calculate the corresponding transition rates. Then we continue toward analyzing specific quantum systems and discussing the corresponding experimental realizations.  First, in section \ref{sec:twostate} we review modulation effects in two-level systems by considering coherent and incoherent modulations of the transition energy and by introducing the concepts of sidebands, dynamical localization, Landau-Zener-St\"uckelberg-Majorana interference, dephasing, and effects of motional narrowing and averaging. We also discuss the experimental observation of the Berry phase in a superconducting circuit. In section \ref{sec:ho}, we address the harmonic oscillator and consider the creation of squeezed states by parametric modulation and continue the discussion on dephasing effects due to random variations of the frequency. In section \ref{sec:coupled} we proceed to discussing the effects in coupled systems. The paradigmatic model for such a system is the qubit-resonator system (Jaynes-Cummings model), where we show that a certain combination of modulations can be used to create two-qubit gates, or bring the system into the ultrastrong-coupling regime. We continue by examining complex quantum systems, such as, non-harmonic multilevel artificial atoms, many-qubit systems with fixed or tunable couplings, as well as atomic systems. Several recent experiments demonstrate the fertility of ideas related to frequency modulation, which has enabled the simulation of topological transitions both with superconducting circuits and with ultracold gases, and the observation of the effects of motional averaging with thermal atoms as well as with a logical qubit relayed between several physical qubits. Finally, in the last subsection we show how frequency modulation is used in quantum heat engines, and explore the connections with fundamental quantum thermodynamics concepts such as fluctuation theorems. Section~\ref{sec:conc} summarizes the results and presents future prospects.


\section{Background}\label{sec:background}

\subsection{Time-dependent Hamiltonian} \label{sec:timedep}
We study a quantum system where the Hamiltonian $\hat{H}(t)$ depends explicitly on time $t$.
Due to the time-dependence, the instantaneous eigenbasis $\{|\psi_n(t)\rangle\}$ of the Hamiltonian consists of the so-called \textit{adiabatic eigenstates}:
\begin{equation}\label{eq:instSE}
\hat{H}(t)|\psi_n(t)\rangle = E_n(t)|\psi_n(t)\rangle.
\end{equation}
For simplicity, we assume here that the spectrum of the Hamiltonian is discrete and non-degenerate. The relevant problem in \textit{quantum dynamics} is to solve the time-dependent Schr\"odinger equation
\begin{equation}\label{eq:TDSE}
\ii\hbar\frac{\partial}{\partial t}|\Psi(t)\rangle = \hat{H}(t)|\Psi(t)\rangle.
\end{equation}
In the adiabatic basis, the solution can be written as the linear combination
\begin{equation}\label{eq:solTDSE}
|\Psi(t)\rangle = \sum_n c_n(t)
|\psi_n(t)\rangle.
\end{equation}
The key point that separates time-dependent quantum systems from the time-independent ones is that the former can experience \textit{transitions} between the energy eigenstates. By a transition, we mean the following. Assuming that the system starts from one of the eigenstates, i.e. $|\Psi(0)\rangle = |\psi_n(0)\rangle$, the transition probability from $n$ to $m$ is defined as
\begin{equation}
p_{n\rightarrow m}(t) = |c_m(t)|^2 = |\langle \psi_m(t)|\Psi(t)\rangle|^2 = |\langle \psi_m(t)|\hat{U}(t,0)|\psi_n(0)\rangle|^2,
\end{equation}
where $\hat{U}(t,0)$ is the time-evolution operator (we will denote the initial time explicitly in order to distinguish the time-evolution operator from other time-dependent unitary transformations denoted with $\hat{U}(t)$). Thus, the transition probabilities can be found once the instantaneous eigenvalue problem for the Hamiltonian has been solved, and the unitary time-evolution is known. For a time-independent system, the transition probabilities are $p_{n\rightarrow m} = \delta_{nm}$ since the corresponding basis is formed by stationary states.

We solve equation~(\ref{eq:TDSE}) by making a unitary transformation $\hat{U}(t) = \sum_n |\psi_n(t)\rangle \langle n|$ into a static orthonormal basis $\{|n\rangle\}$. As a consequence, the time-dependent Schr\"odinger equation can be written as
\begin{equation}
\ii\hbar\sum_n \dot{c}_n(t) |n\rangle = \sum_n c_n(t)\left\{\hat{U}^{\dag}(t)\hat{H}(t)\hat{U}(t) + \ii\hbar \left[\partial_t\hat{U}^{\dag}(t)\right]\hat{U}(t)\right\}|n\rangle.
\end{equation}
We thus see that in the static basis the time-evolution of the probability amplitudes of the instantaneous eigenstates of $\hat{H}(t)$ is determined by the effective Hamiltonian
\begin{eqnarray}
\hat{H}_{\rm eff}(t) &=& \hat{U}^{\dag}(t)\hat{H}(t)\hat{U}(t) + i\hbar \left[\partial_t\hat{U}^{\dag}(t)\right]\hat{U}(t)\nonumber \\
&=& \sum_n\left[E_m(t)-i\hbar \langle \psi_n(t)|\partial_t \psi_n(t)\rangle\right] |n\rangle\langle n| \nonumber\\
&& - i\hbar \sum_{n\neq m}\frac{\langle\psi_n(t)|\partial_t \hat{H}(t)|\psi_m(t)\rangle}{E_m(t)-E_n(t)}|n\rangle \langle m|.\label{eq:solcoords}
\end{eqnarray}
Above, we have used the relation
\begin{equation}
\langle \psi_n(t)|\partial_t \psi_m(t)\rangle = \frac{\langle\psi_n(t)|\partial_t \hat{H}(t)|\psi_m(t)\rangle}{E_m(t)-E_n(t)},
\end{equation}
which holds for $m\neq n$ and results from differentiating equation~(\ref{eq:instSE}). We thus see that the time-dependence in the Hamiltonian $\hat{H}(t)$ causes non-adiabatic transitions between its instantaneous eigenstates and the strengths of the transitions are given by $\hbar|\langle\psi_n(t)|\partial_t \hat{H}(t)|\psi_m(t)\rangle|/|E_m(t)-E_n(t)|$ with $m\neq n$.

When the system starts from the adiabatic state $|\psi_n(t)\rangle$ and the time-evolution is slow, i.e.
\begin{equation}\label{eq:adiabatic}
\left| \frac{\langle \psi_n(t)|\partial_t \hat{H} (t)|\psi_m(t)\rangle}{E_m(t)-E_n(t)}\right| \ll \frac{|E_m(t)-E_n(t)|}{\hbar} \ \ \textrm{for all } m\neq n,
\end{equation}
one can neglect the non-adiabatic transitions and, consequently, the system follows the instantaneous eigenstate in time~\cite{Schiff}. This is called \textit{adiabatic} time-evolution. Accordingly, the transitions are suppressed and the adiabatic eigenstate $\ket{\Psi(t)}=\ee^{\ii \phi_n(t)}\ket{\psi_n(t)}$ gathers only the phase 
\begin{equation}\label{eq:geometricphase}
\phi_n(t)=-\frac{1}{\hbar}\int_0^t E_n(\tau)d\tau+\gamma_n(t),
\end{equation}
where $t=0$ defines the initial time of the problem.
The phase is expressed as the sum of the dynamic phase and a geometric phase $\gamma_n=\ii\int_0^t \left \langle \psi_n(\tau)|\partial_t \psi_n(\tau)\right\rangle d\tau$, which only depends on the traveled path. When the time-evolution is adiabatic and cyclic the collected geometric phase is called Berry's phase~\cite{Berry84}.

The adiabatic approximation is not valid if $\hat H(t)$ is varied sufficiently fast so that the adiabatic condition~(\ref{eq:adiabatic}) does not hold. In such cases, the population can ``leak out'' from the initial state, resulting in general in $p_{n\rightarrow m}(t)\neq \delta_{nm}$. Based on equation~(\ref{eq:adiabatic}), the strongest deviations from adiabaticity occur between such states $n$ and $m$ that are close in energy, i.e. $E_n(t)\approx E_m(t)$. Often, in this case one can restrict the discussion of the dynamics to the subspace spanned by such two (nearly) degenerate states.

Adiabatic following of an instantaneous eigenstate of $\hat{H}(t)$ is always an approximation in a truly time-dependent system. For some applications (see sections~\ref{sec:sbsfm} and~\ref{s.thermal}) relying, e.g., on the creation of specific target states by adiabatic following, this can turn out to be a real problem as the non-adiabatic processes can restrict the fidelity of the procedure considerably. However, there are several methods to exactly restore the adiabatic time-evolution, generically called shortcuts to adiabaticity~\cite{Torrontegui2013117}. Berry, Demirplak, and Rice~\cite{Demirplak03,Demirplak05,Demirplak08,Berry09} proposed to add a correction $\hat{H}_{\rm CD}(t)$ to the Hamiltonian $\hat{H}(t)$, such that the resulting time-evolution determined by $\hat{H}(t)+\hat{H}_{\rm CD}(t)$ keeps the system exactly in the adiabatic state $|\psi_n(t)\rangle$. One can see immediately from equation~(\ref{eq:solcoords}) that this can be achieved in the static basis by choosing
\begin{equation}
\langle \psi_n(t) | \hat H_{\rm CD}(t) |\psi_m(t)\rangle = \ii \hbar \frac{\langle \psi_n(t)|\partial_t \hat H(t) |\psi_m(t) \rangle}{E_{m}(t)-E_{n}(t)}, \label{eq:CD_formula}
\end{equation}
for $n\neq m$, and by setting the diagonal elements $\langle \psi_{n}(t)|\hat{H}_{\rm CD}(t)|\psi_n(t)\rangle = 0$.

For a major part of this review we restrict to studying systems that have a time-dependent Hamiltonian of the following form,
\begin{eqnarray}
\hat H(t)=\hat H_0(t)+\hat H_{\rm C}+\hat H_{\rm P}(t)\label{eq:modHamtot},\\
\label{eq:modHam}
\hat{H}_0(t) = \hbar\sum_n [\omega_n + \xi_n(t)]|n\rangle\langle n|,\\
\label{eq:modHamC}
\hat{H}_{\rm C} = \hbar\sum_{n,m}\Delta_{nm}|n\rangle\langle m|.
\end{eqnarray}
Here $|n\rangle$ is a set of orthonormal states with index $n=1,2,\ldots,N$. The first term $\hat H_0(t)$ shows that their energies fluctuate $E_n(t)=\hbar[\omega_n+\xi_n(t)]$ because of the time dependent classical modulation $\xi_n(t)$. The coupling term $\hat{H}_{\rm C}$ describes possible static couplings between the states with constant coupling amplitudes $\Delta_{nm}$. The probe term $\hat{H}_{\rm P}(t)$ describes the effect of a sinusoidal field that is used to study the properties of the Hamiltonian $\hat H_0(t)+\hat H_{\rm C}$. The probe $\hat{H}_{\rm P}(t)$ is assumed to be a weak perturbation, but no such assumption is made concerning the  modulation $\xi_n(t)$.
For the probe we use
\begin{equation}\label{eq:probeHam}
\hat{H}_{\rm P}(t)=\hbar g_{\rm P} \cos(\omega t) \hat{V},
\end{equation}
where $g_{\rm P}$ and $\omega$ describe the strength and the angular frequency of the probe, and $\hat{V}$ is a hermitian operator that induces transitions between the states $|n\rangle$. Possible diagonal components in the probe do not cause transitions between the states, and can thus be safely included in $\hat{H}_0(t)$.
On top of the Hamiltonian~(\ref{eq:modHamtot}), we sometimes assume a weak coupling to environment, which allows the system to relax and decohere.

\subsection{Perturbation theory in the interaction picture}\label{sec:ptip}
In this section we derive a formula [equation~(\ref{eq:Trate})] for the transition rate caused by the probe. Here we neglect the coupling term $\hat H_{\rm C}$~(\ref{eq:modHamC}). The case of  a finite $\hat H_{\rm C}$ will be addressed later in section~\ref{sec:scafa}. The standard procedure to calculate the effect of a weak perturbation~\cite{Sakurai} is to write the Hamiltonian in the interaction picture, where the diagonal part is removed. This can be achieved with the time-dependent unitary transformation $\hat{H}^{(\rm I)}=\hat{U}^{\dag}\hat{H}\hat{U} + {\rm i}\hbar\big(\partial_t\hat{U}^{\dag}\big)\hat{U}$, where
\begin{equation}\label{eq:iframe}
\hat{U}(t)=\exp\left(-\frac{\rm i}{\hbar}\int_{0}^t \hat H_0(\tau)d\tau\right)=\exp\left(-{\rm i}\sum_n \left[\omega_n t + \zeta_n(t)\right]|n\rangle\langle n|\right).
\end{equation}
Here $\zeta_n(t)\equiv \int_{0}^t \xi_n(\tau)d\tau$ is the dynamical phase accumulated due to the modulations. This modulation phase turns out to be very important quantity when one determines the probe response of the modulated transition.

After the transformation, the Hamiltonian of the probed and modulated quantum system is
\begin{equation}\label{eq:hintpt}
\hat{H}^{(\rm I)}(t) = \frac{\hbar g_{\rm P}}{2}\left(e^{i\omega t}+e^{-i\omega t}\right)\hat{V}(t).
\end{equation}
In the above and what follows, we denote the Hamiltonians in the interaction picture with (I) because of the explicit time-dependence in the Schr\"odinger picture. For other operators, the explicit time dependence in the interaction picture can be written simply as, e.g., $\hat{V}(t) = \hat{U}^{\dag}(t)\hat{V}\hat{U}(t)$. Now, if the probe is weak compared to the transition energies ($g_{\rm P}\ll |\omega_{m}-\omega_{n}|$), we can solve the time-dependent Schr\"odinger equation using the first order perturbation theory for the time-evolution operator $\hat{U}^{(\rm I)}(t,0)$ generated by the Hamiltonian of equation~(\ref{eq:hintpt}). We obtain in the first-order in the perturbation parameter $g_{\rm P}$:
\begin{equation}
\hat{U}^{(\rm I)}(t,0) = 1 - \frac{\ii}{\hbar} \int_0^t\hat{H}^{(\rm I)}(t') dt' + \mathcal{O}(g_{\rm P}^2).
\end{equation}
Let us first consider transitions from initial state $|\psi(0)\rangle = |n\rangle$. After time $t$ has passed, the time-evolved state can be written in the interaction picture as $|\psi(t)\rangle = \sum_m c_{n\rightarrow m}(t)|m\rangle$, where the amplitude of the transition into state $|m\rangle$ ($m\neq n$) is defined as
\begin{eqnarray}
c_{n\rightarrow m}(t) &\equiv & \langle m|\hat{U}^{(\rm I)}(t,0)|n\rangle \nonumber \\
&=& - \frac{\ii}{\hbar} \int_0^t \langle m|\hat{H}^{(\rm I)}(t') |n\rangle dt'+\mathcal O(g_{\rm P}^2).
\end{eqnarray}
Substituting in the expressions~(\ref{eq:iframe}) and~(\ref{eq:hintpt}) we find that the integrand consists of a sum of oscillating terms. Out of these only the slowly oscillating terms (also called secular terms) lead to a substantial transition probability. Therefore we make the \textit{rotating wave approximation} (RWA), where we neglect the fast oscillating terms.  Except in special cases (such as $\omega\approx 0$), only one of the two terms in equation~(\ref{eq:hintpt}) is important for the transition from one given state to another. Therefore we keep only the term proportional to $e^{-i\omega t}$ but note that the other term can be selected simply by reversing the sign of $\omega$. 

The transition probability from an arbitrary initial state $|\psi(0)\rangle = \sum_n c_n(0)|n\rangle$ into the state $|m\rangle$ can be written as
\begin{equation}
c_{m}(t) \equiv \sum_{n} c_n(0)c_{n\rightarrow m}(t)
\end{equation}
where we can include also the $n=m$ term when the probe is off-diagonal.
Summing over all final states, the total transition probability  is
\begin{eqnarray}
P_\omega(t) &=& \sum_m |c_m(t)|^2 \nonumber \\ &=&  \frac{g_{\rm P}^2}{4} \int_0^tdt_1 \int_0^tdt_2\,  e^{\ii\omega(t_1-t_2)}\left\langle\hat{V}(t_1)\hat{V}(t_2)\right\rangle+\mathcal O \left (g_{\rm P}^3\right),
\end{eqnarray}
where the expectation value is calculated with respect to the density matrix $\hat{\rho}_0=|\psi(0)\rangle\langle \psi(0)|=\sum_{n,m} c_n(0)c_{m}^*(0) |n\rangle\langle m|$.
We make a change to new variables $\tau=t_1-t_2$ and $T=(t_1+t_2)/2$ and define
\begin{eqnarray}
B(T)=\left\{
\begin{array}{ll}
T & {\rm if\ } T < t/2,\\
t-T &  {\rm if\ } T > t/2.
\end{array} \right.
\end{eqnarray}
We get
\begin{eqnarray}
P_{\omega}(t)=\frac{g_{\rm P}^2}{4}\int_0^tdT\int_{-B(T)}^{B(T)}d\tau\, e^{\ii\omega \tau}\left\langle \hat{V}(T+\tau/2) \hat{V}(T-\tau/2)\right\rangle+\mathcal O\left(g_{\rm P}^3\right).
\end{eqnarray}
We are interested in the long-time behavior of $P_{\omega}(t)$. The coupling of the system to the environment (see below) makes the correlation to decay for large $\tau$. In the limit of large $t$  the $T$ integration gives linear growth in time and leads to the transition rate
\begin{equation}\label{eq:Trate}
S(\omega)\equiv\frac{dP_\omega}{dt}= \frac{g_{\rm P}^2}{4} \int_{-\infty}^{\infty}  e^{\ii\omega\tau} \left\langle \hat{V}(\tau)\hat{V}(0)\right\rangle d\tau.
\end{equation}
Here positive $\omega$ corresponds to absorption and negative to emission~\cite{Clerk10}.  We see that the transition rate in the modulated system is given by the Fourier transform of the stationary two-time correlation function of the probe Hamiltonian expressed in the interaction picture. In other words, the transition rate is given by the power spectral density of the probe operator evaluated in the unperturbed system. Notice that in the absence of modulation, the transition rate reduces to that given by Fermi's golden rule~\cite{Sakurai}:
\begin{equation}
S(\omega) = \frac{\pi g_{\rm P}^2}{2}\sum_m |\langle m|\hat{V}|n\rangle|^2 \delta(\omega_m - \omega_n-\omega)
\end{equation}
when $\hat{\rho}_0=|n\rangle\langle n|$.

\subsection{Dynamics of open quantum systems}\label{Sec:dynamicsopen}
There is always some coupling between the studied quantum system and its environment. Because of the coupling, the system and the environment become entangled and the open quantum system cannot be in general described in terms of a single quantum state $|\psi\rangle$. Instead, the standard procedure is to introduce a density operator $\hat\rho=\sum_i p_i|\psi_i\rangle\langle\psi_i|$, where $p_i$ is the probability that the system is in the state $\ket{\psi_i}$. All  measurable information about the physical state of the system is contained in the density operator. For example, the expectation value of an operator $\hat O$ is given by $\langle \hat O\rangle=\mathop{\rm Tr}(\hat\rho\hat O)$.
Typically, one is interested in the time development of expectation values and correlators for operators, \textit{cf.}~equation~(\ref{eq:Trate}). These are solved by finding `the master equation' which gives the time development of the density operator of the system obtained by tracing out the dynamics of the environment.

We sketch here the derivation of the master equation, for more details, please see references~\cite{WallsMilburn, GardinerZoller}. The total Hamiltonian consists of a system $\hat H_{\rm S}$ and an environment $\hat H_{\rm E}$ interacting through $\hat{V}_{\rm SE}$: $\hat{H}=\hat H_{\rm S}+\hat H_{\rm E}+\hat V_{\rm SE}$. In the interaction picture defined by $\hat H_{\rm S}+\hat H_{\rm E}$, the time-evolution of the total density operator $\hat w$ and its formal solution in the first iterative order can be written as
\begin{eqnarray}
&\frac{\D{\hat w^{(\rm I)}(t)}}{\D{t}}=\frac{1}{\ii\hbar}[\hat V_{\rm SE}(t), \hat w^{(\rm I)}(t)], \label{eq:w}\\
&\hat w^{(\rm I)}(t)=\hat w^{(\rm I)}(0)+\frac{1}{\ii \hbar}\int^t_0 [\hat V_{\rm SE}(\tau), \hat w^{(\rm I)}(\tau)] \D{\tau}. \label{eq:witer}
\end{eqnarray}
We iterate the formal solution~(\ref{eq:witer}) to the second order in the coupling $\hat V_{\rm SE}$. We assume that the interaction is non-diagonal $0={\rm Tr}_{E}[\hat V_{\rm SE} \hat w(0)]$ and that the environment is large and fast with respect to system dynamics $\hat w(t)\approx \hat \rho(t)\otimes \rho_{\rm E}$. As a result, we obtain the Born-Markov master equation:
\begin{equation}
  \frac{\D{\hat \rho^{(\rm I)}(t)}}{\D{t}}=-\frac{1}{\hbar^2}\int_0^\infty{\rm Tr}_{E}\left(\left[\hat V_{\rm SE}(\tau), \left[\hat V_{\rm SE}(t-\tau), \hat \rho^{(\rm I)}(t)\otimes\hat \rho^{(\rm I)}_{\rm E}\right]\right]\right) \D{\tau}.~\label{eq:BM-master}
\end{equation}
Here, the density operators for the system and the environment are $\hat \rho=\mathop{\rm Tr}_{\rm E} \hat w$ and $\hat \rho_E$, respectively, where $\mathop{\rm Tr}_{\rm E}$ denotes the trace over the environment.

When deriving the master equation for a system that is explicitly time dependent $\hat H_{\rm S}=\hat H_{\rm S}(t)$, one needs to notice that there can be interference between dynamics introduced by the driving and the environment~\cite{Pekola10,Salmilehto10, Salmilehto11}. Especially, when the typical time scale $\tau_{\rm S}$ of the system modulation is comparable with the correlation time of the environment $\tau_{\rm E}$, the effect is pronounced~\cite{Mukamel78, JianLi12}. In the limit $\tau_{\rm E}\ll \tau_{\rm S}$, the two dynamics separate and one can introduce a master equation with instantaneous time-dependent or time-averaged decay rates. In this review, we focus on the quantum effects of the frequency modulation in the system itself and omit deeper discussions on the dynamics of open quantum system in the presence of frequency-modulation.

Conventionally, the environment is modeled as a large set of harmonic oscillators $\hat H_{\rm E}=\sum_j \hbar \omega_j \hat a^\dag_j \hat a_j$, where $\omega_j$ and $\hat a_j$ are the frequency and the annihilation operator for the microscopic oscillator $j$. We denote the Pauli spin matrices with $\hat \sigma_i$, where $i=x,y,z$, and $\hat \sigma_-$ is the annihilation operator for the qubit. Let us consider a qubit that interacts with this environment through the coupling
\begin{equation}
  \hat{V}_{\rm SE}=\hbar \sum_j g_j   \left(\hat{a}_j+\hat{a}^\dagger_j \right)\hat X=\hbar \sum_j g_j   \left(\hat{a}_j\hat \sigma_++\hat{a}^\dagger_j \hat \sigma_-\right), \label{eq:coupling}
\end{equation}
where $g_j$ is the interaction strength with the $j$th environmental oscillator, $\hat X$ is a system operator, and in the second equality we have assumed that $\hat X=\hat \sigma_x$ and applied the rotating wave approximation. For an environment at an effective zero temperature, the master equation~(\ref{eq:BM-master}) can be written for $\hat X=\hat \sigma_x$ and in the secular approximation as
\begin{eqnarray}
\frac{\D{\hat{\rho}}}{\D{t}}=-\frac{\ii}{\hbar}[\hat{H},\hat{\rho}]+\mathcal{D}\left(\sqrt{\Gamma_1}\hat \sigma_-\right)\hat\rho,
\label{eq:mast1dis}\end{eqnarray}
where the dissipator $\mathcal{D}(\hat c)\hat \rho=\hat c\hat \rho\hat c^\dagger-\frac{1}{2}\{\hat c^\dagger \hat c, \hat \rho\}$. This equation describes a qubit with an energy relaxation rate $\Gamma_1$, determined by the coupling strength and the density of the states of the environment. Essentially, an excitation decays in a characteristic relaxation time $T_1=1/\Gamma_1$ and the energy is released to the environment. In addition to the energy relaxation, there is usually a loss of coherence that originates from an unwanted coupling in the longitudinal direction, formally $\hat X = \hat \sigma_z$ in equation~(\ref{eq:coupling}). This yields an additional pure dephasing term
\begin{eqnarray}
\frac{\D{\hat{\rho}}}{\D{t}}=-\frac{\ii}{\hbar}[\hat{H},\hat{\rho}]+\mathcal{D}\left(\sqrt{\Gamma_1}\hat \sigma_-\right)\hat \rho +\frac 1 2 \mathcal{D}\left(\sqrt{\Gamma_\phi}\hat \sigma_z\right)\hat \rho, \label{master.sin.rwa}
\end{eqnarray}with the pure dephasing rate $\Gamma_\phi$. We also define the dephasing rate $\Gamma_2=\Gamma_1/2+\Gamma_\phi$. Similarly, one can derive a master equation for a harmonic oscillator with a dissipator $\mathcal{D}(\sqrt{\kappa}\hat a)$ for cavity decay with the rate $\kappa$ by photon losses.

In equation~(\ref{eq:coupling}), we made the rotating wave approximation before the trace by neglecting the fast-rotating terms, such as $\hat a_j^\dagger \hat \sigma_+$, in the system-environment coupling $\hat{V}_{\rm SE}$. Another possibility would have been to neglect the fast-rotating terms in the master equation of the reduced density operator obtained after the trace. Depending on the desired information about the open system there can, in general, be a difference between the two ways of introducing the rotating wave approximation~\cite{Fleming10}. In this review we are interested in the accuracy of the relaxation rates in the weak coupling regime, where there is no significant difference between the two approximations. However, in the strong coupling limit, the antirotating terms become significant and they can introduce interesting quantum phenomena ranging from quantum irreversibility and chaos~\cite{Bonci91, Emary03} to dynamical Casimir effect~\cite{Dodonov01, Liberato07, Wilson11, Lahteenmaki13}. For the balance of the paper, we concentrate on the dynamics of a frequency-modulated system and, among others, in section~\ref{Sec:cohmod} we discuss the applicability of RWA in the case of coherent modulation of energy levels.


\section{Two state system}\label{sec:twostate}

In this section we consider the case when the relevant state space consists of only two basis states. This is often a good approximation in cases where the transition energy between the two states differs from the transition energies to other states. Such nonlinear systems occur naturally in atomic and molecular systems, and thus the two state model has a long history in NMR~\cite{Abragam} and in quantum optics~\cite{EberlyAllen87}. More recently artificial two state systems have been realized using mesoscopic structures, and especially superconducting circuits allow great flexibility in the experiments as their parameters can be widely varied in the fabrication process and during the measurement~\cite{Clarke08}. A two state system is also known as two level system, or a qubit, meaning a quantum bit. As long as the upper state probability is small, the behaviour of a two-state system is identical to the two lowest levels of a harmonic oscillator. Difference arises when the third and higher levels of the harmonic oscillator are excited, whereas ``saturation'' takes place in the two-state system.

We study the effect of level spacing modulation in a two state system. We apply the general Hamiltonian~(\ref{eq:modHamtot})-(\ref{eq:probeHam}) but limit the discussion first to the case without the static coupling, $\hat H_{\rm C}=0$.  The Hamiltonian is written as 
\begin{eqnarray}\label{ham.first}
\hat{H}(t)&=&\frac{\hbar}{2}\big[\omega_0+\xi(t)\big]\hat{\sigma}_z +\hbar g_{\rm P}\cos(\omega t)\hat{\sigma}_x. 
\end{eqnarray}
The eigenstates of $\hat{\sigma}_z$ are called diabatic states. The diabatic states corresponding to eigenvalues +1 and -1  are denoted with $|\uparrow\rangle$ and $|\downarrow\rangle$, respectively.
The first term in~(\ref{ham.first}) implies that their time dependent energy difference is $\hbar\omega_0+\hbar\xi(t)$. The second, probe term generates transitions between the diabatic states.

\begin{figure}[t!]
\begin{center}
\includegraphics[width=0.5\linewidth]{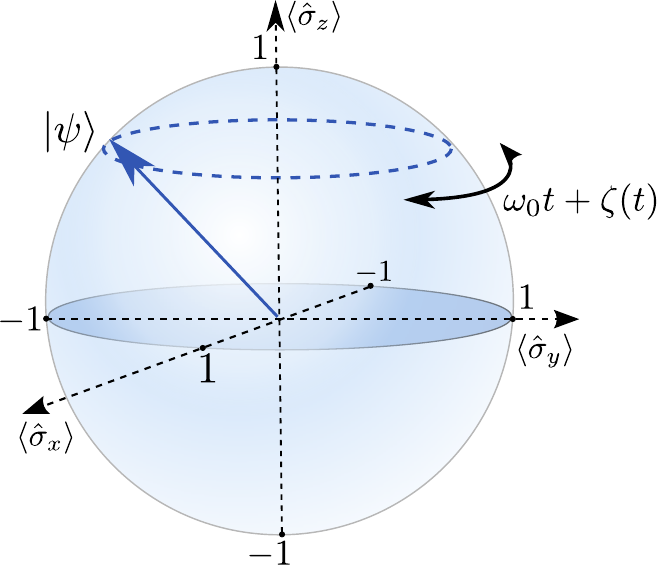}
\caption{Bloch sphere representation of the state of a two-level system with modulated transition frequency.} \label{fig:blochsphere}
\end{center}
\end{figure}

One way to visualize the state of the system is by mapping it onto a two dimensional surface called Bloch's sphere (figure~\ref{fig:blochsphere}).
In this representation, the state is expressed by the expectation values of the $\hat{\sigma}_i$ operators, which give the coordinates of the state vector on three orthogonal axes corresponding to  $i=x,y,z$. 
The diagonal $\hat \sigma_z$ part of the Hamiltonian~(\ref{ham.first})  leads to a rotation of an arbitrary state $|\psi\rangle$ around the $\langle \hat \sigma_z\rangle$-axis by the azimuthal angle $\omega_0 t+\zeta(t)$, where
\begin{equation}\label{eq:zetatss}
\zeta(t)\equiv \int_0^t \xi(\tau)d\tau
\end{equation}
is the time-dependent phase caused by the modulation. Bloch's sphere has unit radius for a normalized, pure quantum state.

In the next section, we will discuss the response of the modulated transition to the probe in terms of the time-dependent perturbation theory. In section~{\ref{Sec:cohmod}}, we apply the perturbative result to the case of coherent modulation. The problem of coherent continuous modulation will be discussed both in the case of weak and strong amplitude modulation. In the former case, the relevant equations are greatly simplified with the application of the RWA but, in the latter, one has to introduce the so-called Floquet approach so that the contributions from the terms neglected in the RWA (so-called generalized Bloch-Siegert shifts) are properly included. Also, a complementary discussion in terms of discretized time-evolution will be given 
in section \ref{Sec:disctimeevo}. Section~\ref{Sec:stochmod} is devoted to the application of the perturbative result into the case of incoherent modulation.

\subsection{Probe spectrum}\label{sec:probetl}

We analyse the two-state Hamiltonian~(\ref{ham.first}) in terms of time-dependent perturbation theory presented in section\ \ref{sec:ptip}. For a two-state system, the unitary transformation~(\ref{eq:iframe}) to the interaction picture takes the form
\begin{equation}
\hat{U}(t)=e^{-\ii\hat{\sigma}_z[\omega_0 t + \zeta(t)]/2}.
\end{equation}
It is useful to define qubit raising and lowering operators, $\hat{\sigma}_+\equiv |\uparrow\rangle\langle \downarrow|$ and $\hat{\sigma}_-\equiv |\downarrow\rangle\langle \uparrow|$, respectively. In the interaction picture they get the time dependence 
\begin{equation}\label{eq:Ispm}
\hat{\sigma}_{\pm}(t) = \hat{U}^\dagger \hat{\sigma}_{\pm}\hat{U}=e^{\pm \ii [\omega_0 t + \zeta(t)]}\hat{\sigma}_{\pm}.
\end{equation}
Corresponding to~(\ref{eq:hintpt}), the Hamiltonian~(\ref{ham.first}) is transformed to
\begin{equation}\label{eq:I2stateHam}
\hat{H}^{(\rm I)}(t) = \hbar g_{\rm P}\cos(\omega t)\left[A(t)e^{i\omega_0 t}\hat{\sigma}_+ + A^*(t)e^{-i\omega_0 t}\hat{\sigma}_-\right],
\end{equation}
where $A(t)\equiv e^{i\zeta(t)}$ is the dynamical phase factor arising from the modulation.
The transition rate can be adapted from equation~(\ref{eq:Trate}):
\begin{equation}
S(\omega)= \frac{g_{\rm P}^2}{4} \int_{-\infty}^{\infty} e^{\ii\omega t}\left\langle \hat{\sigma}_x(t)\hat{\sigma}_x(0)\right\rangle  dt.
\end{equation}
We concentrate here on the absorptive transitions, which allows us to write the absorption rate in the RWA as
\begin{equation}\label{eq:SpecQubit}
S(\omega)=\frac{g_{\rm P}^2}{4}\int_{-\infty}^{\infty}e^{\ii\omega t}\langle \hat{\sigma}_-(t)\hat{\sigma}_+(0)\rangle \D{t}.
\end{equation}

In order to take into account the coupling with the environment, we need the master equation (section\ \ref{Sec:dynamicsopen}). For a two state system the master equation takes the form~\cite{WallsMilburn}
\begin{eqnarray}
\frac{\D{\hat{\rho}}}{\D{t}}=-\frac{\ii}{\hbar}[\hat{H}(t),\hat{\rho}]+\mathcal{D}\left(\sqrt{\Gamma_1}\hat \sigma_-\right)\hat \rho +\frac 1 2 \mathcal{D}\left(\sqrt{\Gamma_\phi}\hat \sigma_z\right)\hat \rho,
\label{master.sin.rwa2}
\end{eqnarray}
which includes the relaxation with the rate $\Gamma_1$ and the pure dephasing with the rate $\Gamma_\phi$. Similarly as above in subsection \ref{Sec:dynamicsopen}, the total dephasing rate is denoted by $\Gamma_2=\Gamma_1/2+\Gamma_\phi$..

The master equation~(\ref{master.sin.rwa2}) allows us to calculate the two-time correlator by using the quantum regression theorem~\cite{WallsMilburn}. The result can be written~\cite{JianLi12}
\begin{equation}\label{eq:qspec}
S(\omega)=\frac{g_{\rm P}^2}{4}\int_{-\infty}^{\infty}e^{\ii(\omega-\omega_0)t-\Gamma_2|t|}\Big\langle e^{-i\zeta(\tau)}\Big\rangle_{\xi} \D{t},
\end{equation}
where we have added averaging over different realisations of $\xi$, which becomes relevant under incoherent fluctuations. The dynamical phase $\zeta(t)$ in equation~(\ref{eq:zetatss}), accumulated due to the fluctuations, is evidently an important quantity, since it determines the spectrum completely. We will emphasize this property in our discussions in the following sections. In the following, we will consider the spectrum~(\ref{eq:qspec}) in the cases of deterministic (coherent) and random (incoherent) modulations $\xi(t)$, together with related novel experiments.

\subsection{Coherent modulation}
\label{Sec:cohmod}

A coherent modulations of the transition frequency can be further categorised in terms of the continuity of the time-dependent part of the Hamiltonian. Continuous modulations (as opposed to stepwise modulations) form an important class of modulations, especially because atomic or molecular systems are typically studied by perturbing the atomic energy levels with electromagnetic radiation via the dipole moment of atoms~\cite{Sakurai}, leading to sinusoidal time-dependence in the semi-classical approximation.

Coherent modulation of the transition frequency occurs when $\xi(t+2\pi/\Omega)=\xi(t)$, i.e.\ when the modulation is correlated with itself at all times $t$. Such $\frac{2\pi}{\Omega}$ - periodic modulations have attracted a lot of theoretical and experimental interest in the context of two-level systems~\cite{Noel98, Grifoni98,Shevchenko10,Ashhab07,Grossmann91, Metcalfe10, Childress10, Oliver05,Sillanpaa06,Wilson07,Saiko06}. Here, we review the subject, show the sideband formation and demonstrate the longitudinal sideband control of qubit state populations. In addition, the periodic modulation serves as an introduction to the stochastic modulation of the transition frequency considered in section~\ref{Sec:stochmod}.

We assume that the static part $\bar{\xi}\equiv\int_0^{2\pi/\Omega}\xi(t)dt$ has been included into $\omega_0$. Thus, also the dynamical phase becomes periodic, $\zeta(t+2\pi/\Omega)=\zeta(t)$, and the dynamical phase factor $A(t)=e^{i\zeta(t)}$ has the Fourier series representation
\begin{equation}
A(t) = \sum_{n=-\infty}^{\infty} \Delta_n e^{\ii n \Omega t},
\end{equation}
where the Fourier coefficients $\Delta_n$ are obtained as
\begin{equation}\label{eq:phaseFC}
\Delta_n = \frac{\Omega}{2\pi}\int_0^{2\pi/\Omega} e^{-\ii n \Omega t}e^{\ii \zeta(t)} dt.
\end{equation}
In the interaction picture, Hamiltonian~(\ref{eq:I2stateHam}) can be written as
\begin{eqnarray}
\hat{H}^{(\rm I)}(t)&=& \frac{\hbar g_{\rm P}}{2}\sum_{n=-\infty}^{\infty}\Delta_n\left[e^{\ii (\omega_0+n\Omega+\omega)t}+ e^{\ii (\omega_0+n\Omega-\omega)t}\right]\hat{\sigma}_+\nonumber\\
&&+\frac{\hbar g_{\rm P}}{2}\sum_{n=-\infty}^{\infty}\Delta^*_n\left[e^{-\ii (\omega_0+n\Omega+\omega)t}+e^{-\ii (\omega_0+n\Omega-\omega)t}\right]\hat{\sigma}_-\label{eq:rotHam}.
\end{eqnarray}
We thus see that the modulations in the transition energy are transformed into an effective transverse multi-photon drive of the qubit. We also notice that the effective coupling strength to a multi-photon mode $m$ is $g_{\rm P}|\Delta_m|$. In the Bloch sphere picture, this transformation corresponds to moving into a frame that rotates around the $z$-axis at the non-uniform and time-dependent angular velocity $\zeta(t)$.

\begin{figure}[t!]
\begin{center}
\includegraphics[width=0.7\linewidth]{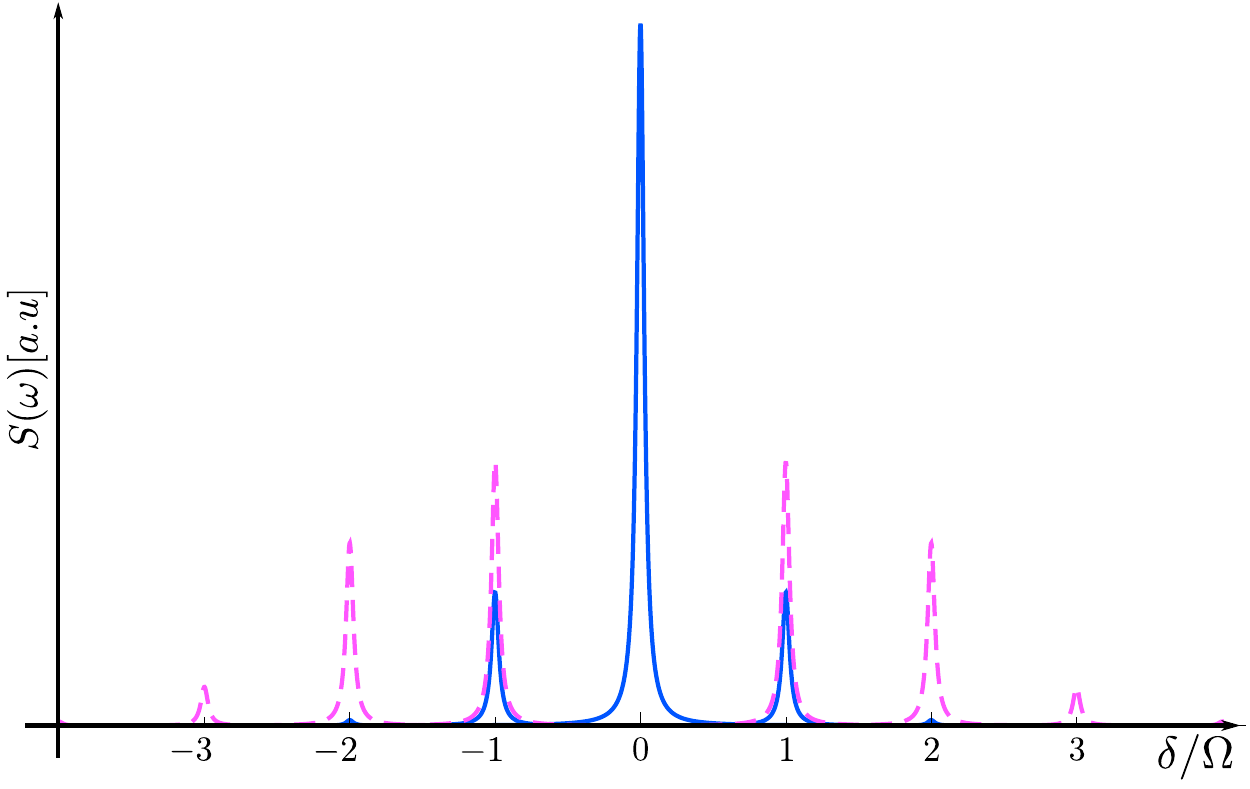}
\caption{Spectrum of the sinusoidally modulated transition as a function of the detuning $\delta$. For the low modulation amplitude $\xi/\Omega = 0.8$, five resonances ($m=-2,-1,0,1,2$) are visible (blue). When the amplitude is increased to $\xi/\Omega=2.4048$ one observes two additional peaks ($m=-3,3$) and the disappearance of the $m=0$ peak due to CDT (magenta). The plot is based on equations~(\ref{eq:rwaspec}) and~(\ref{eq:sinmodj}) and the width of the resonances is determined by $\Gamma_2/\Omega = 0.03$.}  \label{fig:specsimple}
\end{center}
\end{figure}

Analytical calculation of the spectrum for the Hamiltonian~(\ref{eq:rotHam}) is difficult in the general case. Nevertheless, close to a multi-photon resonance, $\omega\approx\omega_0+m\Omega$, when the strengths of the other driving fields are weak enough, $g_{\rm P}|\Delta_n|<\Omega$ for $n\neq m$, one can make the rotating wave approximation (RWA) by neglecting all non-resonant driving fields. The resulting RWA-Hamiltonian can be written as
\begin{equation}\label{eq:rwa}
\hat{H}^{(\rm I)}_{\rm RWA}(t) = \frac{\hbar g_{\rm P}}{2}\left[\Delta_m e^{\ii (\omega_0+m\Omega-\omega)t}\hat{\sigma}_+ +\Delta_m^* e^{-\ii(\omega_0+m\Omega-\omega)t}\hat{\sigma}_-\right].
\end{equation}
As a consequence, one can calculate the spectrum for each mode $m$ separately, and by adding the contributions from all modes, one obtains the so-called multi-photon sideband~\cite{Haroche70} spectrum:
\begin{equation}\label{eq:rwaspec}
S(\omega) = \sum_{m=-\infty}^{\infty}\frac{g_{\rm P}^2}{2}\frac{\Gamma_2|\Delta_m|^2}{(\delta+m\Omega)^2+\Gamma_2^2},
\end{equation}
where $\delta\equiv \omega_0-\omega$ is the detuning between the static qubit and the probe. The sufficient criterion for such resolvable sidebands is that the spacing between the adjacent bands has to be larger than their linewidths, i.e. $\Omega > \Gamma_2$. We also notice that the $m$:th sideband disappears whenever $\Delta_m=0$. This bears similarity to the coherent destruction of tunnelling (CDT), discussed in references~\cite{Grossmann91,Kayanuma94} and observed in numerous experiments with an additional static coupling of stationary states. CDT is closely related to the phenomenon of ``dynamic localization''~\cite{Dunlap86, Grifoni98}. The rotating wave approximation generally becomes insufficient as the probe amplitude $g_{\rm P}$ is increased. As a consequence, the resonance locations are shifted due to the enhanced contributions of the counter-rotating (Bloch-Siegert shift~\cite{BlochSiegert40}) and other multi-mode (generalized Bloch-Siegert~\cite{Tuorila10, Pietikainen16}) terms. However, in the case of direct probing of modulated transition energy, one can typically assume that the probe amplitude is weak and, thus, the RWA is valid. The corresponding shift of the resonance locations (so-called dynamic Stark effect) due to high intensity modulation occurs only when the stationary energy eigenstates experience static coupling, in addition to the time-dependent probe. We will study this in the following section.

In the case of conventional sinusoidal modulation, $\xi(t) = \xi \cos(\Omega t)$, the Fourier coefficients are 
\begin{equation}\label{eq:sinmodj}
\Delta_m= {\rm J}_m(\xi/\Omega),
\end{equation}
where ${\rm J}_m$ are the Bessel functions of the first-kind. Equation~(\ref{eq:rwaspec}) describes the absorbed power from the probe in terms of the transitions between states with coherently modulated transition energy. We see that the modulation creates an infinite set of resonances, spaced by the modulation frequency $\Omega$. The relative magnitudes of these multiphoton resonances are given by the corresponding Bessel functions ${\rm J}_m$. In figure~\ref{fig:specsimple}, we have plotted the sinusoidal modulation spectrum~(\ref{eq:rwaspec}) as a function of the detuning $\delta$ for two values of modulation amplitude $\xi$. Clearly, as the modulation amplitude is increased new multiphoton resonances become visible in the spectrum. Also, we observe the disappearance of the $m=0$ peak for $\xi/\Omega=2.4048$ which is the first zero of ${\rm J}_0$.

\begin{figure}[t!]
\begin{center}
\includegraphics[width=0.7\linewidth]{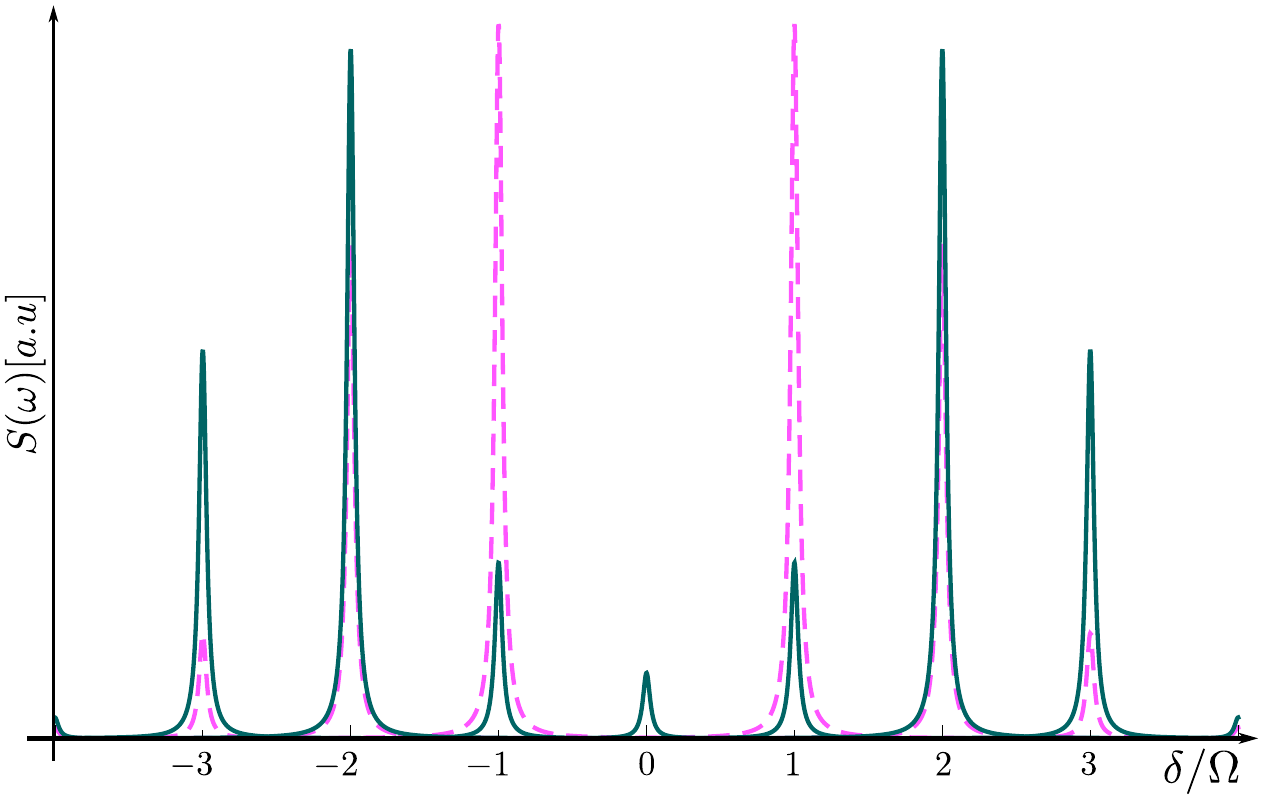}
\caption{Comparison between the spectra for sinusoidally [magenta, equation~(\ref{eq:sinmodj})] and latching [green, equation~(\ref{eq:latchcoupl})] modulated transition as a function of the detuning $\delta$. The parameters are the same as those for the strong modulation spectrum in figure~\ref{fig:specsimple}.} \label{fig:speclatch}
\end{center}
\end{figure}

The modulation of the transition frequency of a qubit does not need to be a simple sine or cosine as a function of time: any other shape can be used in principle~\cite{Dignam02, Blattmann15}. 
So far there have been experiments with square pulses in transmons~\cite{Silveri15} and with bichromatic modulation in double quantum dots~\cite{Forster15} and in a superconducting flux qubit~\cite{Gustavsson13}. In the latter case, the goal was to simulate universal conductance fluctuations in weakly-disordered mesoscopic systems. For the square pulses, i.e. periodic latching modulation, $\xi(t) = \xi \textrm{ sgn}\left[\cos(\Omega t)\right]$ and the sideband amplitudes can be written as~\cite{Silveri15, Eckardt09}
\begin{equation}\label{eq:latchcoupl}
\Delta_m = \frac{2}{\pi}\frac{\Omega \xi}{\Omega^2m^2-\xi^2}\sin\left(\frac{\pi m}{2}-\frac{\pi \xi}{2\Omega}\right).
\end{equation}
We have compared the spectra resulting from the above two types of modulations in figure~\ref{fig:speclatch}. One observes that the resonance heights are clearly dependent on the detailed form of the modulation. Especially, the CDT location $\xi/\Omega=2.4048$ of the $m=0$ resonance of the sinusoidal modulation spectrum is shifted to higher amplitudes $\xi$ due to the rescaling by $\pi/2$ in equation~(\ref{eq:latchcoupl})~\cite{Silveri15}. This effect has been measured with a superconducting qubit whose transition energy was modulated with a square pulsed magnetic flux through its SQUID loop, created by an arbitrary current waveform generator for a bias coil.

In the vicinity of the sidebands $\omega=\omega_0+m\Omega$, $m=0,\pm 1, \pm 2, \ldots$, the general Hamiltonian~(\ref{eq:rotHam}) can be transformed back into the Schr\"odinger picture: 
\begin{equation}
\hat{H}_{\rm RWA}=\frac{\hbar}{2}\left[\left(\delta+m\Omega\right)\hat{\sigma}_z+g_{\rm P}(\Delta_m\hat{\sigma}_++\Delta_m^\star \hat{\sigma}_-)\right]\label{Ham.rwa}.
\end{equation}
Thus, we see that the qubit energies are dressed by those of $m$ modulation photons and a probe photon. The terms in equation~(\ref{Ham.rwa}), thus, describe the two nearly resonant longitudinally doubly dressed states and their coupling, respectively. Considering the dynamics in the Bloch sphere representation, the Hamiltonian~(\ref{Ham.rwa}) generates precession around the vector ${\boldmath \Omega}=\left(g_{\rm P} \textrm{Re}(\Delta_m), -g_{\rm P} \textrm{Im}(\Delta_m), \delta+m\Omega\right)$ with the Rabi frequency
\begin{equation}
g_{\rm d}^{(\rm m)}=\sqrt{(\delta+m\Omega)^2+g_{\rm P}^2|\Delta_m|^2}. \label{rabi:sin:mod}
\end{equation}
In other words, by applying the transverse driving field at the frequency $\omega$ equal to the energy difference of the longitudinally dressed states $\omega_0\pm m \Omega$, coherent oscillations with the rate $g_{\rm d}^{(\rm m)}$ are observed between the qubit eigenstates. This is a demonstration of sideband control of the single qubit states using the frequency modulation, see also references~\cite{Beaudoin12, Strand13} for a similar qubit-oscillator state control.

\subsubsection{Static coupling and Floquet approach}\label{sec:scafa}

In the more general case, the stationary energy eigenstates experience also a constant coupling with strength $\Delta$, in addition to the harmonic probe. This interaction can be modelled with the Hamiltonian [see equations~(\ref{eq:modHamtot})-(\ref{eq:modHamC})]
\begin{equation}\label{eq:probedLZS}
\hat{H}(t) = \frac{\hbar}{2}\left[\omega_0+\xi(t)\right]\hat{\sigma}_z + \frac{\hbar \Delta}{2}\hat{\sigma}_x + \hbar g_{\rm P} \cos(\omega t)\hat{\sigma}_x. 
\end{equation}
This kind of additional static coupling appears naturally in several types of experimental realisations, reaching from superconducting qubits to quantum dots and nuclear magnetic resonance (NMR). It is difficult to write the above Hamiltonian in the form of equations~(\ref{eq:modHamtot})-(\ref{eq:modHamC}). Also, a direct transformation into the interaction picture is complicated since the unprobed Hamiltonians at different times do not commute. However, one can apply the general procedure presented in section~\ref{sec:background} by noticing that the Hamiltonian without the probe,
\begin{equation}\label{eq:probedLZS2}
\hat{H}_0(t)=\frac{\hbar}{2}\left[\omega_0+\xi(t)\right]\hat{\sigma}_z + \frac{\hbar \Delta}{2}\hat{\sigma}_x,
\end{equation}
is time-periodic with the period $2\pi/\Omega$. As a consequence, one can apply the Floquet method~\cite{Shirley65,Zeldovich67,Grifoni98,Chu04}. 

Due to periodicity, the time-dependent Schr\"odinger equation is solved by~\cite{Hausinger10}
\begin{equation}\label{eq:Floqstate}
|\psi_{\alpha}(t)\rangle = |u_{\alpha}(t)\rangle e^{-\ii \varepsilon_{\alpha}t/\hbar},
\end{equation}
where $|u_{\alpha}(t)\rangle$ has the same periodicity as the Hamiltonian and, thus, can be written as a Fourier series. The quasienergies $\varepsilon_{\alpha}$ are the eigenvalues of the so-called Floquet Hamiltonian $\hat{H}_0^{\rm F}\equiv \hat H_0(t)-\ii \hbar \partial_t$:
\begin{equation}\label{eq:quasieval}
\hat{H}_0^{\rm F}(t)|u_{\alpha}(t)\rangle = \varepsilon_{\alpha}|u_{\alpha}(t)\rangle.
\end{equation}
The periodicity of $\hat{H}_0(t)$ is reflected in the fact that the state $|u_{\alpha,n}(t)\rangle\equiv \exp(\ii n \Omega t)|u_{\alpha}(t)\rangle$ gives physically equivalent state to equation~(\ref{eq:Floqstate}), but with shifted quasienergy $\varepsilon_{\alpha,n}\equiv \varepsilon_{\alpha}+\hbar n \Omega$. As a consequence, $\varepsilon_{\alpha}=\varepsilon_{\alpha,0}$ and $|u_{\alpha}(t)\rangle = |u_{\alpha,0}(t)\rangle$. By extending the stationary state basis with that of time-periodic functions, we obtain a composite space spanned by $\{|\sigma,n\rangle\ | \sigma=\mathbin\uparrow,\hspace{-.0em}\downarrow; \ \langle t |n\rangle = e^{\ii n \Omega t}, n\in \mathbb{Z}\}$. In this so-called Sambe space~\cite{Sambe73}, the components of $\hat{H}_0^{\rm F}(t)$ and $|u_{\alpha}(t)\rangle$ are time-independent and can be organized as a matrix $\hat{\mathcal{H}}_{\rm F}$ with its eigenvector $|\alpha\rangle$ and eigenvalue $\varepsilon_{\alpha}$: $\hat{\mathcal{H}}_{\rm F}|\alpha\rangle=\varepsilon_\alpha|\alpha\rangle$. Thus, we see that the time-dependent Schr\"odinger equation is transformed into an eigenvalue equation of an infinite dimensional time-independent Hamiltonian with an infinitely repeating energy structure. For a two-level system, the repeating block consists of two energy eigenstates with the repetition period of $\hbar\Omega$. Due to the similarities with the one-dimensional Bloch's problem of solid state physics, these blocks are sometimes referred to as Brillouin zones. As the Floquet Hamiltonian $\hat{\mathcal{H}}_{\rm F}$ is diagonal and time-independent in the quasienergy basis $|\alpha\rangle$, we can calculate the probe absorption spectrum as in equation~(\ref{eq:Trate}).

The calculation of the spectrum still requires the solution of the quasienergy eigenvalue equation~(\ref{eq:quasieval}). The solution is unavoidably numerical in the general case, and we limit the discussion here to the case where the coupling $\Delta$ is small and one can, therefore, treat it as a perturbation. We solve the eigenvalue equation in the absence of the coupling ($\Delta=0$) and then calculate perturbative corrections. We use so-called nearly degenerate generalized Van Vleck (GVV) perturbation theory. The GVV theory finds a perturbative transformation that decouples a nearly degenerate manifold from the other, far off-resonant, states. The perturbation parameter is the ratio of the coupling strength and the energy gap between the manifolds. After the decoupling, the transformed manifold can be diagonalized with standard methods. Formally, GVV has been discussed on a general level in references~\cite{Certain70,Aravind84}, and in the context of the present case in references~\cite{Son09,Hausinger10}. Also, reference~\cite{Ho85} contains a discussion on the application of GVV in a multimode Floquet problem. 

When $\Delta=0$, the eigenstates of $\hat{H}_0^{\rm F}$ can be written as
\begin{equation}
|u_{\pm,n}^0(t)\rangle = |\uparrow\downarrow\rangle \exp\left[\mp \ii \zeta(t)/2 + \ii n \Omega t\right],
\end{equation}
with the corresponding quasienergies  $\varepsilon_{\pm,n} = \pm \omega_0/2+n\Omega$. The superscript indicates the eigenstates of the unperturbed Hamiltonian. In the Sambe space, one can write the above quasienergy states as
\begin{equation}\label{eq:quasibasis}
|u_{\pm,n}^0\rangle =  \sum_{\ell} \Delta_{\pm(\ell-n)}^{(*)}\left(\xi/2\right)|\uparrow\downarrow,\ell\rangle,
\end{equation}
where $\xi$ is the amplitude of the modulation and the complex conjugation of $\Delta$ applies for the minus sign case. Next, we write the coupling term in this basis by noticing that the non-zero matrix elements of $\Delta\hat{\sigma}_x$ can be written as
\begin{equation}
\Omega_{m} \equiv \Delta\langle u_{-,m}^0|\hat{\sigma}_x|u_{+,0}^0\rangle = \Delta \sum_{\ell}\Delta_{m-\ell}(\xi/2)\Delta_{\ell}(\xi/2) = \Delta \Delta_{m}(\xi),
\end{equation}
because $\Delta_{\ell}(\xi)$ are Fourier coefficients. This can be calculated once the Fourier coefficients~(\ref{eq:phaseFC}) of the dynamical phase factor are known. For the special case of sinusoidal modulation, one obtains~\cite{Son09,Hausinger10}
\begin{equation}\label{eq:djn}
\Omega_{n}=\Delta {\rm J}_{n}(\xi/\Omega).
\end{equation}
As a consequence, Hamiltonian $\hat{\mathcal{H}}_{\rm F}$ can now be written in basis~(\ref{eq:quasibasis}) as
\begin{equation}\label{eq:HFloq}
\hat{\mathcal{H}}_{\rm F} = \sum_{\sigma=\pm,n}\varepsilon_{\sigma,n} |u_{\sigma,n}^0\rangle\langle u_{\sigma,n}^0| + \sum_{n,m}\left[\Omega_{n-m}|u_{-,n}^0\rangle\langle u_{+,m}^0| + \textrm{h.c.}\right],
\end{equation}
where h.c. stands for hermitian conjugate.

Correspondence with the RWA result~(\ref{eq:rwaspec}) of the uncoupled case can be obtained by considering the Floquet Hamiltonian~(\ref{eq:HFloq}) in the vicinity of the multi-photon resonance $\omega_0\approx m\Omega$. Assuming that the coupling between the qubit states, characterized by $\Delta$, can be assumed small, one can use the GVV perturbation theory to find the eigenstates and the corresponding eigenenergies. 

\begin{figure}[t!]
\begin{center}
\includegraphics[width=0.7\linewidth]{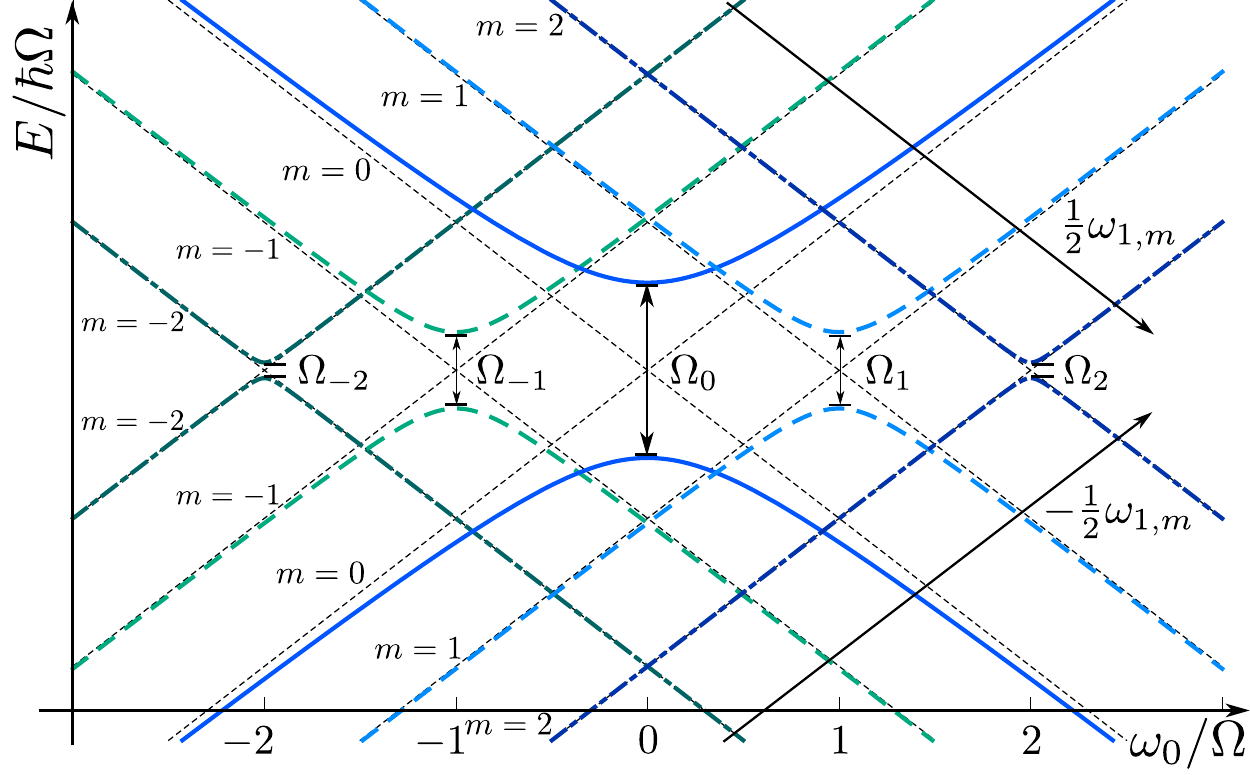}
\caption{Quasienergies $\omega_{1,m}$ for $m=0$ (solid blue), $\pm 1$ (dashed green), $\pm 2$ (dash-dotted dark green) as a function of the bare transition frequency $\omega_0$. We have used the parameter values: $\Delta/\Omega = 0.7$ and $\xi/\Omega = 0.8$. } \label{fig:quasienergies}
\end{center}
\end{figure}

When $\omega_0\approx m\Omega$, the eigenstates $|u_{-,n+m}^0\rangle$ and $|u_{+,n}^0\rangle$ are nearly degenerate. In the weak coupling limit, one can neglect couplings to all other eigenstates. This is called the first order GVV perturbation theory which is equivalent to the rotating-wave approximation made in equation~(\ref{eq:rwa}). Consequently, the Floquet Hamiltonian~(\ref{eq:HFloq}) reduces to a $2\times 2$-matrix
\begin{equation}\label{eq:truncFloq}
\hat{\mathcal{H}}_{\rm F} \approx \frac{\hbar}{2}(\omega_0 -m\Omega)\hat{\sigma}_z^{\rm F} + \frac{\hbar}{2}\left[\Omega_m\hat{\sigma}_+^{\rm F} + \textrm{h.c}\right],
\end{equation}
where $\hat{\sigma}_z^{\rm F}\equiv |u_{+,k}^0\rangle\langle u_{+,k}^0| - |u_{-,k+m}^0\rangle\langle u_{-,k+m}^0|$ and $\hat{\sigma}_+^{\rm F}\equiv |u_{+,k}^0\rangle\langle u_{-,k+m}^0|$.
Thus the eigenenergies, or the so-called quasienergies, in the RWA for the zeroth Brillouin zone are $\pm\frac12\omega_{1,m}\equiv\pm\frac12\sqrt{(\omega_0-m\Omega)^2+|\Omega_m|^2}$~\cite{Ashhab07,Son09,Hausinger10,Oliver05,Wilson07}. The corresponding normalized eigenvectors are then
\begin{eqnarray}
|u_{-,m}^1\rangle &=& \cos\left(\frac{\theta}{2}\right)|u_{-,k+m}^0\rangle -e^{-i\phi}\sin\left(\frac{\theta}{2}\right)|u_{+,k}^0\rangle,\label{eq:trunc1}\\
|u_{+,m}^1\rangle &=& \sin\left(\frac{\theta}{2}\right)|u_{-,k+m}^0\rangle +e^{-i\phi}\cos\left(\frac{\theta}{2}\right)|u_{+,k}^0\rangle,\label{eq:trunc2}
\end{eqnarray}
respectively. In the above, we have defined $\tan \theta =|\Omega_m|/(\omega_0-m\Omega)$ with $\theta\in [0,\pi]$, and $\phi \equiv \textrm{arg } \Omega_m \in [0,2\pi]$. In figure~\ref{fig:quasienergies}, we show a typical quasienergy structure using sinusoidal modulation. We see that the longitudinal driving lifts the degeneracies at the multi-photon resonance locations resulting in characteristic avoided crossing structure. The minimum gaps are given by the Fourier coefficients $|\Omega_m|$. 

In the Sambe space, the transformation into the interaction picture is straightforward in the sense that the Hamiltonian $\hat{H}_0^{\rm F}$ is time-independent. Therefore, the absorption transition rate can be calculated using equation~(\ref{eq:Trate}). The probe Hamiltonian in the interaction picture has to be written as
\begin{equation}
\hat{V}(t) = e^{\ii \hat{H}_0^{\rm F}t/\hbar}  \hat{V} e^{-\ii \hat{H}^{\rm F}_0 t/\hbar}.
\end{equation}
Similar to the uncoupled case~(\ref{eq:rwaspec}), we consider the case of resolved sidebands. In the vicinity of a multi-photon resonance $\omega_0\approx m\Omega$, one can represent the operator $\hat V$ in the eigenbasis of the truncated RWA Hamiltonian~(\ref{eq:truncFloq}). As a consequence, the case with $\hat V=\hat \sigma_x$ can be written as 
\begin{equation}\label{eq:xprobeFloq}
\hat{V}_x^{\rm F}= |\Delta_m|\left[\frac{|\Omega_m|}{\omega_{1,m}}\hat{\sigma}_z^{\rm F} + \frac{\omega_0-m\Omega}{\omega_{1,m}}\hat{\sigma}_x^{\rm F}\right],
\end{equation}
where F denotes the representation in the eigenbasis of the truncated Hamiltonian~(\ref{eq:truncFloq}). We thus see that the probe actually acquires a diagonal part when written in the eigenbasis of the RWA Floquet Hamiltonian. Additionally, we notice that the off-diagonal part vanishes at resonance, which leads to the disappearance of the spectral line. If the coupling $\Delta$ is small, the region where the absorption peak is weak is, nevertheless, narrow. On the other hand, by coupling the probe to $\hat V=\hat{\sigma}_z$ one obtains by, again, writing the probe in the RWA basis~\cite{Tuorila10, Silveri13}:
\begin{equation}\label{eq:zprobeFloq}
\hat{V}_z^{\rm F} = \frac{\omega_0-m\Omega}{\omega_{1,m}}\hat{\sigma}_z^{\rm F} - \frac{|\Omega_m|}{\omega_{1,m}}\hat{\sigma}_x^{\rm F},
\end{equation}
where we have used Parseval's identity which states that $\sum_k |\Delta_k|^2=1$. With both types of probe coupling, the longitudinal terms do not induce transitions, and can be neglected in the weak probe limit. Longitudinal probing~(\ref{eq:zprobeFloq}) has been used in measurements of modulated superconducting qubits with static coupling~\cite{Tuorila10}.

\begin{figure}[t!]
\begin{center}
\includegraphics[width=0.7\linewidth]{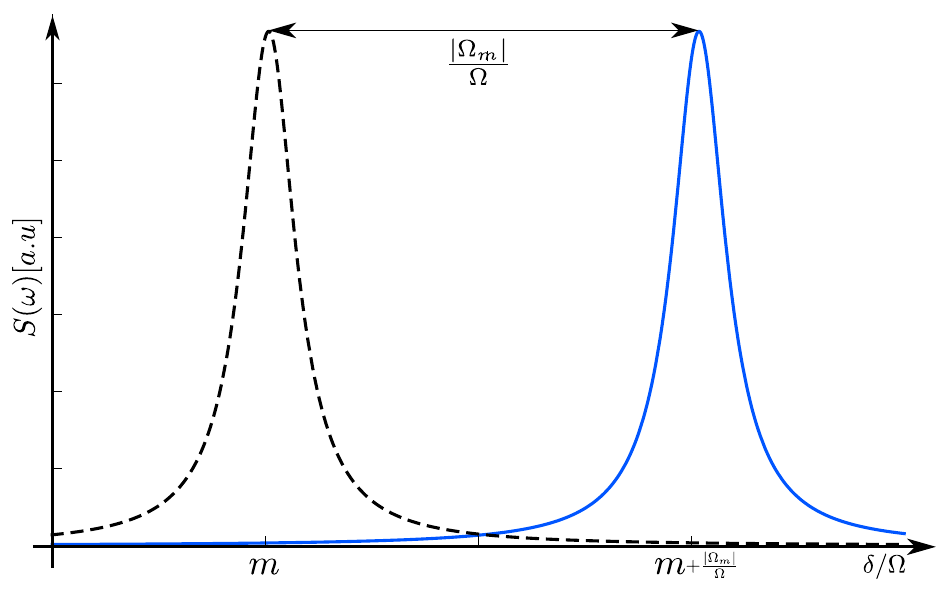}
\caption{Quasienergy (longitudinal) probe spectrum for the $m=1$ resonance of the sinusoidally modulated transition as a function of the detuning $\delta$. The maximal shift occurs at $\xi/\Omega = 1.8412$, which is the first maximum of the Bessel function $J_1$ (blue curve). This is compared with the uncoupled case which can be retrieved by large amplitude modulation (black dashed curve).} \label{fig:specquasi}
\end{center}
\end{figure}

The absorption spectrum in the resolved-sideband limit can now be written as 
\begin{equation}\label{eq:specDelta}
S(\omega) = \sum_{m=-\infty}^{\infty}\frac{g_{\rm P}^2}{2}\frac{\Gamma_2|\langle u_{+,m}^1|\hat{V}^{\rm F}|u_{-,m}^1\rangle|^2}{(\omega_{1,m}-\omega)^2+\Gamma_2^2},
\end{equation}
where $\hat{V}^{\rm F}=V^{\rm F}\hat{\sigma}_x^{\rm F}$ and $V^{\rm F}=|\Delta_m|\frac{\omega_0-m\Omega}{\omega_{1,m}}$ or $V^{\rm F}=\frac{|\Omega_m|}{\omega_{1,m}}$ for transverse and longitudinal probes, respectively. We see immediately that for the transverse probe~(\ref{eq:xprobeFloq}) in the limit of $\Delta\rightarrow 0$ we recover the spectrum~(\ref{eq:rwaspec}). Notice that instead of bare atomic energies, spectrum~(\ref{eq:specDelta}) gives information on the quasienergy structure of a strongly modulated system with an off-diagonal static coupling. As an example, we consider here the case of longitudinal probing of a bare qubit which is in $m^{\rm th}$ multi-photon resonance, i.e. $\omega_0 = m\Omega$. We see that the probe absorption peak is shifted from $\delta=0$ to $\delta=|\Omega_m(\xi/\Omega)|$. One can recover the uncoupled multi-photon resonance locations only asymptotically when $\xi/\Omega \rightarrow \infty$. This is depicted in figure~\ref{fig:specquasi} for sinusoidal modulation of the coupled transition.
In the more general case of non-resonant bare qubit frequency ($\omega_0\neq m\Omega$), the shift of the resonance is given by $\omega=\sqrt{(\omega_0-m\Omega)^2 + |\Omega_m|^2}$.

In the above, we have approximated that the relaxation rates between the quasienergy states are characterised by those of the non-modulated system. Corrections to the resolved sideband approximation and to the relaxation rates can be obtained by applying the so-called Floquet-Born-Markov formalism which combines the Floquet formalism and detailed coupling to the environment~\cite{Blumel89,Blumel91,Hausinger10,Grifoni98,Wilson10,BreuerPetruccione,Goorden04, Hone09, Gasparinetti13}. Notice also that in the RWA, and by neglecting the diagonal probe in equation~(\ref{eq:xprobeFloq}), we obtain in the frame rotating at $\omega$:
\begin{equation}\label{eq:couplRWA}
\hat H_0^{\rm F} = \frac{\hbar}{2}\left[\omega_{1,m}-\omega\right]\hat{\sigma}_z^{\rm F} - \frac{\hbar g_{\rm P}}{2}V^{\rm F}\hat{\sigma}_x^{\rm F},
\end{equation}
which reduces to equation~(\ref{Ham.rwa}) when $\Delta=0$ for a transverse probe.

Sideband resolved multiphoton resonances similar to~(\ref{eq:specDelta}) have been observed in terms of Ramsey interference fringes in Rydberg atoms~\cite{Baruch92,Sirko92}, in quantum dots as resolved sideband emission due to surface acoustic waves~\cite{Metcalfe10}, and in numerous experiments on superconducting qubits~\cite{Oliver05,Wilson07,Gunnarsson08,Silveri15} with microwave modulation. Also, other experiments~\cite{Sillanpaa06,Izmalkov08,Shevchenko08} can be interpreted in terms of non-resolvable Floquet theory, if the dissipation is taken into account with a proper care. The characteristic of these experiments is that the heights of the multi-photon resonances are modulated by the corresponding Fourier coefficients $\Delta_m$ of the dynamical modulation phase factor $e^{i\zeta(t)}$, which are dependent on the modulation amplitude $\xi$. These multi-photon resonances and the accompanying modulations can also be interpreted as resulting from quantum interference, see section~\ref{Sec:disctimeevo}.

\subsubsection{Dynamic Stark effect and basis dependence of RWA}

It should be emphasized that the above RWA result is strongly basis dependent~\cite{Silveri12}. The general problem of finding an optimal basis for RWA has not been addressed, up to our best knowledge. Corrections to the RWA result can be obtained by including the second order corrections in the GVV calculation. As a consequence, the $m$-photon resonance is shifted by
\begin{equation}
\delta_m = \frac{1}{2}\sum_{\ell\neq m} \frac{|\Omega_{\ell}|^2}{\omega_0-\ell\Omega}, \label{eq:BSshift}
\end{equation}
resulting in the quasienergy $\omega_2 \equiv \sqrt{(\omega_0+\delta_m-m\Omega)^2+\Omega^2}$. This shift originates from the influence of the off-resonant states to the nearly degenerate pair of states. The validity criterion for the RWA can be written as $|\Omega_{\ell}|\ll |\omega_0-\ell\Omega|$ which should hold for all values of $\ell\neq m$. Notice that this is more general than the original Bloch-Siegert shift~\cite{BlochSiegert40}, $|\Omega_m|^2/[2(\omega_0+m\Omega)]$, which includes only the counter-rotating correction $\ell=-m$. In general, the deviation between the true modulated transition energy and the RWA result $\omega_1$ is called the generalized Bloch-Siegert shift~\cite{Tuorila10}. The deviation between the uncoupled multi-photon resonance $\omega_0=m\Omega$ and the true modulated resonance is called dynamic Stark shift. The conventional Bloch-Siegert shift has been observed in strongly modulated Rydberg atoms~\cite{Fregenal04}. The generalized Bloch-Siegert correction was needed in the studies of radiative shifts of magnetic resonances~\cite{CohenTannoudji73} and of superconducting qubits with highly nonlinearly coupled sinusoidal modulation~\cite{Tuorila10}. Also, the Bloch-Siegert effect has been observed at quantum level in circuit-QED~\cite{FornDiaz10}.
\begin{figure}[t!]
\begin{center}
\includegraphics[width=0.7\linewidth]{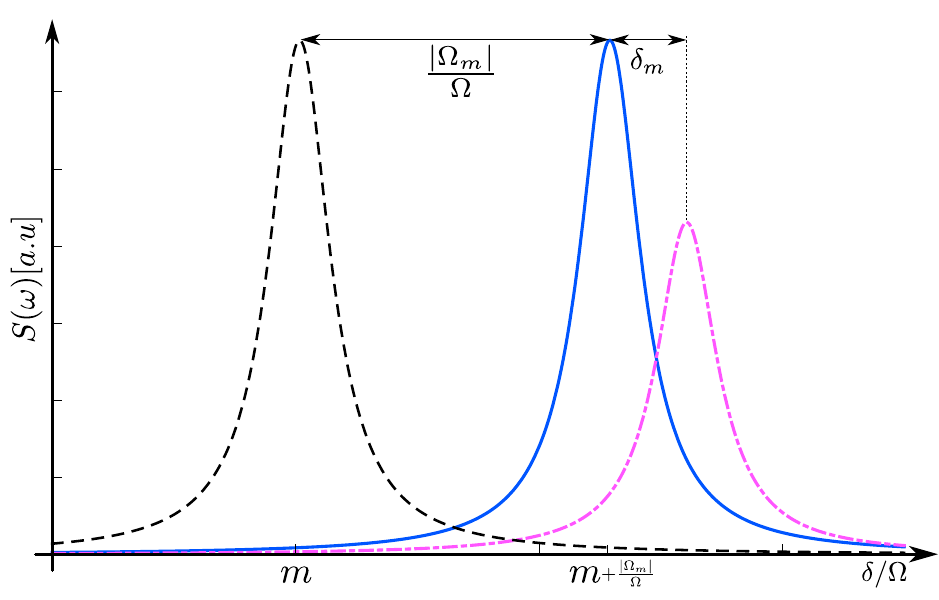}
\caption{Second order generalized Bloch-Siegert shift. The quasienergy resonance is shifted by $\delta_m$ due to the second order couplings to non-resonant states. This plot is based on equation~(\ref{eq:BSshift}).} \label{fig:specBS}
\end{center}
\end{figure}
In figure~\ref{fig:specBS}, we plot the second order correction to the quasienergies. In addition to the location of the resonance, we observe that the magnitude of the resonance is changed due to the non-resonant corrections.

\subsubsection{Steady state population}

Additional insight into the spectrum can be gained by studying the steady state population of the qubit. Also, in some experiments~\cite{Oliver05}, the steady state population is directly measured instead of the probe absorption spectrum. The results in the absence of static coupling can be recovered by setting $\Delta=0$, and by considering transversely coupled probe.

When the coupling to the probe is weak but non-negligible, the sideband control has to be revealed by studying the steady state population of the qubit~(\ref{ham.first}). If the excitation drive $\hbar g_{\rm P} \cos(\omega t)\hat{\sigma}_x$ is in resonance with a longitudinal sideband transition, then the steady state population would deviate from the ground state. To find the steady state population, the master equation~(\ref{master.sin.rwa2}) for the Hamiltonian~(\ref{Ham.rwa}) can be solved analytically around every resolvable resonance~\cite{EberlyAllen87}. By summing the contributions of all independent resonances one results in 
\begin{eqnarray}
P_{\rm e}&=&\frac{g_{\rm P}^2}{2}\sum_{m=-\infty}^\infty \frac{\frac{\Gamma_2}{\Gamma_1} |\Delta_m|^2}{\Gamma_2^2+(\omega_{1,m}-\omega)^2+\frac{g_{\rm P}^2\Gamma_2}{\Gamma_1}\left|\Delta_m\right|^2}\label{Pe_gen}
\end{eqnarray}
for  the steady state occupation probability for the excited state $|u_{+,m}^1\rangle$ of the qubit~\cite{Oliver05, JianLi12, Silveri15}. Thus, as a consequence of multiphoton absorption from the probe, the occupation of the excited state increases. Also, the absorptive transition rate~(\ref{eq:rwaspec}) is obtained from the steady state population as $S(\omega)=\Gamma_1 P_{\rm e}$, which is a restatement of energy conservation. In the absence of coupling and when the probe frequency is zero ($\omega=0$), one obtains the familiar results of the Bloch equation for a qubit with off-diagonal static coupling $g_{\rm P}$ on each resolvable sideband. By including the probe explicitly into the master equation, we also observe that the widths of the sidebands are broadened, compared with the decoherence rate $\Gamma_2$, due to the power broadening by the strong transverse driving with the effective amplitude $g_{\rm P}\Delta_m$~\cite{EberlyAllen87}, which produces a linewidth $\lesssim\sqrt{\Gamma_2^2+g_{\rm P}^2|\Delta_m|^2\Gamma_2/\Gamma_1}$. Also, when $\Delta_m=0$ the contribution to the excited state population due to the corresponding multiphoton resonance disappears, which is conventionally referred to as coherent destruction of tunneling (CDT).

We presented here a discussion that includes the probe, but similar results can be obtained for the excited state population $|u_{+,m}^0\rangle$ in the absence of probe by setting $g_{\rm P}=\Delta$ and $\omega=0$. In the following, we give an alternative interpretation to the origin of the increased excited state population in terms of Landau-Zener-St\"uckelberg (LZS) interference~\cite{Landau32,Zener32,Stueckelberg32}.

\subsection{Discretized periodic time-evolution}
\label{Sec:disctimeevo}
Here we consider the generic LZS-Hamiltonian
\begin{equation}\label{eq:HamLZS}
\hat{H}_{\rm LZS}(t) = \frac{\hbar}{2}\left[\omega_0+\xi(t)\right]\hat{\sigma}_z + \frac{\hbar\Delta}{2}\hat{\sigma}_x.
\end{equation}
This Hamiltonian is exactly the unprobed part of~(\ref{eq:probedLZS}), and can also be obtained from the probed and modulated uncoupled Hamiltonian~(\ref{ham.first}) by making a transformation into a frame rotating with $\omega$ and by denoting $g_{\rm P}=\Delta$. In such case, one talks about photon-assisted LZ-processes between modulation dressed states~\cite{JianLi12}.

We are interested in the time-evolution operator $\hat{U}(t,0)$ which describes the dynamical change of the state vector: $|\psi(0)\rangle\rightarrow |\psi(t)\rangle$. We represent the state in the instantaneous, i.e. adiabatic, eigenbasis $\{|\psi_-(t)\rangle,|\psi_+(t)\rangle\}$ of Hamiltonian~(\ref{eq:HamLZS}), where $H_{\rm LZS}(t)|\psi_{\pm}(t)\rangle = \pm(\hbar/2)\omega(t)|\psi_{\pm}(t)\rangle$ and
\begin{equation}
\omega(t) = \sqrt{\left[\omega_0+\xi(t)\right]^2+\Delta^2}
\end{equation}
is the adiabatic transition frequency. When $\xi(t)$ changes slowly, the system undergoes a free adiabatic time-evolution determined by the unitary operator
\begin{equation}
\hat{U}_{\varphi}\equiv e^{-\ii \varphi (t,0)\hat{\sigma}_z} = \left(\begin{array}{cc}
e^{-\ii \varphi(t,0)} & 0\\
0 & e^{\ii \varphi(t,0)}
\end{array}\right),
\end{equation}
where $\varphi(t,t_0)=\frac12 \int_{t_0}^{t}\omega(t')dt'$. When the adiabatic condition~(\ref{eq:adiabatic}) is not fulfilled, e.g. when the system travels fast across an avoided crossing, a non-adiabatic transition can occur.
Often, the effects of such an event can be captured by the unitary operator~\cite{Ashhab07,Shevchenko10,Silveri15}
\begin{equation}\label{eq:transitionU}
\hat{U}_{\rm T} = \left(\begin{array}{cc}
\sqrt{1-p}e^{-\ii \tilde{\phi}_{\rm S}} & \sqrt{p}\\
-\sqrt{p} & \sqrt{1-p}e^{\ii \tilde{\phi}_{\rm S}}
\end{array}\right),
\end{equation}
where $\sqrt{p}\equiv |\langle \psi_+(t)|\psi(t)\rangle|$ is the transition probability amplitude for the transition $|\psi(0)\rangle \rightarrow |\psi_+(t)\rangle$. Similarly, $\sqrt{1-p}\equiv |\langle \psi_{-}(t)|\psi(t)\rangle|$. We have also taken into account the possibility of non-adiabatic phase shift $\tilde{\phi}_{\rm S}$ during the transition. In a Landau-Zener tunnelling process, this phase shift is determined by the Stokes phase~\cite{Child71,Kayanuma97,Wubs05} $\phi_{\rm S}=\tilde{\phi}_{\rm S} + \pi/2$, and defined as the difference between the dynamic phases collected by the diabatic and adiabatic states during the transition process (see the following section). 
This model allows the separation of the free adiabatic time-evolution from the non-adiabatic transition processes. In consequence, the time-evolution operator of a transition process occurring within the interval $[0,t_2]$ can be discretized formally as
\begin{equation}
\hat{U}(t_2,0) \equiv \hat{U}_{\varphi_2}\hat{U}_{\rm T}\hat{U}_{\varphi_1},
\end{equation}
where $\varphi_i = \varphi(t_i,t_{i-1})$ with $t_0=0$. We emphasize that, although the non-adiabatic transition process is modeled here with $\hat{U}_{\rm T}$ as an instantaneous process occurring at $t=t_1$, physically it has a finite duration. 

We are interested in the periodic modulation of the transition frequency. For a single-period evolution, the time-evolution operator is simply a product of two single-transition operators~\cite{Shevchenko10}:
\begin{eqnarray}
&&\hat{U}_{\rm St} = \hat{U}_{\varphi_4}\hat{U}_{\rm T}\hat{U}_{\varphi_3}\hat{U}_{\varphi_2}\hat{U}_{\rm T} \hat{U}_{\varphi_1} =\left(\begin{array}{cc}
\alpha & \gamma \\
-\gamma^* & \alpha^*
\end{array}\right), \label{eq:multipassLZ}\\
&& \alpha = (1-p)e^{-\ii \zeta_+}-pe^{-\ii\zeta_-},\\
&& \gamma = 2\sqrt{p(1-p)}\sin\left(\Phi+\phi_{\rm S}\right)e^{\ii (\varphi_4- \varphi_1)},\\
&& \zeta_+ = \varphi_1+\Phi+\varphi_4+2\phi_{\rm S}-\pi,\\
&& \zeta_- = \varphi_1+\varphi_4-\Phi,
\end{eqnarray}
where $\tilde{\phi}_{\rm S}=\phi_{\rm S}-\pi/2$, $T=t_4-t_0$ is the period and $\Phi \equiv \varphi_2+\varphi_3$ is the adiabatic phase collected in between the transitions. We note that one can use the same transition matrix $\hat U_{\rm T}$ for the traversal across the avoided crossing to both directions in equation~(\ref{eq:multipassLZ}). This is because the operator is defined in the energy eigenbasis by using a convention where the excited state at far left coincides with the ground state at far right of the avoided crossing. As a result, the excited-state population after a single period can be written as
\begin{eqnarray}
P_{\rm St} &=& |\langle \psi_+(t_4)|\psi(t_4)\rangle|^2 = |\langle \psi_+(t_1)|\hat{U}_{\rm St}|\psi_-(t_1)\rangle|^2 \\
&=& 4p(1-p)\cos\left(\Phi+\phi_{\rm S}\right),
\end{eqnarray}
since $t_4\sim t_1$ due to the periodicity. We thus see that the excited-state population oscillates as a function of the phase $\Phi+\phi_{\rm S}$. These are called St\"uckelberg oscillations~\cite{Stueckelberg32} of the transition probability. The oscillations can be interpreted physically as quantum interference between the two paths along which the system can evolve. The constructive interference occurs when $\Phi+\phi_{\rm S} = n\cdot 2\pi$, with $n\in \mathbb{Z}$. St\"uckelberg oscillations have been observed in inelastic scattering cross-sections of atoms~\cite{Nikitin72} and in microwave excitations of Rydberg atoms~\cite{Baruch92,Sirko92}. The analogous interference phenomenon appearing in Mach-Zehnder interferometers has been observed in references~\cite{Ji03,Pezze07}.

In the case of repeated driving periods, one can observe interference also between the periods. The situation is analogous to a multi-pass Mach-Zehnder interferometer~\cite{Oliver05}. The discretized time-evolution operator over $N$ periods can be calculated by diagonalizing the single-period unitary time-evolution and by taking its $N$th power. By averaging over many periods, we  obtain the modulated St\"uckelberg, i.e. steady state, population of the excited state~\cite{Shevchenko10, Tuorila13}:
\begin{equation}
P_+ = \frac{P_{\rm St}}{\sin^2 \eta},
\end{equation}
where
\begin{equation}
\sin^2\eta = P_{\rm St} + \left[(1-p)\sin \zeta_+ - p \sin \zeta_-\right]^2.
\end{equation}
We thus see that constructive interference occurs when $P_{\rm St}$ has a maximum and $\sin^2\eta$ has a minimum. The resulting resonance conditions can be written as
\begin{eqnarray}
\zeta_+-\zeta_- &=& k\times 2\pi,\\
\zeta_+ &=&\ell \times \pi, \ \ \ \textrm{(constructive interference)}\\
\zeta_- &=&m\times\pi,
\end{eqnarray}
respectively. The destructive interference occurs at the zeroes of $P_{\rm St}$:
\begin{equation}
\zeta_+-\zeta_- = (2k+1)\times\pi \ \ \ \textrm{(destructive interference)}.
\end{equation}
Above, all constants $k,\ell,m$ are integers.

One gains insight on the locations of the multiphoton resonances observed in~(\ref{Pe_gen}) by considering the locations of constructive interference. When $p\ll 1$, the constructive interference is determined by $\zeta_+$. Due to the small probability amplitude for a transition, the system follows the adiabatic state. Thus, the phase $\zeta_+$ is often referred to as the adiabatic phase. When $p\sim 1$, the constructive interference location is determined by the dynamical phase $\zeta_-$, which is therefore called the diabatic phase. By assuming that the coupling $\Delta$ is small, we obtain
\begin{equation}
\zeta_- = \frac{\pi}{\Omega}\omega_0,
\end{equation}
which implies the familiar multiphoton resonance condition $\omega_0=m\Omega$. This is natural since, by neglecting $\Delta$, the adiabatic eigenstates are actually the same as the stationary basis states. The above discussion also clarifies the concept of the dynamic Stark shift, which in the case of finite $\Delta$ can be seen to arise from the fact that the transitions due to the probe are not between the stationary uncoupled states $|u_{\pm,m}^0\rangle$ but, instead, between the (adiabatic) eigenstates of the coupled Hamiltonian. Further progress requires additional knowledge on the transition amplitude $\sqrt{p}$.

\subsubsection{Landau-Zener transition probability}

The value of the probability amplitude $\sqrt{p}$ depends on the explicit form of the modulation $\xi(t)$. Quite often in the case of continuous modulation, the modulation amplitude is so large that the diabatic states become degenerate at some time $t=t_0$ of the period, i.e. $\omega_0 + \xi(t_0)=0$. Typically, the degeneracy is reached twice during the period. In the vicinity of the degeneracy, the Hamiltonian of the system can be approximated as
\begin{figure}[t!]
\begin{center}
\includegraphics[width=0.7\linewidth]{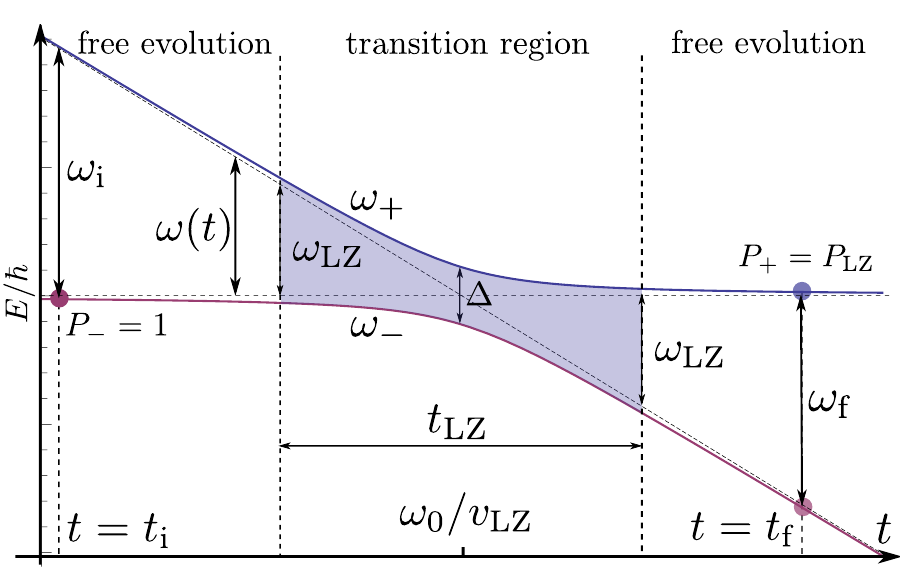}
\caption{Schematic of an LZ process at an avoided crossing. Notice that for the plot we set the zero of energy at $-\omega(t)$.} \label{fig:lz}
\end{center}
\end{figure}

\begin{equation}
\hat{H}(t) = \frac{\hbar\omega_0}{2} \hat{\sigma}_z + \frac{\hbar \Delta}{2} \hat{\sigma}_x + \frac{\hbar v_{\rm LZ}t}{2}\hat{\sigma}_z,
\end{equation}
where $v_{\rm LZ} = d\xi(t_0)/dt$ is the rate of change of the transition energy at the degeneracy point, and without loss of generality we can assume that $\Delta>0$. This is a prototype example of the so-called Landau-Zener (LZ) Hamiltonian~\cite{Landau32,Zener32,Stueckelberg32,Majorana32}: it describes an avoided crossing of two adiabatic energy states $|\pm\rangle$, whose eigenfrequencies can be written into the form
\begin{equation}
\omega_{\pm}(t) = \pm \frac{1}{2}\sqrt{\omega^2(t) + \Delta^2},
\end{equation}
where $\omega(t)=\omega_0 + v_{\rm LZ}t$ and the crossing of the diabatic states occurs with the rate $v_{\rm LZ}$. 

The LZ-scheme is depicted in figure~\ref{fig:lz}. The basic problem is to study a system that is initially in the state $|\psi_-(t_{\rm i})\rangle$ at $\omega_{\rm i}\equiv \omega(t_{\rm i})$, travels across the avoided crossing, and finishes in the state $|\psi_+(t_{\rm f})\rangle$ at $\omega_{\rm f}\equiv \omega(t_{\rm f})$. When $t_{\rm i}=-\infty$, one can find an analytic asymptotic solution of the transition probability $p_{\rm LZ} \equiv p_{-\rightarrow +}(t_{\rm f}=\infty)$ between the adiabatic states. The solution to this so-called Landau-Zener problem is~\cite{Landau32,Zener32,Stueckelberg32,Majorana32}
\begin{equation}
p_{\rm LZ} = \exp\left[-2\pi \gamma\right],
\end{equation}
where $\gamma\equiv \Delta^2/(4|v_{\rm LZ}|)$. This is the celebrated Landau-Zener formula. The adiabaticity criterion~(\ref{eq:adiabatic}) can be written into the form $\Delta^2/|v_{\rm LZ}| \gg 1$. Clearly, in the adiabatic limit $p_{\rm LZ}\approx 0$. In the opposite limit, i.e. when $\Delta^2/|v_{\rm LZ}| \ll 1$, the system follows its diabatic eigenstate and $p_{\rm LZ}\approx 1$.

In addition to the LZ-probability, one can find the whole quantum mechanical state at $t_{\rm f}=+\infty$ for an arbitrary initial state, expressed analytically in terms of Weber functions. However, the derivation is rather involved and we encourage the reader to refer to references~\cite{Shevchenko10,Tuorila} for a proof with modern notation. We also point out that the solution for the LZ-problem can be found with contour integration without solving the Schr\"odinger equation directly~\cite{Wittig05,Ho14}. Here, we give the resulting unitary transition matrix
\begin{equation}\label{eq:lztrans}
\hat{U}_{\rm LZ} = \sqrt{1-p_{\rm LZ}} \exp(-\ii \tilde{\phi}_{\rm S}\hat{\sigma}_z) + \ii \sqrt{p_{\rm LZ}}\hat{\sigma}_y 
\end{equation}
in the adiabatic basis, where $\tilde{\phi}_{\rm S}=\phi_{\rm S}-\pi/2$ and
\begin{equation}
\phi_{\rm S} \equiv \frac{\pi}{4}+ \textrm{arg}\left[\Gamma(1-\ii\gamma)\right]+\gamma(\ln \gamma -1)
\end{equation}
is the so-called Stokes phase which accounts for the dynamical phase collected during the non-adiabatic time-evolution~\cite{Kayanuma97}. We emphasize that by describing the tunneling process with transformation~(\ref{eq:lztrans}) one implicitly assumes that the transition is located exactly at the avoided crossing. In addition, the system undergoes a free adiabatic phase evolution, characterized by the operator
\begin{equation}
\hat{U}_{\varphi} = \exp(-\ii\varphi\hat{\sigma}_z).
\end{equation}
Consequently, the time-evolution of the general LZ process, starting from $t_{\rm i} = (\omega_0-\omega_{\rm i})/v_{\rm LZ}$, and ending at $t_{\rm f}=(\omega_0-\omega_{\rm f})/v_{\rm LZ}$, can be discretized by the unitary time evolution operator
\begin{equation}\label{eq:uniadi}
\hat{U}(t_{\rm f},t_{\rm i})=\hat{U}_{\varphi_{\rm f}}\hat{U}_{\rm LZ}\hat{U}_{\varphi_{\rm i}}.
\end{equation}
In the above, $\varphi_{\rm i} = \int_{t_{\rm i}}^{t_1} [\omega_+(t)-\omega_-(t)]dt$ and $\varphi_{\rm f} = \int_{t_1}^{t_{\rm f}}[\omega_+(t)-\omega_-(t)]dt$. This is an asymptotic result and holds exactly only when $|t_{\rm i}|,t_{\rm f}\rightarrow \infty$. In similar transfer matrix method has been reported by Child in reference~\cite{Child71}. Also, the Weber equation and the Stokes phase appear naturally in studies of scattering from an inverted parabolic potential~\cite{Connor68}. Notice that the transition matrix~(\ref{eq:lztrans}) is exactly of the same form as in equation~(\ref{eq:transitionU}).

In their original papers, Landau, Zener, and others~\cite{Landau32,Majorana32} implicitly assumed that apart from the two adiabatic energy states considered, the other states do not undergo an avoided crossing. However, in the case of superconducting realizations, the energy dispersion is typically periodic with respect to the control parameters, e.g. the offset charge on a superconducting island and the magnetic flux through a superconducting loop. As a consequence, when either of these parameters is modulated strongly one might in fact have traversals over multiple different avoided crossings within a modulation period. A generalization to the single-passage LZ problem can be obtained with the so-called multi-level LZ Hamiltonian
\begin{equation}\label{eq:multiLZ}
\hat{H} = \hbar \hat{\Delta} + \hbar \hat{v}t,
\end{equation}
where $\hbar \hat{\Delta}$ is the Hamiltonian of the time-independent system represented in the basis where the sweep term $\hbar \hat{v}t$ is diagonalized. Brundobler and Elser~\cite{Brundobler93} conjectured that sequential LZ approximation gives an exact result in some multi-level crossing cases. Shytov~\cite{Shytov04} showed that the above conjecture holds for the existent exact solutions for Hamiltonian~(\ref{eq:multiLZ}). However, in this review we concentrate only on the LZ physics of a single avoided crossing. 

\subsubsection{Applicability for crossings with finite duration}

A vast number of avoided crossings of adiabatic energy states occurring in nature can be modelled with a good accuracy in terms of the time-evolution operator given in equation~(\ref{eq:uniadi}). Nevertheless, the LZ transition is not instantaneous, but has a finite duration $t_{\rm LZ}$~\cite{Mullen89, Vitanov96, Garraway97} that scales as
\begin{equation}
t_{\rm LZ} \sim \frac{2}{\Delta}\sqrt{\gamma}\textrm{max}(1,\sqrt{\gamma}).
\end{equation}
This is depicted in figure~\ref{fig:lz} as the shaded region. Consequently, the LZ transition is not located strictly at the avoided crossing but, on the contrary, the so-called transition region $\omega_{\rm LZ}\equiv |v_{\rm LZ}|t_{\rm LZ}/2$ has a finite width:
\begin{equation}
\omega_{\rm LZ} \sim \frac{\Delta}{4\sqrt{\gamma}}\textrm{max}(1,\sqrt{\gamma}).
\end{equation}
Thus, equation~(\ref{eq:uniadi}) is applicable only when $|\omega_{\rm i,f}| \geq \omega_{\rm LZ}$ and, in the adiabatic limit, the LZ formula can be applied whenever $|\omega_{\rm i,f}| \geq \Delta/4$. In the opposite limit, the situation is more involved. Even though $t_{\rm LZ}\rightarrow 0$ when $|v_{\rm LZ}|\rightarrow \infty$, the width of the transition region diverges as $\omega_{\rm LZ}\sim \sqrt{|v_{\rm LZ}|}$. In certain experimentally realisable cases~\cite{Silveri15}, this means that the LZ transition matrix~(\ref{eq:lztrans}) cannot be used and one has to come up with another way of describing the non-adiabatic time-evolution.

\subsubsection{Sudden approximation} \label{sec.multilevel.sudden}

Whenever $\omega_{\rm i}\rightarrow \omega_{\rm f}$ can be taken to be instantaneous ($|v_{\rm LZ}|\rightarrow \infty$), so that the state of the system does have time to adjust itself to the change~\cite{Messiah}, one can replace equation~(\ref{eq:lztrans}) with
\begin{equation}\label{eq:sudden}
\hat{U}_{\rm i \rightarrow f} = \sqrt{1-p_{\rm s}} \hat{\rm I} + \ii \sqrt{p_{\rm s}}\hat{\sigma}_y,
\end{equation} 
where $\sqrt{p_{\rm s}} \equiv \langle \psi_+(t_{\rm f}) |\psi_-(t_{\rm i})\rangle$ and $\sqrt{1-p_{\rm s}}\equiv \langle \psi_-(t_{\rm f})|\psi_-(t_{\rm i})\rangle$, and $p_{\rm s}$ is real because $\hat{H}$ is symmetric. This so-called \textit{sudden approximation} holds whenever~\cite{Messiah,Silveri15}
\begin{equation}
\left(\frac{\omega_{\rm i}-\omega_{\rm f}}{2}\right)^4 \frac{1}{v_{\rm LZ}^2}\frac{\Delta^2}{\Delta^2 + \omega_{\rm i}^2}\ll 1.
\end{equation}
We emphasize that the adiabaticity condition~(\ref{eq:adiabatic}) and the sudden approximation transition matrix~(\ref{eq:sudden}) are valid irrespective of whether the system travels across an avoided crossing or not. On the contrary, the LZ approach relies on the fact that such a crossing lies on the temporal trajectory of the system. The sudden approximation and the LZ formula agree only when the system travels instantaneously across an avoided crossing and $|t_{\rm i}|,t_{\rm f} \rightarrow \infty$. For finite $t_{\rm i}$ and $t_{\rm f}$, one has to rely on the sudden approximation results.

\subsubsection{Dynamic and geometric phases via control of detuning}

Let us assume that the transition frequency of a qubit can be effectively controlled by a coherent transverse drive. In a rotating frame, the effective qubit energy is given by the detuning $\delta_{\rm d}=\omega_0-\omega_{\rm d}$, where $\omega_{\rm d}$ is the drive frequency, and results in the accumulation of phase by $\delta_{\rm d} t$ under the time evolution. One immediate application of this observation is in the realization of phase gates for quantum computing. Another application is quantum simulation: by changing randomly the detuning, one can emulate weak localization, with the detuning playing the role of the phase accumulated by an electron through scattering as it moves in a disordered medium~\cite{Chen2014}. Similar ideas can be used to simulate time-reversal symmetry and universal fluctuations, in this case by driving the system across the Landau-Zener transition using a bichromatic modulation~\cite{Bylander2009,Gustavsson13}.

Even with a fixed but nonzero detuning it is possible to probe fundamental effects such as Berry's phase accumulation for closed adiabatic trajectories in the Hamiltonian parameter space (see equation~(\ref{eq:geometricphase})). As an example, we describe an experimental realisation~\cite{Leek2007} which employs the analogy between a spin-1/2 in a magnetic field of variable direction and the qubit driven by a microwave field with externally-tunable detuning and amplitude. For a drive field with the Rabi coupling strength $\xi_{\rm d}$ and phase $\phi$, we have in general
\begin{equation}
\hat{H}(t) = \frac{\hbar\omega_{0}}{2} \hat{\sigma}_{z}+ \hbar \xi_{\rm d} \cos (\omega_{\rm d}t + \phi) \hat{\sigma}_{x} .
\end{equation}
In the frame rotating at the drive frequency, defined by the unitary operator $\hat{U} (t) = \exp (- i \omega_{\rm d} t \hat{\sigma}_{z} /2 )$, we neglect the fast rotating terms. This yields
\begin{equation}
\hat{H}^{(\rm I)} = \frac{\hbar}{2} \left[ \delta_{\rm d} \hat{\sigma}_{z} + \xi_{\rm d}\cos(\phi) \hat{\sigma}_{x} + \xi_{\rm d}\sin(\phi ) \hat{\sigma}_{y}\right],
\end{equation}
where the detuning is $\delta_{\rm d} = \omega_{0} - \omega_{\rm d}$. This is a Hamiltonian of the type $\hat{H} = -(\hbar /2) \vec{R}\vec{\sigma}$, representing a spin-1/2 in a magnetic field.
By keeping the detuning fixed and changing the phase $\phi$ from $0$ to $2\pi$, the vector $\vec{R}$ describes a cone. The accumulated Berry phase is then equal to the solid angle encompassed by this cone,
\begin{equation}
\gamma_{\rm B} = 2\pi \left( 1 - \frac{\delta_{\rm d}}{\sqrt{\xi_{\rm d}^2 + \delta_{\rm d}^2}} \right).
\end{equation}
This phase was measured~\cite{Leek2007} in a Ramsey interference setup, by first preparing a superposition of $|0\rangle$ and $|1\rangle$ with a $\pi/2$ pulse, followed by adiabatic manipulation of the qubit Hamiltonian, and finally reading the accumulated phase with another $\pi /2$ pulse. In practice, a spin-echo sequence was used which not only reduces the noise caused by the dephasing but also cancels exactly the nonadiabatic dynamical phase.

\subsection{Incoherent modulation}
\label{Sec:stochmod}

It is well known that off-diagonal external fluctuations in a quantum system can induce transitions between energy eigenstates, and thus lead to dissipation. This is known in the literature as the fluctuation-dissipation theorem~\cite{Callen51}. In addition, also the transition frequency can experience incoherent and typically low-frequency fluctuations, which do not lead into resonant energy exchange between the system and its environment but still cause losses in the quantum coherence due to the random and unequal phase evolution of the eigenstates. We will refer to this as dephasing or classical noise and emphasize its distinction with respect to decoherence, which is seen as a consequence of environmental entanglement~\cite{Schlossauer14}. A review of quantum noise has been given in reference~\cite{Clerk10}. 

As the deterministic modulations would serve as a possible generator for quantum gates and quantum simulations~\cite{Cirac95}, the random modulations can, in principle, be used as an analog simulator of atomic dephasing effects caused by different environments~\cite{Saira07}. For example, the dephasing due to quantum entanglement with environment has been simulated with classical fluctuations of parameters in ion traps qubits with engineered reservoirs~\cite{Schneider98,Myatt00}, and with superconducting qubits~\cite{JianLi12} with experimentally controllable noise in the magnetic flux bias. Understanding  the nature and origins of dephasing is essential in all fields of experimental quantum physics, and becomes crucial in the development of quantum information devices~\cite{Makhlin01,Makhlin04,Ithier05}. 

In the conventional theory of classical dephasing noise, the essentially weak and linear quantum mechanical coupling to an external bath of fluctuators is approximated as a classical incoherent modulation of the energy levels of the studied system. The spectrum for an atom (or an atomic ensemble) whose transition frequency changes randomly is known since the early days of NMR physics~\cite{Abragam, Anderson54} under the name motional averaging or narrowing due to atomic movement, where the uncertainty of transition frequency is typically caused by temporal variations of the local magnetic field for rotating or moving atoms, or exchange narrowing due to exchange interactions of electronic magnetic moments. Typically, one has a large number of particles and the experimentalist can only indirectly attempt to change the properties of the external fluctuation processes, generally by modifying temperature or pressure. 

Different methods to suppress dephasing have been developed since the invention of the spin-echo technique~\cite{Hahn50}, which uses designed pulse sequences to eliminate the variations in the local external magnetic environment. Similar procedure has been developed for a generic quantum mechanical two-level system, i.e. a qubit, where the dephasing effects arising from coupling to a thermal bath of harmonic oscillators is suppressed by repeated time-reversal operations on the coupled system and bath. This is referred to as quantum bang-bang control~\cite{Viola98}. Quantum bang-bang is a special case of dynamical decoupling~\cite{Viola99}, where the irreversible open system evolutions are manipulated with external controllable interactions. In addition, the dephasing can be suppressed by actively monitoring and conditionally correcting the quantum state with quantum error-correction codes~\cite{Shor95,Ekert96}. Also, suppression of qubit dephasing by superconducting qubit motion has been observed~\cite{averin}.

\subsubsection{Generating function}

We assume that the energy levels are modulated with an incoherent $\xi(t)$. The modulated spectrum~(\ref{eq:qspec}) is again determined by the phase factor
\begin{equation}\label{eq:phasediff}
A(t) = \langle e^{\ii \int_{0}^t \xi(\tau)d\tau}\rangle =\langle e^{\ii\int_{0}^t \xi(\tau)d\tau}\rangle_{\xi} = \int e^{\ii\int_{0}^t \xi(\tau)d\tau} P[\xi(t)]d\xi,
\end{equation}
where the first expectation value is calculated over the whole ensemble. In addition to averages over an ensemble of particles, one can study individual transitions and averages over different realisations of incoherent modulations, denoted as $\langle\cdot\rangle_{\xi}$ in the second equality. The third equality gives the means to calculate the average in terms of the probability distribution $P[\xi(t)]$ of different realisations of $\xi(t)$. The connection to the dephasing of a two-level system can be obtained by considering the decay of the off-diagonal density matrix element $\hat{\rho}_{01} = \rho_{01} |0\rangle\langle 1|$:
\begin{equation}
\rho_{01}(t) \equiv \langle \hat{\rho}_{01}(t) \rangle_{\xi} = \rho_{01}(0) e^{-\ii\omega_0 t} A^*(t),
\end{equation}
where $\hat{\rho}(t)$ is the representation of the density operator in the interaction picture. Later, we will show that with certain assumptions for the noise process, $A(t)$ can be written into the typical form of a dephasing element: $A(t) \sim e^{-|t|\Gamma_{\phi}}$, where the $\Gamma_{\phi}$ is the effective dephasing rate. 

Equation~(\ref{eq:phasediff}) describes classical noise as phase diffusion due to the random fluctuations. It is the characteristic functional of the stochastic process $\xi(t)$ and, thus, also its moment and cumulant generating function~\cite{Kubo62,vanKampen,Stepisnik99}:
\begin{eqnarray}
A(t) &=& \sum_{n=0}^{\infty} \frac{\ii^n}{n!}\int \cdots \int \langle \xi(t_1)\cdots \xi(t_n)\rangle_{\xi} dt_1\cdots dt_n\\ 
&=& \exp\left[\sum_{n=1}^{\infty} \frac{\ii^n}{n!}\int \cdots \int \langle \xi(t_1)\cdots \xi(t_n)\rangle_\xi^{\rm c}dt_1\cdots dt_n\right],
\end{eqnarray}
where $\langle \xi(t_1)\cdots \xi(t_n)\rangle_{\xi}$ and $\langle \xi(t_1)\cdots \xi(t_n)\rangle_{\xi}^{\rm c}$ are the $n$th moment and cumulant of the random process $\xi(t)$, respectively. Without  loss of generality we can assume that the average modulation over different realisations of fluctuations vanishes, i.e. $\langle \xi(t)\rangle_{\xi} = 0$. We will also simplify our discussion by assuming that the noise processes are stationary:
\begin{equation}
\langle \xi(t_1+\tau)\cdots \xi(t_n +\tau)\rangle = \langle \xi(t_1)\cdots \xi(t_n)\rangle,
\end{equation}
i.e. they are not affected by a shift in time. The above condition should hold for all $n$, $\tau$ and $t_1, \ldots, t_n$. The detailed modelling of the environment can be done in a couple of ways. Near the thermodynamic equilibrium, many different types of fluctuations can be treated as Gaussian. Such Gaussian models of the fluctuations can be treated universally with a linear coupling to a bath of harmonic oscillators~\cite{PhysRealQC}, which in the case of a single transition is the so-called spin-boson model~\cite{Leggett87,Caldeira81,Weiss}. However, not all noise sources are Gaussian. Typical counter-examples arise when either the bath~\cite{Prokofev00} or the coupling to the bath~\cite{Makhlin04} is nonlinear. Inherently non-Gaussian random telegraph noise (RTN) of the transition energy is created when the environment goes through random and discrete frequency fluctuations~\cite{Anderson54,Abragam}.  Nevertheless, the central limit theorem states that when there are several RTN fluctuators the collective statistical behaviour should recover the Gaussian form. Gaussian and random-telegraph noises belong to the small set of stochastic processes that are experimentally relevant and also allow analytic solutions. Generally, one has to rely on numerical solutions of the stochastic master equation for the system~\cite{BreuerPetruccione,GardinerZoller}.

\subsubsection{Gaussian modulation}

The characteristic feature of a Gaussian random process is that the cumulants above $n=2$ are zero. When $\langle \xi(t)\rangle_{\xi} = 0$, we have that $\langle \xi(t_1)\xi(t_2)\rangle_{\xi}^{\rm c}=\langle \xi(t_1)\xi(t_2)\rangle$~\cite{Kubo62}, and the random process is characterized solely by the second order moment. As a consequence, the probe spectrum of the incoherently modulated transition is completely determined once the noise power spectral density
\begin{equation}\label{eq:noisespec}
S_{\xi}(\omega) = \int_{-\infty}^{\infty} \langle \xi(t)\xi(0)\rangle e^{\ii\omega t} dt
\end{equation}
is known. Spectral density encapsulates the physical properties of the environment that are responsible for the incoherent modulations~\cite{Schlossauer14}. Thus, in order to find the experimentally relevant form for the power spectral density of the classical dephasing noise, one has to have a detailed microscopic model for the processes that create the fluctuations. Prototype examples include the spin-boson model~\cite{Leggett87}, spin-fluctuator model~\cite{Paladino02, Shnirman05, Beaudoin15}, and quasiparticle tunnelling~\cite{Zanker15}.

The dynamical phase factor can be written in terms of the spectral density as
\begin{eqnarray}
A(t) &=& \exp\left[-\frac{1}{2}\int_0^t\int_0^t \langle \xi(t_1)\xi(t_2)dt_1dt_2\right]\\
&=& \exp\left[-\frac{1}{2}\int_{-\infty}^{\infty} \frac{d\omega}{2\pi}S_{\xi}(\omega)\frac{\sin^2(\omega t/2)}{(\omega/2)^2}\right],
\end{eqnarray}
where in the second equality we have substituted the inverse Fourier transform of equation~(\ref{eq:noisespec}), and calculated the time-integrals. This is a standard result in the theory of decoherence.

Consider then times much larger than the coherence time $t_{\rm c}$ of the autocorrelation function, i.e. $t\gg t_{\rm c}$. As a consequence, the function $\sin^2(\omega t/2)/(\omega t/2)^2$ is strongly peaked at $\omega\approx 0$, thus we see that its behaviour is of window-type. Accordingly, the off-diagonal element of the density operator can be written into the form
\begin{equation}
\rho_{01}(t) = \rho_{01}(0)e^{-\ii\omega_0 t -|t|\Gamma_{\phi}},
\end{equation}
where we have defined the decoherence rate in terms of the zero frequency fluctuations as
\begin{equation}
\Gamma_{\phi} = \frac12 S_{\xi}(\omega=0),
\end{equation}
provided that the spectral density is defined at $\omega=0$. This loss of quantum coherence occurs due to the pure dephasing of the energy levels. In addition, another contribution to the decoherence rate $\Gamma_2$ comes from the transverse fluctuations~\cite{Makhlin01}.

On the other hand, let us consider $1/f$-noise, with spectral density $S(\omega)\sim 1/\omega$, which is the experimentally relevant dephasing noise in the case of solid-state qubits~\cite{Rabenstein04, Nakamura02,Yoshihara06}. By restricting the discussion to time scales $t<\omega^{-1}$ where $S(\omega)$ is negligible, we can make a short-time expansion for the window-function and obtain
\begin{equation}
\rho_{01}(t)=\rho_{01}(0)e^{-\ii\omega_0 t - (t\Gamma_{\phi}^{1/f})^2},
\end{equation}
where
\begin{equation}
\Gamma_{\phi}^{1/f} = \left[\int_{-\infty}^{\infty}\frac{d\omega}{4\pi}S_{\xi}(\omega)\right]^{\frac12}.
\end{equation}
Such nonlinear exponential decay is characteristic of $1/f$-noise~\cite{Ithier05}. However, it is not at all certain that real experimental $1/f$-noise is Gaussian~\cite{Bergli09}.

\subsubsection{Increasing the dephasing time of qubits by motional narrowing}

Consider now what happens when instead of a single qubit we have a system of $N$ physical qubits, positioned at different locations.
The Hamiltonian of such system, in the absence of driving or any other interactions, can be written as
\begin{equation}
\hat{H}(t) = \frac{\hbar}{2}\sum_{j=1}^{N}[ \omega^{(j)}_{0} + \xi_{j}(t)] \hat{\sigma}^{(j)}_{z} ,
\end{equation}
where $j$ is the qubit index and $\omega^{(j)}_{0}$ is the bare transition frequency.

Now, one can relay a logical qubit between the physical qubits $j$ by using SWAP gates, with the logical qubit spending an equal time $t/N$ at each physical qubit~\cite{averin}.
The swapping operation is assumed to be faster than the single-qubit dephasing time and has unit fidelity. In this case, a state $\alpha_{j}|0\rangle_{j} + \beta_{j}|1\rangle_{j}$ imprinted on the qubit $j$ is moved to the qubit $k$, which was initially in the state $\alpha_{k}|0\rangle_{k} + \beta_{k}|1\rangle_{k}$,
\begin{eqnarray}
&&(\alpha_{j}|0\rangle_{j} + \beta_{j}|1\rangle_{j})\otimes (\alpha_{k}|0\rangle_{k} + \beta_{k}|1\rangle_{k}) \stackrel{SWAP}{\longrightarrow} \nonumber \\
&&(\alpha_{j}|0\rangle_{k} + \beta_{j}|1\rangle_{k})\otimes (\alpha_{k}|0\rangle_{j} + \beta_{k}|1\rangle_{j}).
\end{eqnarray}
Let us assume now that there is no correlation between the noises at different qubits, i.e. $\langle \xi_{j}(0)\xi_{k}(t) \rangle = 0$ if $k\neq j$, while the noise of each qubit has a power spectral density $S_{\xi}^{(j)}(\omega )$.
This allows us to factorize the averages,
\begin{equation}
\rho_{01}(t) = \Pi_{j=0}^{N}\langle e^{-\ii \int_{0}^{t} dt' \xi^{(j)} (t')} \rangle_{\xi^{(j)}},
\end{equation}
and further
\begin{equation}
\rho_{01} (t) = \rho_{01} (0) e^{-\frac{1}{\pi} \int d\omega \frac{\sin^2 \frac{\omega t}{2 N}}{\omega^2} \sum_{j=1}^{N}S_{\xi}^{(j)}(\omega )}.\label{sinc}
\end{equation}
We can already notice that the decay of the off-diagonal matrix element contains a quite nontrivial dependence on $N$. To see this more clearly, we consider the low-frequency 1/f noise. The function $\sin^2 (\omega t /2 N)/\omega^{2} = (t/2N)^2 {\rm sinc}^{2}(\omega t /2N)$ appearing in equation~(\ref{sinc}) is of window type, with width given by the first-order zeroes $\pm 2 \pi N/t$. Thus, if the noise $S(\omega ) \sim 1/\omega $ is negligible at $\omega = \pm 2 \pi N/t$, we can expand the window function around small $\omega$, obtaining
\begin{equation}
\rho_{01} (t) = \rho_{01} (0) e^{-t^{2}/\tau_{\rm deph}^{2}},
\end{equation}
with
\begin{equation}
\tau_{\rm deph} = \sqrt{2 N/\sigma},
\end{equation}
where $\sigma = (2\pi )^{-1}\int d\omega S_{\xi}^{(j)}(\omega )$ is the low-frequency  noise power of each physical qubit. In other words, when the number of physical qubits is increased to $N$, the dephasing time is increased by a factor of $\sqrt{N}$ compared to that of a single qubit. This phenomenon is qualitatively similar to the motional narrowing effect described in more detail in the following section.
Let us summarize the mechanism here: for low-frequency noise the dephasing depends quadratically on time, therefore
each physical qubit contributes to the dephasing by an amount proportional to $(t/N)^2$. The total dephasing coefficient in the exponent is the sum of all $N$ contributions from each physical qubit, and as result the scaling of the exponent is as $-t^2 /N$.

This prediction has been recently tested experimentally by measuring Ramsey interference fringes for a logical qubit moved between $N=2$ and $N=3$ physical qubits (phase qubits and X-mons)~\cite{averin}.
Ramsey experiments can indeed provide a direct measurement of $\tau_{\rm deph}$. In Ramsey experiments, a $\pi /2$ pulse is applied to the qubit to the $|0\rangle \rightarrow |1\rangle $ transition putting these two states in an equal superposition, then the qubit is left to evolve freely during a time $t$, and finally another $\pi /2$ pulse is applied to read the phase difference accumulated between the two states during the time interval $t$. In this case the envelope of the oscillation in the population measured in the experiments is of the form
\begin{equation}
\exp\left[ -\frac{1}{2}\Gamma_{1} t - \left(\frac{t}{\tau_{\rm deph}}\right)^2 \right],
\end{equation}
where $\Gamma_{1}$ is the single-qubit relaxation rate. The relaxation rate can be measured in a separate experiment, for example by exciting the qubit with a $\pi$ pulse and measuring the exponential decay of the population in the state $|1\rangle$. The Gaussian decay rate with time constant $t_{\rm deph}$ can therefore be extracted from the experiment.

\subsubsection{Random telegraph noise and motional effects}

When the incoherent modulation is non-Gaussian, the contributions from higher order correlation functions become relevant. Typical examples arise in real and artificial atomic systems with environments whose fluctuations result in random jumping between two different values of the transition energy, leading into characteristic phenomena of motional averaging and narrowing~\cite{Abragam,Anderson54} of spectral lines as a function of the jumping amplitude. Such processes occur naturally in atomic ensembles and condensed-matter systems which experience variations in the electronic state populations, chemical potential, molecular conformation, effective magnetic fields, lattice vibrations, microelectric fields producing dynamic (ac) Stark shifts and so on~\cite{Abragam, Anderson54, Jiang08, Kohmoto94, Berthelot06, Sagi10}. Also, discrete and random modulations of the transition frequency arise when one studies $1/f$-noise caused by random two-level fluctuators in superconducting qubits~\cite{Gassmann02,Burkard09}. Closely related phenomena include the Dyakonov-Perel effect~\cite{Dyakonov72}, the Dicke line narrowing of Doppler spectra~\cite{Dicke53}, and the quantum Zeno effect~\cite{Milburn88}.

The incoherent modulation of the transition energy between two discrete values can be modelled as a random-telegraph noise (RTN) created by a symmetric two-valued Poisson fluctuation process. The statistical properties of the RTN are, thus, described by the factorization rule~\cite{Wodkiewicz84}
\begin{equation}
\langle \xi(t_1)\xi(t_2)\cdots \xi(t_n)\rangle_{\xi}=\langle \xi(t_1)\xi(t_2)\rangle_{\xi}\langle \xi(t_3)\cdots\xi(t_n)\rangle_{\xi}.
\end{equation}
Thus, the RTN is inherently non-Gaussian. Nevertheless, a symmetric ($\langle \xi(t)\rangle_{\xi} = 0$) RTN is completely defined by the  second order moment (autocorrelation function)
\begin{equation}\label{eq:RTNnoise}
\langle \xi(t) \xi(0)\rangle_{\xi} = \xi^2e^{-2\Omega |t|},
\end{equation}
where $\xi$ is the fluctuation amplitude and $\Omega$ is the average fluctuation frequency. A Gaussian process with similar autocorrelation function is called Ornstein-Uhlenbeck process, and used to characterize Brownian motion~\cite{Ornstein30}.

A possible generalization for the symmetric random telegraph noise would be that one allows different correlations $\expes{\xi(t_{n+1}) \xi(t_n)}$ between the sequential values of a given realisation of the noise, which could be modelled, for example, so that the values of the process $\xi(t)$ would be sampled from a stationary distribution $P(\xi)\D{\xi}$ instead of choosing $\pm\xi$ one after another. Another generalization would be that the dwelling (waiting) time $t=t_{n+1}-t_n$ would be distributed non-exponentially. Experimentally, one can attempt to control the noise processes and correlations by invoking the methods used in a superconducting qubit measurement~\cite{JianLi12} where the RTN of the transition energy was simulated by creating random magnetic flux pulses with an arbitrary waveform generator. In such case, the generalizations need only reprogramming of the waveform generator, and it could enable, e.g., simulation of motional effects for continuous heavy-tailed stationary distributions~\cite{Sagi11}. Nevertheless, we consider here the pure random telegraph noise. 

\begin{figure}[t!]
\begin{center}
\includegraphics[width=0.7\linewidth]{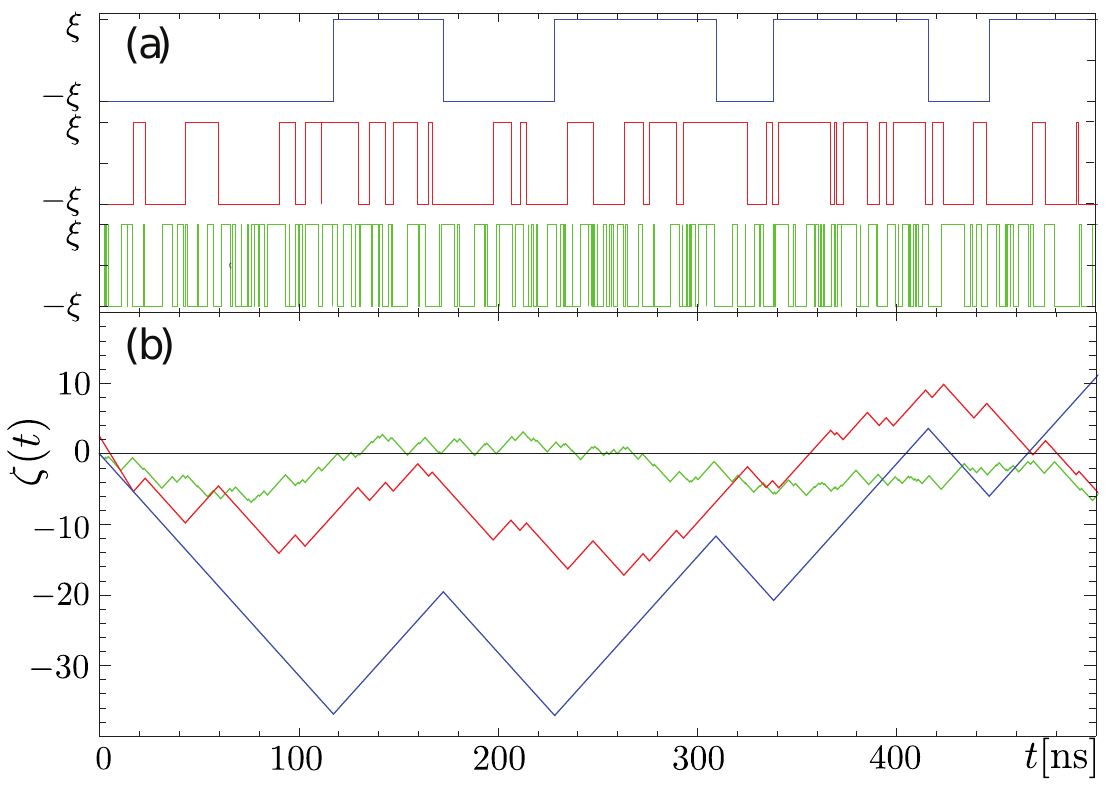}   
\caption{\label{u_fig}(a) Sample trajectories of $\xi(t)$ at jumping frequencies $\Omega=10$ MHz (blue), $100$ MHz (red), and $400$ MHz (green), shifted vertically for clarity.  (b) The corresponding trajectories of the accumulated phase $\zeta(t)=\int_0^t\xi(\tau)\D{\tau}$ with $\xi/2\pi=50$ MHz.} 
\end{center}
\end{figure}

Without going deeper into the qubit dynamics, two intuitive asymptotic limits can be found by a simple argumentation. First, in the fast jumping limit, the energy fluctuations $\Delta E=\hbar\xi$ occur so frequently, with average spacing $\Delta t=\Omega^{-1}$, that the ability to resolve them is fundamentally limited by the energy-time uncertainty relation~\cite{Sakurai}: $\Delta E \Delta t=\hbar\xi/\Omega\gtrsim \hbar$. Then, external driving or a measuring pulse observes that the transition frequency is dynamically averaged to $\omega_0$, although the system is at any given time in either one of the states $\omega_0\pm\xi$ and spends no time in between. This is referred to as motional averaging. In contrast, when the mean time $\Delta t=\Omega^{-1}$ between the jumping events is long enough so that $\xi/\Omega\gg 1$, one can consider the system as a statistical average over two systems with transition frequencies $\omega_0\pm\xi$. This is denoted as the slow jumping limit. In this limit, one can simply ignore the dynamics of the $\xi(t)$-process as measurement or control pulses needed to resolve the energy difference $\Delta E=\hbar \xi$ are shorter than $\Omega^{-1}$. In the cross-over region, neither the statistical nor dynamical averaging works and one has to take into account the frequency fluctuations explicitly. 

Quantitatively, the accumulated phase factor $A(t)=e^{i\zeta(t)}$ is evidently, again, an important quantity. Sample paths of $\zeta(t)=\int_0^t\xi(\tau)d\tau$ are shown in figure~\ref{u_fig}. In the slow jumping limit (blue, $\Omega\ll\xi$), the slopes are well resolved between the jumping events. In contrast, in the fast jumping limit (green, $\Omega\gg\xi$), the trajectory becomes diffusive and the slopes are indistinguishable. In this limit, the spectral broadening, i.e., the uncertainty of the frequency becomes proportional to the diffusion coefficient $\xi^2/\Omega$ of the process~\cite{Dykman10}. For a physical insight, one can draw an analogy between the $\xi(t)$-process and a randomly sign-changing velocity of a 1D particle, which implies that the accumulated phase $u(t)$ is analogous to the position of the 1D particle. Then, the transition from the slow jumping to the fast jumping limit is analogous to the transition from the ballistic to the diffusive evolution under an increasing rate of elastic collisions, i.e., the rate of sign changes in the 1D velocity~\cite{Sagi11}.

\begin{figure}[t!]
\begin{center}
\includegraphics[width=0.7\linewidth]{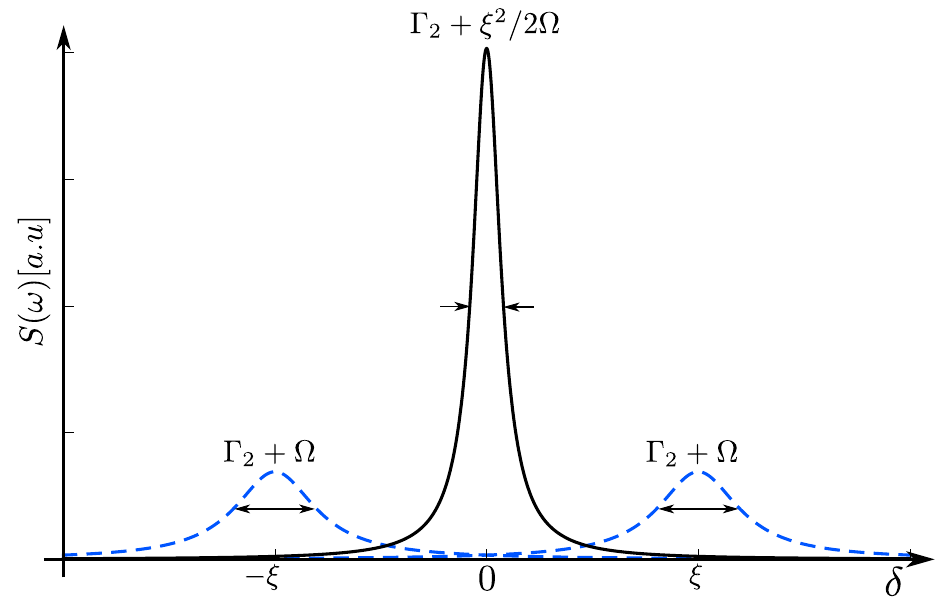}
\caption{Schematic of the motional effects. In the slow jumping limit ($\Omega\ll \xi$, the qubit absorbs energy at the frequencies $\omega = \omega_0\pm \xi$ (blue). The linewidth is given by $\Gamma_2+\Omega$. When the jumping frequency is increased to $\Omega\gg\xi$, the two resonance peaks is subjected to motional averaging and merge into a single peak at $\omega=\omega_0$. Also, the width of the peak experiences motional narrowing, and is given by $\Gamma_2+\xi^2/2\Omega$.} \label{fig:specMA}
\end{center}
\end{figure}

For the RTN process $\xi(t)$, equation~(\ref{eq:qspec}) gives (see references~\cite{Abragam, Wodkiewicz84, Anderson54, JianLi12})
 \begin{equation}
 S(\omega)= \frac{g_{\rm P}^2}{2}\frac{2\JumpR\Dev^2+\Gamma_2\left[(\Gamma_2+2\Omega)^2 +\delta^2+\Dev^2\right]}{[\Dev^2-\delta^2+\Gamma_2(\Gamma_2+2\Omega)]^2 +4(\Gamma_2+\JumpR)^2\delta^2} \label{eq:spect-mot-aveg-full},
\end{equation}
where $\delta=\omega_0-\omega$. At the asymptotic limits, the full spectrum reduces to the forms
\begin{eqnarray}
S_\pm(\omega)&=&\frac{g_{\rm P}^2}{4}\frac{\Gamma_2+\Omega}{\left[\delta\pm\xi\right]^2+\left(\Gamma_2+\Omega\right)^2}, \ \Omega\ll\xi, \\
S(\omega)&=&\frac{g_{\rm P}^2}{2} \frac{\Gamma_2+\xi^2/2\Omega}{\delta^2+\left(\Gamma_2+\xi^2/2\Omega\right)^2}, \ \Omega\gg\xi,\label{fast-spec}
\end{eqnarray}
which confirm the intuitive argumentation at the slow and fast jumping limits. In the slow jumping limit, the qubit absorbs energy from the probe at $\omega=\omega_0\pm\xi$ and the decoherence rate is increased by $\Omega$, in agreement with the mean lifetime $1/\Omega$ of the energy levels $\omega_0\pm\xi$ (exchange broadening~\cite{Anderson54}). In the fast jumping limit, the qubit absorbs at the average frequency $\omega=\omega_0$  with the excess $\xi^2/2\Omega$ of the decoherence rate. In the fast jumping limit, we see that the decoherence rate is inversely proportional to the fluctuation frequency. This is depicted in figure~\ref{fig:specMA} and referred to as motional narrowing~\cite{Anderson54,Abragam}.

The effects of power broadening due to, e.g., the probe signal can be taken into account by calculating the excited state population in the steady state with stochastic master equation~\cite{JianLi12,SilveriPhD}. This can be done analytically by employing the final value theorem, and the result can be written as
\begin{equation}
P_{\rm e} = \frac{g_{\rm P}^2\Omega[2\xi^2\Omega+\Gamma_2(\Omega_{\rm P}^2+4\Omega^2)}{D},
\end{equation}
where $g_{\rm P}$ is the probe amplitude, $\Omega_{\rm P} \equiv \sqrt{\delta^2+g_{\rm P}^2}$, and
\begin{eqnarray}
D=4\Gamma_1\xi^2\Omega\delta|\delta| + 2\Gamma_1 \xi^4\Omega &+& \xi^2\left[g_{\rm P}^2\Gamma_1\Gamma_2 +4\Omega^2(g_{\rm P}^2+2\Gamma_1\Gamma_2)\right]\nonumber\\ &&+2\Omega(\Omega_{\rm P}^2+4\Omega^2)\left[g_{\rm P}^2\Gamma_2+\Gamma_1[\Gamma_2^2 +\delta^2]\right].
\end{eqnarray}


\section{Harmonic oscillator}\label{sec:ho}
The harmonic oscillator is one of the most commonly employed model in physics, both in its classical and  quantum versions. The generic Hamiltonian of the harmonic oscillator is $H=\frac{1}{2m}p^2 +\frac12kx^2$. Here we have used the canonical variables $x$ and $p$, which can correspond to a particle's coordinate and momentum, and the parameters $k$ and $m$, which can correspond to the spring constant and the mass of the particle. The eigenfrequency of the oscillator is $\omega_0=\sqrt{k/m}$. After quantization, the Hamiltonian can be written $\hat H=\hbar\omega_0(\hat a_0^\dagger\hat a_0+\frac12)$, where $\hat a_0^\dagger$ and $\hat a_0$ are creation and annihilation operators. The energy spectrum consists of equally spaced levels $E_n=\hbar\omega_0(n+\frac12)$ with $n=0,1,\ldots$.

Besides the two-state system, the harmonic oscillator is another basic model for studying frequency modulation. The frequency can be modulated by changing  nonproportionally  either $m$ or $k$ resulting in a different eigenfrequency $\omega_1=\omega_0 (1-\frac{\Delta m}{2m}+\frac{\Delta k}{2k})$ to the first order in the displacements $\Delta m$ and $\Delta k$. Measuring the changes in the oscillator frequency can be used as an ultra-sensitive probe to various physical phenomena. These type of approaches have demonstrated remarkable single-atom~\cite{jensen_atomic-resolution_2008}, single-molecule~\cite{burg_weighing_2007, naik_towards_2009, hanay_single-protein_2012}, single-spin~\cite{rugar_single_2004} and sub 100 nm imaging~\cite{mamin_nuclear_2007} resolutions. We concentrate here on reviewing some quantum phenomena of frequency modulations of an oscillator, that is, squeezing, dephasing, and effects of the quantum vacuum and amplification. 

\subsection{Squeezing by sudden frequency modulation}\label{sec:sbsfm}
Let us consider a harmonic oscillator with the Hamiltonian
\begin{equation}
  \hat{H}_0=\hbar \omega_0 \Big( \hat{a}_0^\dagger \hat{a}_0 +\frac 12 \Big), \label{eq:H0}
\end{equation}
and assume that the frequency of the oscillator changes suddenly from $\omega_0\to \omega_1$ resulting in a new Hamiltonian
\begin{equation}
  \hat{H}_1=\hbar \omega_1 \Big( \hat{a}_1^\dagger \hat{a}_1 +\frac 12 \Big). \label{eq:H1}
\end{equation}
With the term `sudden' we mean a change occurring in the time scale much smaller than any natural timescale of the system, such as the oscillation periods $2\pi/\omega_{i}$, see also section~\ref{sec.multilevel.sudden}. In a sudden change, there is no time evolution during the transition, that is, the wavefunction remains the same while only the Hamiltonian changes $\hat H_0\to \hat H_1$~\cite{Schiff,Messiah}. For example, the ground state of the Hamiltonian $\hat H_0$ can be a non-trivial excited state of the Hamiltonian $\hat H_1$.

Let us study the transition $\hat H_0\to \hat H_1$ and assume, for simplicity, that the frequency change occurs due to a change in the spring constant. Notice that in equations~(\ref{eq:H0})-(\ref{eq:H1}) the creation and annihilation operators are not identical before and after the change since they depend explicitly on the eigenfrequency of the oscillator as
\begin{equation}
  \hat a_i= \sqrt{\frac{m \omega_i}{2 \hbar}}\left ( \hat x + \ii \frac{\hat p}{m \omega_i} \right) \label{eq:ai},
\end{equation}
written with the help of the position $\hat x$ and momentum $\hat p$ operators. Since the quantum wavefunction stays unchanged also physical observables such as position $\expes{\hat x}$ and momentum $\expes{\hat p}$ cannot change either during the transition. This gives a justification to solve the relation between old and new bosonic creation and destruction operators:
\begin{equation}
  \hat a_1=\frac{\omega_1+\omega_0}{2\sqrt{\omega_0\omega_1}} \hat a_0 + \frac{\omega_1-\omega_0}{2\sqrt{\omega_0\omega_1}} \hat a^\dagger_0=\cosh r \hat a_0-\sinh r \hat a_0^\dagger .\label{eq:sqa}
\end{equation}
This Bogoliubov transformation is in fact a squeezing transformation $\hat a_1=\hat{S}(r)\hat a_0 \hat{S}^\dag(r)$ with the unitary operator $\hat{S}(r)=\exp\left(\frac{r}{2} [(\hat{a}_0^\dagger)^2-\hat{a}_0^2]\right)$ and the squeezing parameter $r=\ln \sqrt{\frac{\omega_0}{\omega_1}}$. After a sudden change of the frequency, the quantum state of the oscillator expressed in the energy eigenbasis of the new Hamiltonian is a squeezed state~\cite{graham_squeezing_1987, kiss_time_1994, abdalla_squeezing_1993}, denoted formally as  $\ket{\psi_1}=\hat S(r)\ket{\psi_0}$. Squeezed states are a particular case of Gaussian (the Wigner function is Gaussian) continuous-variable states; operating with Gaussian states is theoretically a great simplification, owing to the development of theoretical tools~\cite{RevModPhys.84.621} and simple analytical formulae for almost any quantity of interest({\it e.g.} fidelity~\cite{PhysRevA.61.022306}, entanglement~\cite{Simon2000,Duan2000}, and distance measures~\cite{PhysRevA.58.869}). Assuming $\omega_0>\omega_1$, the uncertainty of the other quadrature is squeezed by the factor of $\exp(-r)=\sqrt{\omega_1/\omega_0}$, the other increased by the factor of $\exp(r)=\sqrt{\omega_0/\omega_1}$, and the product still satisfies the Heisenberg uncertainty principle, see more in reference~\cite{WallsMilburn}.

From a practical point of view, it is quite likely that the amount of squeezing production one can achieve in a single frequency change is limited. To overcome this limitation, it is natural to ask whether it would be possible to generate squeezing by repeating a cycle where the sudden change $\omega_0\to \omega_1$ is followed by a slow, adiabatic change $\omega_1\to \omega_0$. Notice that one cannot make another sudden change back since it would invert the squeezing achieved during the first part of the cycle. Furthermore, the squeezing occurs along a fixed direction of the $X$-quadrature in the phase space. When considering cyclic squeezing it is important to notice that the squeezed state rotates in the phase space in between of the squeezing events by the linear part of Hamiltonian. For most efficient cyclic squeezing production, the period of the frequency modulation cycle needs to be timed so that sudden change of the frequency incrementally increases the squeezing always along the same direction. Squeezing production via this route was originally studied and proposed in references~\cite{janszky_strong_1992, averbukh_enhanced_1994} in a general setting. Recently, new theoretical interest has arisen in this problem~\cite{zagoskin_heat_2012, zagoskin_controlled_2008} motivated by superconducting circuit technology where the superconducting oscillator parameters are rapidly and accurately tunable.

Let us now generalize the squeezing transitions for the continuously modulated harmonic oscillator $\hat H(t)=\hbar \omega(t) (\hat a^\dag_\omega(t) \hat a_\omega(t) +\frac 1 2 )$, where the annihilation $\hat a_\omega(t)$ and creation operators $\hat a^\dag_\omega(t)$ refer to operators between adiabatic Fock states. As discussed above, when the frequency changes the wavefunction experiences squeezing, that is, transitions between the adiabatic Fock states. To include this explicitly in the Hamiltonian, we make the inverse squeezing transformation $\hat a=\hat S^\dag(t) \hat a_{\omega}(t) \hat S(t)$ which is a formal solution of $\hat a_{\omega}(t)=\hat S(t)\hat a \hat S^\dag(t)$. This is equivalent to the transformation into a static orthonormal basis (see section \ref{sec:timedep})~\cite{Law94}. Here $\hat a$ is an annihilation operator for the states in the static basis corresponding to a fixed frequency $\widetilde{\omega}$ and $\hat S(t)=\exp\left(\frac{r(t)}{2} \left[(\hat{a}^\dagger)^2-\hat{a}^2\right]\right)$ where $r(t)=\ln \sqrt{\frac{\widetilde{\omega}}{\omega(t)}}$. This results in
\begin{eqnarray}
\hat{H}(t)&=\hat S^\dag(t)\left[\hbar \omega (t) \left(\hat a_\omega(t)^\dagger \hat a_\omega(t)+\frac 1 2 \right)\right]\hat{S}(t) +\ii \hbar [\partial_t \hat{S}^\dag(t)]\hat{S}(t) \\
&=\hbar \omega(t)\Big(\hat a^\dagger \hat a +\frac 12\Big) - \frac{\ii \hbar}{4} \frac{ \D{\ln \omega(t)}}{\D{t}}\left[\hat a^2- (\hat a^\dagger)^2\right]. \label{eq:hamsq}
\end{eqnarray}
This form clearly shows that rapid changes in $\omega(t)$ will generate squeezing through the latter term in the Hamiltonian.

For a complementary point of view, let us consider the squeezing production from an opposite direction. While the fast frequency changes produce useful effects such as squeezing, there are situations where we would like to avoid as much as possible these additional excitations. This is the case of quantum engines (discussed further in section~\ref{s.thermal}), where the harmonic oscillator plays often the role of the working fluid. To achieve this, we can use the method of transitionless driving presented in section \ref{sec:timedep}. The calculations are straightforward~\cite{Muga2010,Deng13}, however, to illustrate the power and simplicity of this formalism we give a detailed calculation here. The frequency modulated harmonic oscillator can be written as
\begin{equation}
\hat H(t) = \frac{\hat p^2}{2m} + \frac{1}{2} m \omega^2(t)\hat x^2,
\end{equation}
where $\hat x = \sqrt{\hbar /(2m\omega)}(\hat a + \hat a^{\dag})$, $\partial_t \hat H = m \omega \dot{\omega}\hat x^2 = \frac 1 2 \hbar \dot{\omega} (\hat a + \hat a^{\dag})^2$ and $E_{n}(t) - E_{k}(t) = \hbar \omega(t) (n -k)$.  Inserting these results into equation~(\ref{eq:CD_formula}) we find that the matrix elements for the counterdiabatic correction $\hat H_{\rm CD}$ of the harmonic oscillator are
\begin{equation}
\langle \psi_k(t) | \hat H_{\rm CD}(t) |\psi_n(t)\rangle = \ii \hbar \frac{\dot{\omega}(t)}{4\omega(t)}\left\{ \begin{array}{cc} \sqrt{n(n-1)} & {\rm if}~ n=k+2 ,\\ -\sqrt{(n+1)(n+2)} & {\rm if}~ n=k-2 ,\\ 0 & {\rm otherwise}, \end{array} \right.
\end{equation}
which leads to
\begin{equation}
\hat H_{\rm CD}(t)= \frac{\ii \hbar}{4} \frac{\rm{d} \ln \omega (t)}{\rm{d} t}\left[ \hat a^2 - (\hat a^\dag)^2\right]=-\frac 1 4 \frac{\rm{d} \ln \omega (t)}{\rm{d} t}\left(\hat x \hat p+ \hat p \hat x\right). \label{CD_harmonic}
\end{equation}
Note now the similarity between this result and equation~(\ref{eq:hamsq}); indeed, equation~(\ref{CD_harmonic}) shows that in order to maintain an adiabatic transitionless evolution for the harmonic oscillator, one has to add a term in the Hamiltonian that would cancel exactly the squeezing-producing Hamiltonian in equation~(\ref{eq:hamsq}). It is also instructive to calculate the Heisenberg equations of motion $\dot{\hat x}=\frac{\ii}{\hbar} [\hat H_0(t)+\hat H_{\rm CD}(t), \hat x]$. One notices that the correction term $\frac \ii \hbar [ \hat H_{\rm CD}(t), \hat x]$ in the right-hand side cancels the left-hand side term originating in the explicit time dependence of $\omega$ in $\hat x = \sqrt{\hbar /(2m\omega(t))}(\hat a + \hat a^{\dag})$. Doing similarly also for $\hat p$ results in the usual $\dot{\hat a} = -\ii \omega (t) \hat a$ as desired.

\subsection{Effects of the quantum vacuum and amplification by parametric modulation}
A transmission line is described as an ensemble of harmonic modes associated to each wave vector. If the line contains tunable elements such as SQUID loops, then it is possible to change rapidly the frequency of these modes. In the Lagrangian of the transmission line, this means changing the inductance per unit length, while the capacitance per unit length remains constant. If this change is realized over a distance comparable to the wavelength of the mode, one can regard this as a change in the speed of propagation or in the index of refraction. If the change is local, over a distance much shorter than the wavelength, then it can be assimilated to a change in a boundary condition of the field. Using this observation, one can create parametric perturbations and observe effects that are analogous to those expected in vastly different systems: from vacuum instability due to strong electric fields (Schwinger effect) to the Hawking radiation emitted at the event horizon of black holes~\cite{Nation12}. An example of such an effect that has been experimentally observed is the dynamic Casimir effect ~\cite{Wilson11, Lahteenmaki13}. The (static) Casimir effect is the attraction between two plates of metal that realize boundary conditions for the electromagnetic field due to different vacuum energies of the field between and outside the plates. One wonders what happens if the plates move: in this case, real photons are generated from vacuum fluctuations. To realize this experimentally, one could use either a single modulated SQUID at the end of the line ~\cite{Wilson11} or many SQUIDs that form a leaky cavity~\cite{Lahteenmaki13}: in both cases, this creates an effective tunable boundary condition for the incoming vacuum-state field. The observed photons are two-mode squeezed, as one expects from theoretical considerations similar to those leading to equation~(\ref{eq:hamsq}).

A related development is the recent surge of interest in using similar devices as parametric amplifiers~\cite{Lahteenmaki14}. In this case, the input field is not the vacuum but a real signal that we wish to amplify. In nanoelectronics, these amplifiers have important advantages over semiconductor-based HEMT (high-electron-mobility transistor) amplifiers: the added noise referred to the input is near the limit permitted by quantum physics and equals 1/2 for large gain, thus enabling a consistent increase in the signal-to-noise ratio in sensitive measurements. Currently they are indispensable in the research on superconducting qubits. Several designs based on modulated SQUIDs in ring configuration~\cite{Bergeal10,Roch12} or arrays of SQUIDs were proven successful~\cite{Castellanos-Beltran07,Castellanos-Beltran08}, and wideband traveling-wave versions based on the phase matching of the four-wave mixing by using resonant inclusions were developed recently~\cite{OBrien14}. An important direction previously unexplored with microwave photons is the use of the Gaussian states generated by parametric processes for quantum processing of information~\cite{Lloyd:Braunstein:1999,RevModPhys.77.513,RevModPhys.84.621}. To achieve this one should be able to create complex entangled quantum states (cluster states) and implementing quantum gates, for example using the measurement-based (one-way) approach to quantum computing. So far a first step has been made into this direction by pumping a resonant circuit at two different frequencies ({\it i.e.} producing a ``double dynamical Casimir effect''), which results in the creation of a three-mode states~\cite{Paraoanu:coherence}  that display genuine tripartite entanglement~\cite{Bruschi2016}.

\subsection{Dephasing by stochastic frequency changes}
Let us now turn our attention to the stochastic variation of the frequency of the oscillator, $\omega(t)=\omega_0+\xi(t)$. Here, $\xi(t)$ denotes the displacement of the frequency from the mean value $\omega_0$. Furthermore, $\xi(t)$ is assumed to be a stochastic process whose time-evolution is characterized by probabilistic laws. In the perspective of Hamiltonian~(\ref{eq:hamsq}), every time there is a change in the stochastic displacement $\xi(t)$ one observes also squeezing production. But, because the squeezing is in random directions with variable magnitudes, it averages out. Hence, it is safe to neglect the squeezing part of equation~(\ref{eq:hamsq}) and consider the intuitive time-dependent Hamiltonian,
\begin{equation}
  \hat{H}(t)=\hbar [\omega_0+\xi(t)] \Big( \hat a^\dagger \hat a +\frac 12\Big). \label{eq:flucham}
\end{equation}
The literature of random changes in the frequency of a harmonic oscillator has a long history since it is  easy to imagine various uncontrolled physical process, which can be modeled as stochastic modulations of the oscillator frequency. Just to mention a few, stochastic frequency fluctuations can be caused by changes in the interacting environment~\cite{Anderson54, anderson_exchange_1953}, fluctuations of spurious two-level systems~\cite{gao_noise_2007}, attachment and diffusion of molecules or atoms~\cite{jensen_atomic-resolution_2008,Dykman10,atalaya_diffusion-induced_2011, hanay_single-protein_2012}, or uncontrolled transitions in a coupled ancillary qubit~\cite{LaHaye09, reagor_quantum_2015}.

Consider a quantum superposition of two Fock states $\ket{\psi}=(\ket{n}+\ket{n+1})/\sqrt{2}$ that evolves under the fluctuating Hamiltonian~(\ref{eq:flucham}). By ignoring the phase evolution by the mean $\omega_0$, we get that the time-evolved state is $\hat\rho(t)=\ket{\psi(t)}\bra{\psi(t)}$
\begin{eqnarray}
\hat{\rho}(t)=\frac{1}{2}\Big(&\ket{n}\bra{n}+\ee^{\ii \int_0^t \xi(\tau)\D{\tau}}\ket{n+1}\bra{n}\nonumber\\&+\ee^{-\ii \int_0^t \xi(\tau)\D{\tau}}\ket{n}\bra{n+1} +\ket{n+1}\bra{n+1}\Big),
\end{eqnarray}
where we clearly observe that the dynamical phase factor $A(t)=\exp\big(\ii \int_0^t \xi(\tau)\D{\tau} \big)$ has again an important role. As we have no knowledge on the specific realization of the stochastic process the best we can do is to average $A(t)$ over all possible realizations resulting in
\begin{eqnarray}
\hat{\rho}(t)=\frac{1}{2}\Big(&\ket{n}\bra{n}+\expes{\ee^{\ii \int_0^t \xi(\tau)\D{\tau}}}_\xi \ket{n+1}\bra{n}\nonumber\\&+\expes{\ee^{-\ii \int_0^t \xi(\tau)\D{\tau}}}_\xi\ket{n}\bra{n+1} +\ket{n+1}\bra{n+1}\Big),\label{eq:rhodephasing}
\end{eqnarray}
which shows that the averaging (denoted by $\expes{\cdot}_\xi$) leads to dephasing since $\hat{\rho}$ is a mixed state when $|\expes{\exp\big(\ii \int_0^t \xi(\tau)\D{\tau}\big)}_\xi| < 1$. The above discussion can be generalized to
\begin{eqnarray}
\hat{\rho}(t)=\sum_{nm}c_{nm}\expes{\ee^{(n-m)\ii \int_0^t \xi(\tau)\D{\tau}}}_\xi\ket{n}\bra{m},
\label{eq:rhodephasing_gen}
\end{eqnarray}
showing that phase coherence  decays faster for pairs with larger energy difference.

Let us consider a simple example demonstrating the loss of phase coherence and the role of temporal fluctuations in $\xi(t)$. Assume that $\xi(t)$ has no temporal fluctuations or the fluctuations are so slow that in the relevant time scale of the experiment they can be neglected. Then we need to know only a static distribution of $\xi$ which is assumed to be a Gaussian distribution centered at zero and with variance $\delta^2$. The static distribution characterizes the frequency variation of an oscillator between different experimental runs or the statistical static variation in an ensemble of oscillators. The resulting phase factor is $\expes{\ee^{\ii \int_0^t \xi(\tau)\D{\tau}}}_\xi=\int_{-\infty}^\infty P(\xi)\ee^{\ii \xi t } \D{\xi}=\exp(-\delta^2 t^2/2)$ which implies that in the limit $t\to \infty$ the state~(\ref{eq:rhodephasing}) approaches a fully mixed state because of loss of phase coherence by ensemble averaging. We now assume that $\xi(t)$ is a genuine stochastic process with temporal fluctuations, but at every instant of time it has a frequency distribution which is Gaussian centered at zero and with variance $\delta^2$. In section~\ref{Sec:stochmod}, we saw that for the Gaussian modulations of a two-level system, the fluctuation leads to motional narrowing, that is, the phase factor decaying slower, $\exp(-t \Gamma_{\phi})$, in the presence of fluctuations. Sagi \textit{et al.}\ in reference~\cite{Sagi11} show that this is generally true regardless of the type of the temporal fluctuations if the marginal frequency distribution is Gaussian. They also show that for heavy-tailed frequency distributions temporal fluctuations are expected to result in motional broadening which is that the phase factor decays faster than in the static case.


\section{Coupled, multilevel, and many-body systems}
\label{sec:coupled}

Frequency modulation can be realized in a variety of complex quantum systems such as coupled qubits and resonators, and atomic gases. As such, it is a vast topic, and we do not aim to give an exhaustive review on its developments here. We will first concentrate on two prototype examples, the Jaynes-Cummings and radiation pressure Hamiltonians, then consider other systems including non-interacting and interacting many-body systems consisting of identical particles, and finally discuss level modulation in the context of quantum thermodynamics. The examples are chosen because of their broad applicability to multitude of different physical realisations, and also because of their timely character in the modern literature. 

\subsection{Modulated Jaynes-Cummings Hamiltonian}
The Jaynes-Cummings (JC) model~\cite{WallsMilburn} is the backbone for the study of the intertwining of matter and light. It has found applications in a myriad of different physical systems, ranging from quantum optics and quantum electrodynamics (QED) to ion traps, vibronic transitions in molecules, and solid-state quantum information science. In its simplicity, the JC model couples a two-level atom (transition frequency $\omega_0$) to a quantized electromagnetic field mode (frequency $\omega_{\rm c}$). The coupling, characterized by the coupling strength $g$, allows the coherent exchange of an energy quantum between the atom and the radiation field. Quantum effects are expected to be seen in the strong coupling limit, where $g>\kappa$, with $\kappa$ being the cavity dissipation rate. In addition, the coupling of the two-level atom with the environment is modeled by the coupling constant $\gamma$.

The typically weak interaction can be strengthened by confining the atom and the field inside an optical or microwave cavity, which decreases the mode volume and, therefore, increases the coupling~\cite{Walther06}. In optical realizations, the strength of the interaction is typically determined as $g=\mathcal{E}_{\rm rms} d/\hbar$, where $d$ is the atomic dipole moment  and $\mathcal{E}_{\rm rms}$ the root mean squared zero-point fluctuations of the electric field in the cavity (see more in reference~\cite{WallsMilburn}). In modern transmon qubit-based superconducting circuit configurations, analogously, the coupling strength describes how strongly the voltage of the microwave cavity or transmission line couples to the Cooper-pair charge of the qubit~\cite{Blais04, Wallraff04, Koch07}. This setup allows a broad control over all of the parameters of the JC problem~\cite{Schoelkopf08}. This circuit version of the more conventional cavity QED has been dubbed in the literature as circuit QED.

The Jaynes-Cummings Hamiltonian is given by
\begin{equation}\label{eq:JC}
\hat{H} = \hbar \omega_{\rm c} \hat{a}^{\dag}\hat{a} + \frac{\hbar\omega_0}{2}\hat{\sigma}_z + \hbar g ( \hat{a}^{\dag}\hat{\sigma}_- + \hat{a}\hat{\sigma}_+),
\end{equation} 
where we have neglected the energy related to the zero-point fluctuations of the cavity. The above Hamiltonian is obtained from the more general Rabi Hamiltonian by making a RWA where the terms $\hat{a}^{\dag}\hat{\sigma}_+$ and $ \hat{a}\hat{\sigma}_-$ have been neglected. These terms that do  not conserve the excitation number are negligible with naturally occurring coupling rates ($g\ll \omega_{\rm c}$) in NMR, cavity QED, or in standard experiments in circuit QED. With an innovative sample design, the ultrastrong coupling can be reached in a circuit QED setups~\cite{Niemczyk10, FornDiaz10, yoshihara16, forn-diaz16}. In addition to direct engineering, the ultrastrong coupling can be achieved with a quantum simulation~\cite{Ballester12} in a JC system with two properly chosen external fields.

When the cavity is driven strongly, the JC Hamiltonian reduces to the transversely driven qubit Hamiltonian, which results in Rabi oscillations of the excited state population of the atom~\cite{Cohen}. As a consequence, the emission spectrum of the atom displays the so-called Mollow triplet~\cite{Mollow69}. In this review, we concentrate on the situation where the coupling between the cavity and the atom is at the quantum level, and either the qubit or the cavity frequency is modulated. We build our discussion on the two previous sections, which dealt with each of the modulated subsystems separately.

\subsubsection{Spectrum of the Jaynes-Cummings Hamiltonian}

\begin{figure}[t!]
\begin{center}
\includegraphics[width=0.7\linewidth]{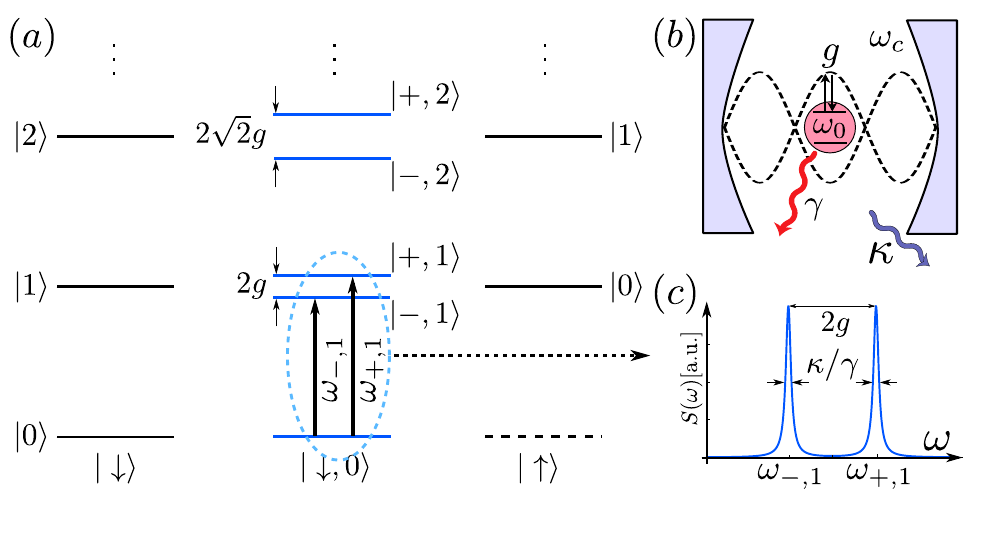}
\caption{\label{fig:vacuumRabi}(a) Energy level diagram of the JC Hamiltonian. We show with black the uncoupled basis states for the qubit (left to right) and for the cavity (down to up). The JC energies are shown in blue. The vertical dots emphasize the continuation of the energy levels with increasing $n$. (b) Sketch of the JC interaction between the two-level atom and the cavity field. (c) Probe spectrum displaying the vacuum Rabi mode splitting in the strong coupling regime. The widths of the peaks are determined by $\kappa$ or $\gamma$, depending on to which degree of freedom the probe couples.}
\end{center}
\end{figure}

The eigenproblem for the non-modulated JC Hamiltonian is readily solved by analytic means. As a result, one obtains the characteristic ladder of energies given by
\begin{equation}
E_{\pm, n} = n\hbar\omega_{\rm c} \pm \frac{\hbar}{2}\sqrt{\delta_{\rm c}^2+4ng^2},
\end{equation}
where $\delta_{\rm c}\equiv \omega_0-\omega_{\rm c}$ is the detuning between the qubit and cavity frequencies, and $n$ is a positive integer. The states appear as pairs $|\pm,n\rangle$, but the ground state $|\downarrow,0\rangle$ is a single state with energy $E_0 = -\hbar\omega_0/2$. The doublets are created by the interaction $g$, which is most clearly seen in resonance ($\delta_{\rm c}=0$) as the removal of the degeneracy between the states $|\uparrow,n\rangle$ and $|\downarrow,n+1\rangle$. The resulting energy levels are depicted in figure~\ref{fig:vacuumRabi}.
The probe spectrum~(\ref{eq:Trate}) for the non-modulated JC Hamiltonian is readily solved. As a consequence, the bare qubit transition is split and observed in the spectrum as two Lorentzians at $\omega=\omega_{\pm,1}\equiv (E_{\pm,1}-E_0)/\hbar$, separated by the vacuum Rabi frequency $2g$. The necessary condition to resolve the doublet is that the strong coupling regime $g>\kappa$ is achieved. This phenomenon of vacuum Rabi mode splitting~\cite{Sanchez-Mondragon83,Agarwal84} was first observed with ensembles of atoms in optical cavities~\cite{Raizen89,Zhu90,Thompson92}, and later in semiconductor heterostructures~\cite{Weisbuch92,Wang95,Pau95} and with an ensemble of Rydberg atoms in a superconducting high-Q microwave cavity~\cite{Bernardot92}. Finally, also the single atom interplay with the cavity field was observed at quantum level with Rydberg atoms~\cite{Brune96}. Vacuum Rabi oscillations and the level splitting were observed in a circuit QED system~\cite{Wallraff04,Bishop08}, and have been demonstrated since in various experiments, by either spectroscopic means and/or in the time domain. Vacuum Rabi mode splitting is, at least in some sense, analogous to the normal mode splitting of two coupled classical oscillators. Thus, it has been pointed out~\cite{Carmichael96}, that in order to show the quantum character of the interaction one should demonstrate the $\sqrt{n}$ scaling of the doublet splittings. This was achieved with circuit QED~\cite{Fink08} and in a quantum-dot/microcavity system~\cite{Kasprzak10}.

\subsubsection{Modulation of the atomic frequency}

A working quantum computer requires means to transfer information between qubits. One of the most promising realisations for this is the cavity-QED setup~\cite{Cirac95,Blais07}, where the cavity acts as a storage and a mediator of information between qubits. The unitary time-evolution operator related to such an elementary two or many qubit process is called a quantum gate. In the cavity QED scheme, multiqubit gates can be constructed by coupling the qubits to a joint cavity where multiqubit interactions can be produced by real or virtual excitations between the qubits, created by sequences of rapid control pulses~\cite{Strand13}. It was discussed in the context of trapped ion qubits that also sideband transitions could be used for such purposes~\cite{Cirac95}. Conventionally, such sidebands are created by transverse driving on either the qubit~\cite{Liu07,Leek09} or the cavity~\cite{Chiorescu04, Blais07,Wallraff07}. However, sidebands created in such way unavoidably rely on two-photon processes which are proportional in $g^2$ and, thus, result in slow two-qubit gates.

Recently, it was proposed that the sidebands created by the longitudinal coherent modulation $\xi(t)\hat{\sigma}_z/2$ (see also section~\ref{Sec:cohmod}) with $\xi(t) = \xi\cos(\Omega t)$ are of first-order in $g$ and, therefore, can be used to create faster two-qubit gates~\cite{Beaudoin12}. In reference~\cite{Beaudoin12}, one can find a detailed discussion on the dynamics related to a red sideband process. They consider the error coming from the interaction with a spectator qubit and show how a controlled-NOT gate can be created in presence of a second excited state. Also, a suggestion for physical implementation with transmon~\cite{Koch07} qubits is given. Experimental demonstration of sideband transitions in a transmon system can be found in reference~\cite{Strand13}.

Typically, the interactions between the environment and the qubit-cavity system are treated separately when the coupling is weak, i.e. when  $g<\kappa,\gamma$~\cite{HarocheRaimond}. In the context of quantum information processing, and physics in general, one wishes to have as strong coupling as possible. When the coupling is ultra-strong, $g>\omega_{\rm c},\omega_0$, the RWA applied in the derivation of the JC Hamiltonian becomes insufficient. As a consequence, one should use the Rabi Hamiltonian in the calculation of the eigenstates and the corresponding energies of the qubit-cavity system. Especially, the inclusion of the terms that do not conserve the excitation number causes the Rabi ground state to deviate from the JC ground state. Thus, the dissipation model relaxing the system towards the JC ground state at zero temperature unavoidably brings the system out of its Rabi ground state. Reference~\cite{Beaudoin11} develops a Lindbladian master equation for the approximately diagonalized Rabi Hamiltonian and uses it to solve the issues arising from the application of the JC master equation for the Rabi Hamiltonian. As a result of this, the peaks in the vacuum Rabi splitting spectrum   become asymmetric, i.e. have different widths and heights due to the different dissipation rates for the two transitions.

\subsubsection{Modulation of the cavity frequency}

Modulations of the cavity frequency can possibly also be used to create entanglement between two qubits coupled to the same cavity~\cite{Xie07}. On a more fundamental level, the frequency modulation of the cavity can be created, e.g., by allowing one of its mirrors to vibrate. We assume here again that the modulation created by the vibrations is sinusoidal: $\xi(t)=\xi\cos(\Omega t+\theta)$. When the modulation can be assumed slow, we can apply the adiabatic approximation (see section~\ref{sec:background}) and as a result the modulated cavity Hamiltonian can be written as $\hbar\omega(t)\hat{a}^{\dag}\hat{a}$. The cavity QED setup with the modulated cavity was studied in the JC approximation in references~\cite{Law95,Yang04,Janowicz98}, and also in a Kerr-like medium in reference~\cite{Wang10}.

The vacuum Rabi peaks can be further split into two doublets by the frequency modulation of the cavity~\cite{Law95}. Assuming that the system at the initial time $t=0$ is in the atomic excited state, there is also the important effect of the phase $\theta$ of the modulation. For resonant modulation $\Omega=2g$ the spectrum can be written into the form~\begin{eqnarray}
S(\omega) &=& N\bigg| \frac{\alpha}{\Gamma_1+\ii(\delta+g-\xi/4)}+\frac{\beta}{\Gamma_1+\ii(\delta+g+\xi/4)}\nonumber\\
&&+\frac{\alpha^*}{\Gamma_1+\ii(\delta-g-\xi/4)}+\frac{\beta^*}{\Gamma_1+\ii(\delta-g+\xi/4)}\bigg|^2,
\label{evrfsd}\end{eqnarray}
where $\delta = \omega_0-\omega$, $N$ is a normalization constant and
\begin{equation}
\alpha=1-e^{-\ii\theta}, \ \ \beta = 1+e^{-\ii\theta}.
\end{equation}
\begin{figure}[t!]
\begin{center}
\includegraphics[width=0.7\linewidth]{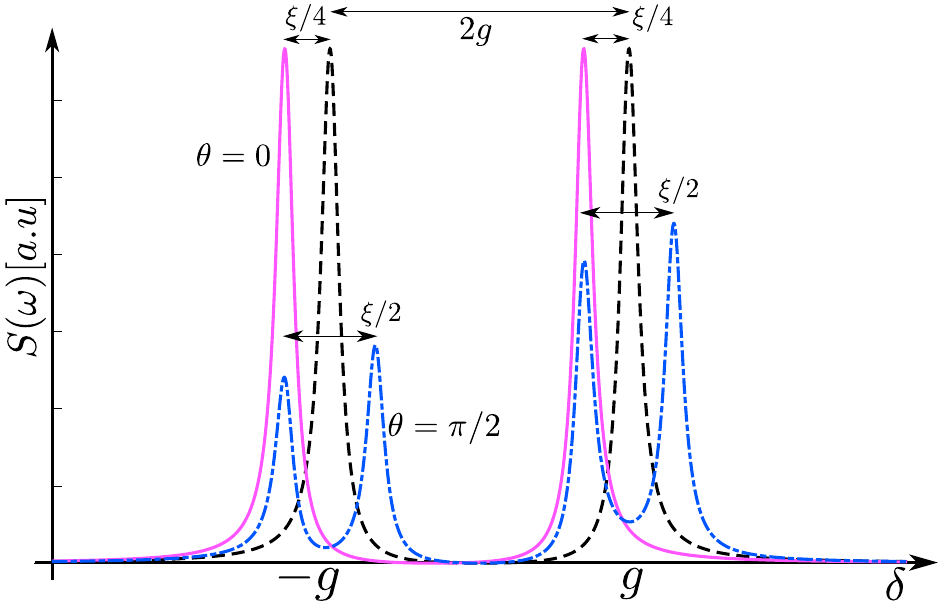}
\caption{\label{fig:vacuumRabisplit} Probe spectrum displaying the split vacuum Rabi peaks and the dependence on the phase $\theta$ of the modulation. When $\theta=0$ the vacuum Rabi peaks experience only a BS-type shift (magenta) with respect to the non-modulated peaks [dashed, c.f. figure~\ref{fig:vacuumRabi}(c)]. When $\theta=\pi/2$ the vacuum Rabi peaks are split because the JC eigenstates $|\pm,1\rangle$ are dressed by the modulation. The widths of the peaks are determined by the bare qubit relaxation rate $\Gamma_1$. The plot is based on equation~(\ref{evrfsd}).}
\end{center}
\end{figure}
In figure~\ref{fig:vacuumRabisplit}, we plot the vacuum Rabi doublets with the values $\theta=0,\pi/2$ for the modulation phase. When $\theta=0$ the vacuum Rabi oscillations are shifted in frequency by $\xi/4$ due to the BS-type correction (see also section~\ref{Sec:cohmod}). In the case of $\theta=\pi/2$, the modulation dresses the JC-eigenstates $|\pm,1\rangle$, resulting in four quasienergy states and corresponding four different spectral peaks. In addition, the two Rabi doublets are asymmetric due to the interference between the doubly dressed states.~\cite{Law95}

Quantum fluctuations in the cavity occupation $n$ lead to photon shot noise~\cite{Schuster05,Gambetta06,Sears12}. As a consequence,
the qubit linewidth is broadened and, thus, represents the cavity backaction on the qubit. In references~\cite{Schuster05,Gambetta06,Sears12}, the transmon setup with photon shot noise was considered in the dispersive strong coupling regime where $\kappa< g\ll|\omega_0-\omega_{\rm c}|$. Due to the shot noise $\delta n(t)$, the qubit gains a relative phase
\begin{equation}
\phi(t) = \frac{2g}{\omega_0-\omega_{\rm c}}\int_0^t d\tau \delta n(\tau).
\end{equation}
By assuming that the noise process is Gaussian, with photon correlation function (white photon noise)
\begin{equation}
\langle \delta n(t)\delta n(0)\rangle = \bar{n} e^{-\kappa |t|/2},
\end{equation}
one obtains Lorentzian qubit resonance which is broadened as $\Gamma_2 \rightarrow \Gamma_2 + 2\theta_0^2\bar{n}\kappa$, where $\theta_0 = 2g^2/(\kappa(\omega_0-\omega_{\rm c})$~\cite{Schuster05,Gambetta06}. In the above, we assume that the mean number of cavity quanta $\bar{n}$ is small. On the other hand, in the large-$\bar{n}$ limit one observes a spectrum which is a convolution of a Lorentzian and Gaussian lineshapes, instead of a pure exponential decay. As a consequence, the resonance width scales as $\sqrt{\bar{n}}$~\cite{Gambetta06}.

\subsubsection{The simulated ultrastrong coupling by frequency modulation}
In addition to direct sample engineering, the ultrastrong coupling in a JC system can be achieved by quantum simulation~\cite{Ballester12}, using two properly chosen driving fields. In reference~\cite{Ballester12}, the driving fields were both transverse, but a similar effect can also be accomplished by employing longitudinal coupling for one of the fields (see supplement of reference~\cite{JianLi12}). The starting point is the doubly modulated and strongly coupled JC Hamiltonian:
\begin{eqnarray}
\hat{H}(t)&=&\hbar\omega_{\rm c} \hat{a}^\dagger\hat{a} + \frac{\hbar}{2}\left[\omega_0 + \xi \cos(\Omega t)\right]\hat{\sigma}_z \nonumber \\
&&+\hbar g (\hat{a}^{\dag}\hat{\sigma}_-+\hat{a}\hat{\sigma}_+ )+\hbar G \cos(\omega t) \hat{\sigma}_{x}. \label{ham.total_res_q}
\end{eqnarray}
In an interaction picture, by choosing $\Omega=G$ and by making two RWAs in properly chosen rotating frames~\cite{JianLi12}, the above Hamiltonian can be written into the form
\begin{equation}
\hat{H}^{(\rm I)} = \hbar\tilde{\omega}_{\rm c}\hat{a}^{\dag}\hat{a} +\frac{\hbar\tilde{\omega}_0}{2}\hat{\sigma}_z + \hbar \tilde{g}(\hat{a}^{\dag}+\hat{a})\hat{\sigma}_x.
\end{equation}
In the above Hamiltonian, we have defined $\tilde{\omega}_{\rm c}\equiv\omega_{\rm c}-\omega$, $\tilde{\omega}_0\equiv \xi/2$, and $\tilde{g}\equiv g/2$. One sees that the Hamiltonian is of the form of the Rabi Hamiltonian. The qubit and cavity frequencies can be controlled with the longitudinal modulation amplitude and transverse drive frequency, respectively. Especially, the ratio $\tilde{g}/\tilde{\omega}_{\rm c}\gtrsim 0.1$ is reachable with experimentally realisable parameters~\cite{JianLi12} and, thus, the above scheme has a prospect of being used as an analog simulator of cavity QED with ultra-strong coupling. Reference~\cite{JianLi12} discusses the above idea in the circuit QED environment, which allows for fast tuning of the effective qubit and cavity frequencies.

\subsection{Modulated radiation pressure Hamiltonian}

\begin{figure}[t!]
\begin{center} 
\includegraphics[width=0.6\linewidth]{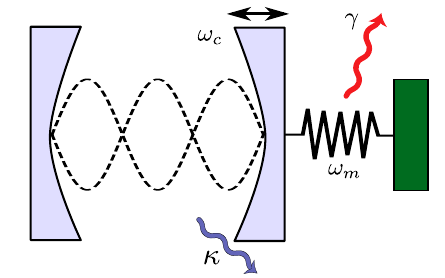}
\caption{\label{fig:optomech} Schematic of the optomechanical setup. One of the mirrors of an optical cavity is made compliant with the mechanical oscillations of a spring. The mechanical motion changes the effective length between the cavity mirrors and, thus, alters the resonance frequency $\omega_{\rm c}$ of the cavity. As a consequence, a radiation pressure coupling is created between the optical and mechanical degrees of freedom. The internal dissipation rates of the optical cavity and the mechanical oscillator are denoted with $\kappa$ and $\gamma$, respectively.} 
\end{center}
\end{figure}

Optomechanics studies the interplay between light and mechanical motion. The standard optomechanical model can be obtained by connecting one of the cavity mirrors to a spring, as depicted in figure~\ref{fig:optomech}. One of most intriguing current goals in optomechanics is achieving the strong coupling regime, where the interplay between the radiation field and mechanical motion would occur at the single-quantum level. This would ultimately allow cooling of the oscillator to the quantum mechanical ground state, achieving truly quantum \textit{mechanical} motion. Different physical realisations and the whole topic is reviewed in reference~\cite{Aspelmeyer14}.

The prototype Hamiltonian of the optomechanical system can be written as
\begin{equation}
\hat{H} = \hbar \omega_{\rm c} \hat{a}^{\dag}\hat{a} + \hbar \omega_{\rm m} \hat{b}^{\dag}\hat{b} + \hbar g \hat{a}^{\dag}\hat{a}(\hat{b}^{\dag}+\hat{b}),
\end{equation}
where $\hat{a}$ and $\hat{b}$ are the annihilation operators for the cavity and mechanical oscillators, respectively. The so-called radiation pressure term $g$ describes the momentum transfer from cavity photons to the mechanical motion. Coherent modulation of the mechanical frequency in the optomechanical setup has been studied in reference~\cite{Farace12}. They have shown that the quantum effects such as squeezing, entanglement, and discord are all enhanced in the steady state when the modulation frequency $\Omega=2\omega_{\rm m}$. This is interpreted as parametric phase locking between the modulation force and mechanical motion. Also, the effects of mechanical frequency modulations on cooling of the mechanical motion have been studied in reference~\cite{Bienert15}. In addition to mechanical modulations, they allowed also the mechanical dissipation $\gamma(t)$ to be dependent on time. They found that additional cooling resonances appear when strong periodic mechanical frequency modulations are applied in the resolved sideband limit~\cite{Bienert15}. In addition, when the modulations are weaker or the cavity linewidths are larger, these resonances overlap and can be influenced by adjusting the pulse shape.

The cavity frequency modulations have also been studied in the setup where two cavities are coupled via photon hopping and mechanical oscillation in the so-called "membrane-in-the-middle" configuration~\cite{Liao15}. When the cavity frequencies are coherently modulated, there exists a set of conditions such that a single photon can displace the mechanical membrane even in the weak coupling regime where $g\ll \omega_{\rm m}$.


\subsection{Motional averaging with hot atomic gases}
A spectacular application of motional averaging is the realization of a surprisingly simple interface between photons and an ensemble of atoms~\cite{polzik}, allowing write/read operations on the collective atomic quantum state. In general, the advantage of using an ensemble is that the interaction between a single photon and a collective mode such as the Dicke state scales as $g\sqrt{N}$, where $g$ is the interaction strength between a single photon and a single atom and $N$ is the number of atoms. For systems such as atoms or spins ({\it e.g.} nitrogen-vacancy centers in diamond), this increases the interaction and thus decreases the write/read time to values below the coherence time~\cite{kurizki}. The experiment by Borregaard et al.~\cite{polzik} uses thermal (room-temperature) atomic vapors in microcells with spin-protecting alkene coating. This contrasts with the usual realization of collective-effects quantum memories in atomic ensembles, which requires typically low temperatures, long coherence times, and localized atoms. Here the idea is that the atoms would cross randomly the laser beam, thus erasing the information about which atom the photon has interacted with.

The atoms have a $\Lambda$-level structure, with excited state $|{\rm e}\rangle$, ground state $|0\rangle$, and a metastable state $|1\rangle$. Initially, the system starts with all the atoms pumped in the state $|0\rangle$; the so-called write process consists of adding a single excitation, creating of a collective state with one excitation, distributed between all the atoms. This state is known as symmetric Dicke state, or W state. W states are also relevant for the foundations of quantum physics, and it can be shown that they are incompatible with hidden-variable theories~\cite{Wme}. To create this state, one drives the $|0\rangle \rightarrow |{\rm e}\rangle$ transition with a weak laser pulse. The interaction part of the Hamiltonian of the system reads
\begin{equation}
\hat{H} = \hbar \sum_{j=1}^{N} \left[ \delta_{\rm d} \hat{\sigma}_{\rm ee}^{(j)} - \left( \frac{\xi_{j}(t)}{2} \hat{\sigma}_{{\rm e}0}^{(j)}
+ g_{j}(t) \hat{a}_{\rm cell} \hat{\sigma}_{{\rm e}1}^{(j)} + {\rm h.c.}\right)\right], \label{eq_totalham}
\end{equation}
where $\delta_{\rm d} = \omega_{{\rm e}0}-\omega_{\rm d}$ is the detuning, $\omega_{\rm d}$ is the frequency of the driving laser, $\xi_{j}$ is the coupling to the laser field, $g_{j}$ is the coupling to the cavity $\hat{a}_{\rm cell}$, and $\hat{\sigma}_{mn}^{(j)} = |m\rangle_{j}\langle n|$ are the creation/annihilation operators associated to the transition $m\leftrightarrow n$ for the atom $j$. Then, the emission of a single photon due to the $|{\rm e}\rangle \rightarrow |1\rangle$ decay heralds the creation of the symmetric Dicke state
\begin{equation}
\frac{1}{\sqrt{N}} \left(|100\ldots0\rangle + |010\ldots0\rangle
+ |001 \ldots 0\rangle + \ldots +|00\ldots01\rangle\right)
\end{equation}
The essential phenomenon employed to create the superposition is the absence of which-atom information: we cannot know, even in principle, that the detected photon has resulted from the decay of a specific atom in the ensemble. In the readout process, the protocol is run backwards: the $|1\rangle \rightarrow |{\rm e}\rangle$ transition is driven by a classical pulse and the photon resulting from the decay $|{\rm e}\rangle \rightarrow |0\rangle$ is recovered.

Next, one studies the Langevin equations corresponding to Hamiltonian~(\ref{eq_totalham}) with decay rates $\gamma$ and
$\kappa_{1}$ for the atoms and the cavity, respectively. These equations can be solved in
the approximation that the interaction between the atoms and the field is a small perturbation.
Under these conditions the Langevin equations can be integrated, and assuming that $\hat{\sigma}_{10}(t)$ is a slow variable one gets in the Heisenberg picture~\cite{polzik}
\begin{equation}
\hat{a}_{\rm cell} (t)=  - \frac{1}{2}\sum_{j=1}^{N} \Theta_{j} (t) \hat{\sigma}_{10}^{(j)}, \label{eq_cell}
\end{equation}
where
\begin{equation}
\Theta_{j}(t) = \int_{0}^{t} {\rm d}t'\int_{0}^{t'} {\rm d}t'' e^{-\frac{\kappa_{1}}{2}(t-t')} e^{-(\frac{\gamma}{2} + i \delta_{\rm d} )(t'-t'')} g_{j}^{*}(t')\xi_{j}(t''). \label{eq_int}
\end{equation}
From equation~(\ref{eq_cell}) one clearly sees that the detection of a photon in the cell would project the state of the atoms onto a symmetric Dicke state, provided that $\Theta_{j} (t)$ is not dependent on the atom $j$. This atom-independence is realized through the averaging of the motion of the atoms as they move into and out of the laser beam due to thermal motion: for each atom, this happens many times during the duration of the laser pulse due to the collision and trajectory reversal at the walls of the cavity. These collisions preserve the phase information.

If the photons are detected only within a very narrow bandwidth, the operator $\hat a$ in time-domain will have a broad distribution. This increases the effect of motional averaging, since it includes more events of passing into the beam in equation~(\ref{eq_int}), resulting in the same value of $\Theta_{j}(t)$ for all the atoms. Since increasing the finesse of the cell cavity is not easy to realize, another solution~\cite{polzik} is to use a filter cavity with decay rate $\kappa_{2}$ placed after the cell cavity. The photons are eventually detected at the output of this cavity. Treating this as a two-cavity cascaded system, the input-output theory gives for the output field~\cite{polzik}
\begin{equation}
\hat{a}(t) = -\frac{\kappa_{2}\sqrt{\kappa_{1}}}{4} \sum_{j=1}^{N} \theta_{j} (t) \hat{\sigma}_{10}^{(j)},
\end{equation}
where 
\begin{equation}
\theta_{j}(t) = \int_{0}^{t} \rm{d}t' e^{-\frac{\kappa_{2}}{2}(t-t')}\Theta_{j}(t') .
\end{equation}
The method is rather general, and can be employed in any many-body system where there exist fluctuations in the coupling.


\subsection{Three-level systems}\label{sec:tls}
When a three-level system with states $|0\rangle , |1\rangle , |2\rangle$ is subjected to irradiation by two fields of amplitude $\xi_{01}$ and $\xi_{12}$ that are nearly resonant with the first two transitions
$|0\rangle \rightarrow |1\rangle$ and  $|1\rangle \rightarrow |2\rangle$, the Hamiltonian reads
\begin{eqnarray}
\hat{H}&=&\sum_{i=0}^{2} \hbar\omega_{i}\hat{\sigma}_{i,i} + \frac{\hbar \xi_{01}}{2}\cos (\omega_{01}^{(\xi)} t)\sum_{i=0}^{1}(\hat{\sigma}_{i,i+1} + {\rm h.c.}) \nonumber \\
& & + \frac{\hbar \xi_{12}}{2}\cos (\omega_{12}^{(\xi)} t)\sum_{i=0}^{1}(\hat{\sigma}_{i,i+1} + {\rm h.c.}),
\end{eqnarray}
where $\hat{\sigma}_{i,j}=|i\rangle \langle j|$ and $\omega_{01}^{(\xi)}, \omega_{12}^{(\xi)}$ are the drive frequencies. Let us now perform the transformation $\hat{H}^{(\rm I)} = \hat{U}^\dag \hat{H} \hat{U} + \ii\hbar(\partial_t \hat{U}^\dagger)\hat{U}$ with
\begin{equation}
\hat{U}(t) = \hat{\sigma}_{0,0} + e^{-\ii \omega_{01}^{(\xi )}t} \hat{\sigma}_{1,1} + e^{-\ii \left(\omega_{01}^{(\xi )} + \omega_{12}^{(\xi )}  \right)t} \hat{\sigma}_{2,2}
\end{equation}
and neglect the fast-rotating terms, which brings us to a multiple-rotating frame with the effective three-level Hamiltonian
\begin{equation}
\label{eq:awg_hamiltonian3states}
H^{(I)} = \frac{\hbar}{2}
\left( \begin{array}{ccc}
0 & \xi_{01} & 0 \\
\xi_{01} & 2\delta_{01} & \xi_{12} \\
0 & \xi_{12} & 2(\delta_{01} + \delta_{12})
\end{array} \right).
\end{equation}
Here the detunings are
\begin{equation}
\delta_{j,j+1} = \omega_{j+1} - \omega_j
- \omega_{j,j+1}^{(\xi )}.
\end{equation}
The Hamiltonian~(\ref{eq:awg_hamiltonian3states}) is the basis of several experiments realized with superconducting qubits at fixed detuning. Notably, using $\xi_{12}$  as a relatively strong coupling field, the transition $|0\rangle \rightarrow |1\rangle $ becomes dark, which can be probed by doing spectroscopy with $\xi_{12}$. The initial $|0\rangle \rightarrow |1\rangle $ spectral line splits into two lines, a process known in atomic physics as Autler-Townes splitting~\cite{AT_wallraff,AT_us,PhysRevB_AT}. This phenomenon has been observed also in time-domain, with $\xi_{12}$ turned on and off~\cite{dynamicalAT}. A related effect due to the coherent simultaneous drive with two fields is the destructive interference on the first excited state, which results in electromagnetically-induced transparency~\cite{coherent_population_trapping,EIT_abdumalikov}. Finally, by adiabatically turning on and off the Rabi frequencies $\xi_{12}$ and $\xi_{01}$ in the so-called counterintuitive order (first $\xi_{12}$, then $\xi_{01}$) it is possible to transfer the population from the state  $|0\rangle$ to the state $|2\rangle$ without populating at all the intermediate state $|1\rangle$. This technique is known in atomic physics as STIRAP (stimulated adiabatic Raman passage)~\cite{reviewSTIRAP,stirap_overlap,Vitanov200155,stirap_Paladino} and has been demonstrated recently in a superconducting circuit~\cite{Kumar16}. Since it is an adiabatic protocol, the fidelity of STIRAP can be improved by applying counterdiabatic corrections, using the general methods presented in section \ref{sec:background}. 

\subsection{Topological transitions}\label{s.Haldanemodel}
In 1988 Duncan Haldane introduced a model that showed for the first time that an external magnetic field is not a necessary condition for the phenomenology of the quantum Hall effect (quantized Hall conductivity $\sigma_{xy}$ and the appearance of topologically protected edge states) to occur in a system. To create topological nontrivial states, the key concept is that of Berry curvature~\cite{Berry84}. This idea has been very fertile, generating in recent years the field of topological insulators~\cite{Hasan10, Qi11}. By engineering time-dependent lattice potentials and detunings, the Berry curvature can be realized in a controlled way and measured not only in natural materials but also in artificial systems such as superconducting qubits and degenerate ultracold gases.

Haldane's model~\cite{Haldane88} consists of a honeycomb lattice, similar to graphene, but with a second-neighbour hopping that includes a phase $\varphi_{\rm H}$ that changes alternatively from one link to another. The latter can be realized by a staggered local magnetic field around the edges of the hexagons, resulting in zero average magnetic flux over the cell. In addition, a difference in the chemical potentials of the two sublattices is assumed, which produces a mass term $m_{0}$.

By expanding the lattice Hamiltonian around the nonequivalent corners ${\bm K}$ and ${\bm K}'$ of the first Brillouin zone, and keeping only the small-momentum contributions $\bm{k}_{\pm}$ measured with respect to these points (upper sign for ${\bm K}$, lower sign for ${\bm K}'$, all vectors in $x$-$y$ plane), the Haldane Hamiltonian becomes
\begin{equation}
\hat{H}_{\pm } = \hbar v_{\rm F} \left( k_{\pm}^{x} \hat{\sigma}_{x} \pm k_{\pm}^{y} \hat{\sigma}_{y} \right) + (m_{0} \mp m_{\rm t} ) \hat{\sigma}_{z} \label{haldane}
\end{equation}
The astute reader familiar with the physics of graphene will immediately recognize the part linear in $k_{\pm}^{x}$ and $k_{\pm}^{y}$
as the  Hamiltonian for Dirac fermions with Fermi velocity $v_{\rm F}$, namely $\hbar v_{\rm F} \bm{k}_{+}\cdot\hat{\bm\sigma}$ for the Dirac cone at ${\bm K}$ and $\hbar v_{\rm F} \bm{k}_{-}\cdot\hat{\bm \sigma}^{*}$ for the Dirac cone at ${\bm K}'$. In addition to this, the Haldane Hamiltonian contains also a mass term $(m_{0} \mp m_{\rm t} ) \hat \sigma_{z}$: as mentioned above, the mass $m_{0}$ arises from the sublattice inversion symmetry breaking, while $m_{\rm t}$ is a mass term resulting from the second-order hopping terms with amplitude $t_{2}$ and variable phase $\varphi_{\rm H}$, $m_{\rm t} = 3\sqrt{3} t_{2} \sin \varphi_{\rm H}$. For this model, Haldane predicted that there should exist two phases: a trivial insulating phase if $m_{0} > |m_{\rm t}|$ and a topologically non-trivial phase for $m_{0} < |m_{\rm t}|$.

\begin{figure}[t!]
  \begin{center}
\includegraphics[width=0.7\linewidth]{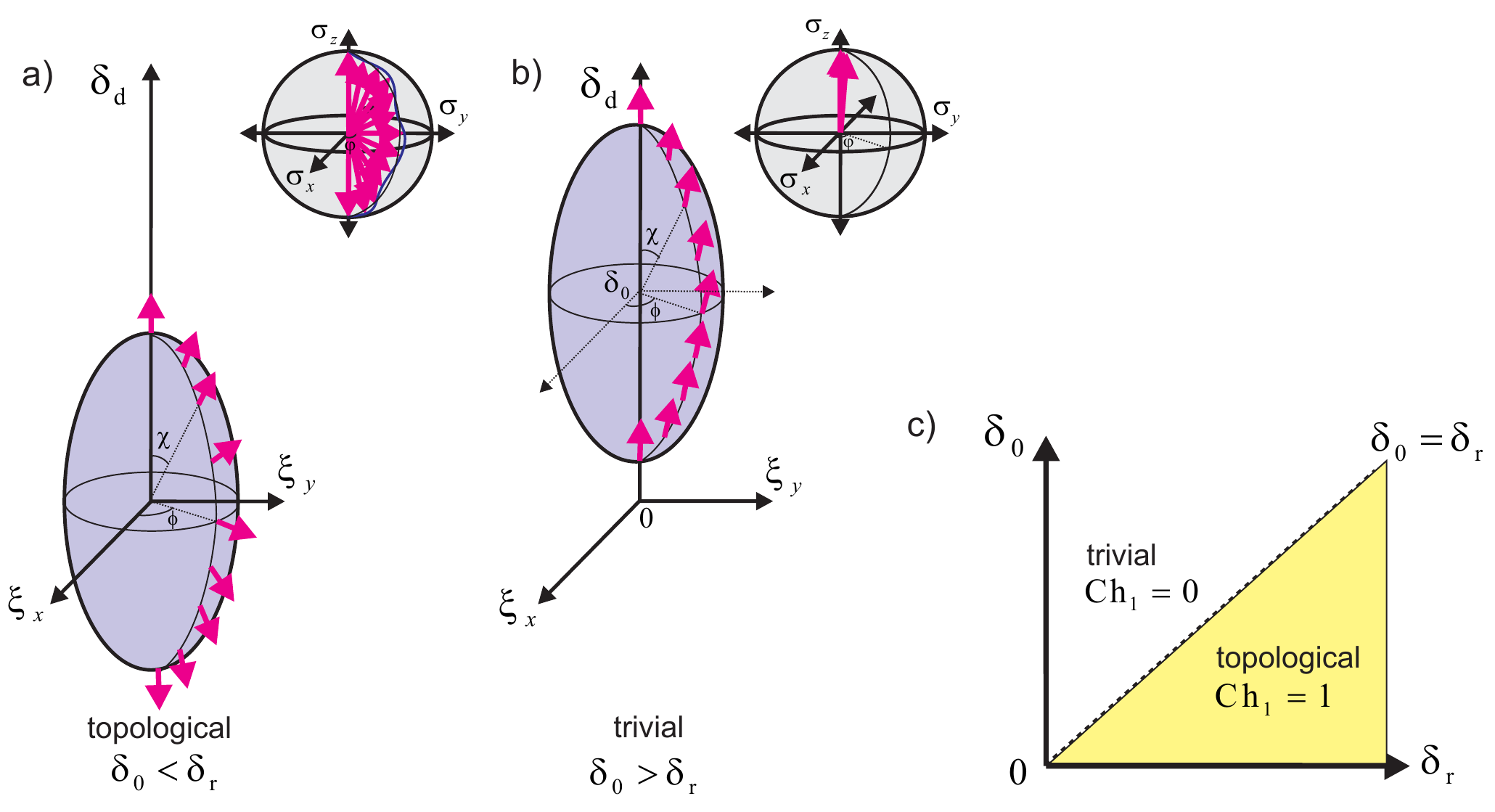}    
  \end{center}
\caption{(color online) Bloch sphere representation corresponding to the quasi-adiabatic dynamics of a two-level system, with $|0\rangle$ as the starting state, for two cases: a) $\delta_{0} = 0$ and b) $\delta_{0} >\delta_{\rm r}$. In the diagrams below, the Bloch vector is represented as an arrow for each point on the surface of the ellipsoid in the coordinates $\xi_{x}$, $\xi_{y}$, and $\xi_{z}$. For a perfect adiabatic transport, the angle $\varphi$ from the $\sigma_{x}$ axis on the Bloch sphere should equal $\phi$ (both are zero in the experiments in references~\cite{Roushan2014,Schroer2014}). Nonadiabaticity produces shifts in the Bloch vectors away from the plane $\varphi =0$, and this leads to a measurable response function which, in the first order in the speed $v_{\chi} = d\chi /dt$, is precisely the Berry curvature. In a) the vector wraps the sphere, and this phase is topologically nontrivial with the first Chern number ${\rm Ch}_{1}=1$. In b) the phase is topologically trivial, ${\rm Ch}_{1} =0$. c) Phase diagram showing the expected transition between the two phases at $\delta_{0} = \delta_{\rm r}$, obtained experimentally in references~\cite{Roushan2014,Schroer2014} by measuring the Berry curvature and using equation~(\ref{ch1}).}
\label{topological_phases}
\end{figure}

Consider now the case of a qubit with Larmor frequency $\omega_{0}$ driven by a microwave fields with frequency $\omega_{\rm d}$ and Rabi coupling $\xi_{\rm d}$. In a rotating frame of the driving field, the Hamiltonian is
\begin{equation}
\hat{H} = \frac{\hbar}{2} \left(\delta_{\rm d} \hat{\sigma}_{z}  + \xi_{\rm d} \hat{\sigma}_{x} \cos \phi + \xi_{\rm d} \hat{\sigma}_{y} \sin \phi \right), \label{qubit_haldane}
\end{equation}
where the detuning is $\delta_{\rm d} = \omega_{0} - \omega_{\rm d}$ and $\phi$ is the phase of the driving field. Already at this point one could see that there is a resemblance to the $k$-space Haldane Hamiltonian~(\ref{haldane}). To make this correspondence more precise, let us introduce a parametrization of $\delta_{\rm d}$ and $\xi_{\rm d}$ in terms of an angle $\chi$, defined by $\delta_{\rm d} = \delta_{0} + \delta_{\rm r} \cos \chi$ and $\xi_{\rm d} = \xi_{\rm r} \sin \chi$. Let us take $\delta_{0}$ and $\delta_{\rm r}$ to be positive for the simplicity of the argument. Then the drives in the $x$ and $y$ directions in equation~(\ref{qubit_haldane}) can be written as $\xi_{x} = \xi_{\rm r} \sin \chi \cos \phi$ and $\xi_{y} = \xi_{\rm r} \sin \chi \sin \phi$. This leads to a parameter surface in the form of an ellipsoid,
\begin{equation}
\frac{(\delta_{\rm d} - \delta_{0})^2}{\delta_{\rm r}^{2}} + \frac{\xi_{x}^{2} + \xi_{y}^2}{\xi_{\rm r}^{2}} = 1.
\end{equation}
Now, the possibility of controlling the detuning $\delta_{\rm d}$ in these systems enables the adiabatic exploration of this ellipsoidal manifold, see figure~\ref{topological_phases}. The experiment starts with $\chi (t=0)=0$, where $\xi_{\rm d} =0$, $\delta_{\rm d} = \delta_{0} + \delta_{\rm r}$, and the qubit is in the eigenstate $|0\rangle$ corresponding to the Bloch vector pointing to the North Pole of the Bloch sphere. Then $\chi$ is increased linearly in time until it reaches the value $\pi$, where again $\xi_{\rm d} =0$ but in this case $\delta_{\rm d} = \delta_{0}-\delta_{\rm r}$. We can see that at this end point in the parameter space $(\chi , \phi )$ the orientation of the Bloch vector depends on whether $\delta_{0} > \delta_{\rm r}$ or $\delta_{0} < \delta_{\rm r}$. In the first case the Bloch vector points to the North, while in the second to the South. Thus, if we follow adiabatically the Bloch vector from $\chi=0$ to $\chi=\pi$, in the first case it will only slightly move around the North Pole, while in the second case it will move from the North to the South, wrapping the entire Bloch sphere. The first case corresponds to a topological trivial phase, while the second to a topological nontrivial phase, see figure~\ref{topological_phases}.

This difference has a precise mathematical characterization by  the so-called first Chern number~\cite{Chern46},
\begin{equation}
{\rm Ch}_{1} = \frac{1}{2\pi} \int_{\cal S} F_{\mu\nu} d S_{\mu\nu}
\end{equation}
where ${\cal S}$ is a closed manifold, $d S_{\mu\nu}$ is an area element of the surface, and $F_{\mu\nu}$ is the Berry curvature. The first Chern number is the quantization number of the transversal conductance in the integral quantum Hall effect, where we have $\sigma_{xy} = (e^2 /h) {\rm Ch}_{1}$, and is also called TKNN invariant~\cite{Thouless82,Kohmoto85} in this context. The Berry curvature is the analog of a magnetic field, being defined as
\begin{equation}
F_{\mu\nu} = \partial_{\mu}A_{\nu} - \partial_{\nu}A_{\mu} ,
\end{equation}
where the Berry connection, defined as
\begin{equation}
F_{\mu\nu} = i \langle \psi |\partial_{\mu} |\psi\rangle
\end{equation}
is the analog of the magnetic vector potential. The origin $\delta_{\rm d} = 0$, $\xi_{x}=\xi_{y}=0$ of the coordinate system in the parameter space, where the eigenvectors become degenerate with eigenvalue zero, is the analog of a magnetic monopole. The phase transition occurs when this point passes though the surface of the ellipsoid. Note that the existence of a topological transition does not depend on having the specific ellipsoid shape of the parameter surface: the important thing is that we have a closed surface and the degeneracy point (the monopole) passes into or outside this surface.

To demonstrate a topological transition one should measure the Berry curvature and calculate the Chern invariant as the system crosses the phase transition. This has been demonstrated in two experiments with superconducting qubits~\cite{Roushan2014,Schroer2014}. Differently from the work on geometric phases, as we show below, the measurement technique in these works is not an interferometric scheme. Rather, it is based on finding the response of the system under small deviations from adiabaticity.

Indeed, it can be shown that the Berry curvature can be found directly from the following expansion~\cite{Gritsev2012}
\begin{equation}
\langle \partial_{\phi} \hat{H} \rangle = \langle \partial_{\phi}\hat{H}\rangle_{0} + v_{\chi}F_{\chi \phi} + {\cal O}(v^2),
\end{equation}
where $v_{\chi} = d\chi /dt$ is a constant in the experiment ($\chi$ is varied linearly in time), and corrections of second or higher order in $v_{\chi}$ are neglected. The quantity $-\langle \partial_{\phi} \hat{H} \rangle$ has the meaning of a force, analogous to the Lorenz force in a magnetic field. Because of the symmetry of the parameter manifold (the ellipsoid) with respect to the angle $\phi$, it is enough to consider $\phi =0$ when calculating the Berry curvature,
\begin{equation}
F_{\chi\phi} = \frac{\langle \partial_{\phi} \hat{H} \rangle \vert_{\phi =0}}{v_{\chi}} = \frac{\xi_{\rm r}\sin\chi }{2 v_{\chi}}\langle \hat{\sigma}_{y}\rangle .
\end{equation}
The quantity $\langle \hat{\sigma}_{y}\rangle$ is obtained by performing quantum tomography at each time, that is, at each point in the parameter space~\cite{Schroer2014}. From the values of the Berry curvature, the Chern number can be calculated by integration, using the symmetry with respect to $\phi$,
\begin{equation}
{\rm Ch}_{1} = \frac{1}{2 \pi}\int_{0}^{\pi}d\chi \int_{0}^{2\pi} d\phi F_{\chi\phi} = \int_{0}^{\pi} F_{\chi\phi} d\theta . \label{ch1}
\end{equation}
Next, the parameter manifold $(\theta,\phi )$ is explored and the Chern number is obtained for several values of the ratio $\delta_{0} /\delta_{\rm r}$. It is found that if $\delta_{0} < \delta_{\rm r}$ the Chern number is close to 1 within two percentages. This is the hallmark of the topological phase. For $\delta_{0} > \delta_{\rm r}$ the Chern number becomes zero, and the system is in a trivial phase.

A step forward toward employing these ideas has been done using two coupled qubits~\cite{Roushan2014}, with an $x-x$ interaction Hamiltonian $\hat{H}_{\rm int} = \frac{\hbar g}{2}(\hat{\sigma}_{x}^1\hat{\sigma}_{x}^2 + \hat{\sigma}_{y}^1 \hat{\sigma}_{y}^2 )$. In this experiment, a nonzero detuning $\delta_{0}$ was applied only to the first qubit: thus, when the qubits were decoupled ($g=0$) the second qubit remained all the time in the topological phase with Chern number 1, while the first qubit was driven across the transition, changing its Chern number from 0 to 1. Thus, the total Chern number of the system varied between 1 and 2. Surprisingly, when the interaction $g$ was turned on, a new phase appeared, characterized by zero Chern number.

In general, the idea of mapping spatial degrees of freedom and quasimomenta of many-body systems into time or frequency-domain could turn out to be an extremely fertile concept. In the future, with better and better control of the experimental systems, we will probably see more and more of what can be called ``Floquet materials''~\cite{OkaAoki09, Kitagawa10, Lindner11, Kitagawa12, Rechtsman13, Rudner13, Jotzu14}\----quantum systems where the band structure is built by time-domain modulation rather than spatial periodicity.



\subsection{Cold atomic gases}\label{s.coldgases}
In the previous sections we have studied systems with discrete energy levels. Such cases are well known in few-particle systems but occur also in systems having macroscopic number of particles. In particular, we have discussed superconducting qubits, where the macroscopic wave function has discrete states when sufficiently constrained spatially. Similar macroscopic wave functions occur also in Bose condensed dilute gases. Below we consider some example cases of driving cold gases in optical lattices~\cite{Eckardt_16}.

Let us consider an ultracold gas of bosonic atoms. By using standing waves of laser light, it is possible to form a periodic potential $V\propto\cos(2\pi x/d)$  for the atoms\----an optical lattice. The period of the lattice is $d=\lambda/2$, where $\lambda$ is the wave length of the light. The low energy states of a particle in the lattice form energy bands. If the temperature of the gas is small compared to the first band gap and the interactions between the atoms sufficiently weak, we need to consider only the lowest band. In the simplest case (small tunneling between the lattice sites), the system can be described by a tight-binding Hamiltonian, where the tunneling is directly only to the neighboring lattice sites. Such a system can be described by the Bose-Hubbard model~\cite{Jaksch05, Bloch08}
\begin{eqnarray}
\hat H_0=-J\sum_{\ell}(\hat b_{\ell}^\dagger\hat b_{\ell+1}+\hat b_{\ell+1}^\dagger\hat b_{\ell})+\frac{U}2\sum_{\ell} \hat n_{\ell}(\hat n_{\ell}-1).
\label{e.bhham}\end{eqnarray}
Here $\hat b_{\ell}$ and $\hat b_{\ell}^\dagger$ are the annihilation and creation operators for the lattice site $\ell$, $J>0$ is the hopping matrix element, $\hat n_{\ell}=\hat b_{\ell}^\dagger\hat b_{\ell}$ is the number operator at site $\ell$, and $U$ parametrizes the on-site interaction. Model~(\ref{e.bhham}) is written here for a simple one-dimensional lattice and for particles without internal degrees of freedom, but can be generalized to more general cases. In the following we concentrate on the superfluid part of the phase diagram of $\hat{H}_0$~(\ref{e.bhham}), where the tunneling dominates the interaction. In the absence of other effects, the atoms Bose condense in the lowest Bloch state corresponding to crystal momentum $k=0$ in the first Brillouin zone $|k|<\pi/d$.

Suppose the optical lattice potential  is made to oscillate,  $V(x)=V_0[x-X_0(t)]$ with a time dependent coordinate $X_0(t)$. Experimentally this can be achieved by changing periodically the phase shift of the counter propagating laser beams. By appropriate transformations this can be seen as a gauge field affecting the momentum $p\rightarrow p-m \dot X_0(t)$, where $m$ is the atomic mass. It can further be transformed to add to the Bose-Hubbard Hamiltonian~(\ref{e.bhham}) the term
\begin{eqnarray}
\hat H_1=-F(t)d\sum_{\ell} {\ell}\hat n_{\ell}.
\label{e.bhham1}\end{eqnarray}
Here $F(t)=-m\ddot X_0(t)$ is a time dependent force that shifts the energies of the levels localized at different lattice sites. This form resembles the  Hamiltonian studied above~(\ref{eq:probedLZS2}): the sites $\ell$ correspond to different states whose energies are modulated in time. The potential can be used to induce different effects.

Let us consider sinusoidal driving,  $F(t)=F_0\sin(\omega t)$. As discussed above in equation~(\ref{eq:djn}), it modifies the effective tunneling,
\begin{eqnarray}
J\rightarrow J_{\rm eff}=J {\rm J}_0(F_0d/\hbar\omega).
\label{e.efftunme}\end{eqnarray}
Here  ${\rm J}_n$  is the Bessel function of order $n$. The modulation also causes coupling to the excited modes of the Bose gas~\cite{Eckardt05}. Choosing the parameters appropriately ($\hbar\omega$~large compared to band width but small compared to band gap), these couplings can be neglected, and thus the only effect that remains is the scaling of the tunneling~(\ref{e.efftunme}). This effect has been studied experimentally in reference~\cite{Lignier07}. The value of $J_{\rm eff}$ was determined by letting the gas to expand along the lattice direction and then recording the spatial distribution. The observations accurately agree with the Bessel function modulation~\cite{Eckardt09}, including the first zeros of $\rm J_0$. The zeros mean vanishing of tunneling, and according to the Bose-Hubbard model~(\ref{e.bhham}) there should be a transition from a superfluid state to Mott insulator phase~\cite{Eckardt05, Zenesini09}. The vanishing of tunneling by modulation in a lattice has been termed ``dynamic localization''~\cite{Dunlap86}. It has long been studied theoretically and experimentally in different systems~\cite{Grifoni98}, for example, in one-dimensional transverse Ising chains~\cite{Das10,Hegde14} and semiconductor superlattices~\cite{Holthaus92}.

A phenomenon more or less similar to dynamical localization occurs in a double-well potential~\cite{Grifoni98} and is discussed in section\ \ref{Sec:cohmod}. This has been realized using single atoms in an optical lattice consisting of double-wells separated by higher barriers~\cite{Kierig08}. Note that this experiment uses single atoms rather than a condensate. The atoms enter perpendicular to the lattice direction $x$ and their  momentum distribution in  the $x$ direction is monitored after they have traversed a finite distance  parallel to the minima of the potential ($y$ direction). Under sinusoidal modulation, suppression of tunneling is observed at Bessel zeros~(\ref{e.efftunme}). If one uses a saw-tooth pulse instead of sinusoidal form, the tunneling is not suppressed~\cite{Kierig08}. This arises from breaking of the symmetry  $F(t+T/2)= -F(t)$, where $T$ is the period of the modulation~\cite{Grifoni98}.

What happens when the sign of the tunneling element changes in crossing a Bessel zero~(\ref{e.efftunme})? In a two-state system this means a change of the coupling phase by $\pi$. In a lattice, this leads to the inversion of the Bloch band, the lowest energy Bloch level becomes the maximum energy level and vice versa. The effect of this on Bose condensation has been studied experimentally in reference~\cite{Arimondo12}. They find indeed that the momentum distribution of the condensate  is peaked at the Brillouin zone boundaries $k=\pm \pi/d$. The ability to engineer the tunneling element can be applied to realize quantum simulations of, for example, magnetism on a lattice~\cite{Struck11}.

A recent trend in cold gases has been to generate artificial gauge fields~\cite{Dalibard15}. This means generating complex-valued $J=|J|e^{\ii\phi}$, where $\phi$ is called the Peierls phase. We have encountered these phases in section~\ref{s.Haldanemodel}, when we discussed the Haldane model. Such a phase appears naturally for a charged particle in magnetic field. As the standard cold atomic gases are uncharged, one has to find alternative other ways to generate the Peierls phase. One  of the methods is by shaking the lattice.  A sinusoidal shaking of the lattice changes the hopping matrix element, but keeps it real valued, c.f. equation~(\ref{e.efftunme}). Complex values are possible for a more complicated time dependence. For example, the modulation period $T$ consists of a period of sine in time $T_1$ and is constant for time $T_2=T-T_1$. The complex valued hopping matrix element enforces the Bose condensation into a state of finite crystal momentum corresponding to lowest average energy, as demonstrated by measurements in reference~\cite{Struck12}. An interesting outlook is to generate a net phase shift going around a plaquette of a 2D lattice, similar to for charged particles in magnetic field. The shaking does not do this for a square lattice, but in more complicated lattices it is possible to generate an effective staggered magnetic field~\cite{Struck13,Jotzu14}. In particular, reference~\cite{Jotzu14} realizes the Haldane model (section\ \ref{s.Haldanemodel}) with fermionic atoms in a shaken lattice.

In order to simulate a uniform (not staggered) magnetic field, the following procedure can be used~\cite{Aidelsburger13,Miyake13}. First one has two orthogonal standing laser beams in directions $x$ and $y$ to generate a 2D lattice potential. In the $x$ direction one applies a constant gradient, for example by gravity or by magnetic dipole force. This blocks the tunneling in the $x$ direction when the energy separation $\Delta$ between neighboring sites in the $x$ direction exceeds $J$. Two additional laser fields (wave vectors $\bm k_1$ and $\bm k_2$)  are used to shake the lattice. Selecting their frequency difference $\omega=\Delta/\hbar$ restores the tunneling in the $x$ direction as a photon-assisted process,
\begin{eqnarray}
 J_{\rm eff}^{(x)}=J\langle e^{\ii(F_0d/\hbar\omega)\sin(\omega t+\phi_0)-\ii\Delta t/\hbar}\rangle=J {\rm J}_1(F_0d/\hbar\omega)e^{\ii\phi_0}.
\label{e.efftunfat}\end{eqnarray}
Here the location dependent phase of the additional lasers $\phi_0(\bm r)=(\bm k_1-\bm k_2)\cdot\bm r$ is imprinted on the tunneling in the $x$ direction and allows to generate a nonzero phase shift in tunneling around a plaquette. This makes it possible to study the interplay of magnetic field and crystal lattice using cold gases in optical lattice, which would require enormous magnetic fields in ordinary crystals.

\subsection{Quantum thermodynamics}\label{s.thermal}
Above we have discussed various cases of frequency modulation in quantum systems starting from pure quantum states. The description used for pure states can be generalized in a straightforward way to an ensemble of states. One simply repeats the calculation for all members of the ensemble and expectation values are obtained as averages over the ensemble. This works at least for small systems where the added computational cost arising from repeating the same calculation for all ensemble states is manageable. With increasing the system size, this task becomes more costly. In this case the problem can sometimes be simplified  using specific properties of the ensemble. Of particular interest is the thermal ensemble, where the density matrix is given by the Gibbs distribution $\hat \rho=\exp[\beta(F-\hat H)]$ with $\beta=1/k_{\rm B}T$. This leads us to the field of quantum thermodynamics. This field has developed rapidly in recent years, see~\cite{Vinjanampathy16,Hanggi15,Millen16} for recent reviews. Besides theory, there are various experimental setups, with prominent effort in Josephson-junction circuits~\cite{Pekola15}. Here we consider two specific topics where level modulation is applied: thermal engines and fluctuation relations.

One key goal of thermodynamics, and also the initial motivation for the field, is to determine the efficiency of thermal engines and refrigerators. In particular, the efficiency $\eta$ of a thermal engine working between temperatures $T_2$ and $T_1<T_2$ is limited by the Carnot efficiency, $\eta\le\eta_{\rm C}=1-T_1/T_2$. Traditionally these considerations have been made for classical systems  and under slow, quasistatic conditions. More recently there has been generalizations to quantum systems~\cite{Scovil59} and to cycles taking place in finite time~\cite{Curzon75}. The interest for the present review is to consider finite time effects in quantum systems. A theoretical example is presented by Rezek and Kosloff~\cite{Rezek06}. They consider the Otto cycle where the working medium consists of harmonic oscillators. The Otto cycle is an idealization of the process employed in the engine developed by N.~A.~Otto in 1860's and nowadays commonly called the gasoline engine. In the adiabatic (in the thermodynamic sense) compression and power strokes the curvature of the harmonic potential is changed leading to modulation of the natural frequency of the oscillators. In the isochoric strokes the oscillators are in contact with reservoirs at temperatures $T_2$ and $T_1$ with a Lindblad type dissipator $\mathcal{D}(\sqrt{k_\downarrow}\hat a)+\mathcal{D}(\sqrt{k_\uparrow}\hat a^\dagger)$. (See definitions in section \ref{Sec:dynamicsopen}.) Because of the relative simplicity of the medium and the strokes, the state of the system obeys a generalized Gibbs distribution, and the dynamics can be solved semianalytically. It is found that in slow operation the efficiency  reaches the maximal efficiency of the classical Otto engine, which is lower than for the Carnot engine. For faster operation the efficiency decreases but the power increases. With increasing speed the power reaches a maximum, after which it starts to decrease because dissipation is increased.   The dissipation arises from two sources: the nonadiabaticity (in quantum sense) of the adiabats and the incomplete thermalization at the isochores. A fundamental difference between quantum heat engines and the classical ones is that in the first case the adiabaticity condition is stronger: not only the reservoirs have to be decoupled, but also the transitions to upper levels due to the modulation of eigenfrequencies should be ideally zero. An option to improve efficiency at a finite operation rate is to use shortcuts to adiabaticity (sections~\ref{sec:timedep}~and~\ref{sec:sbsfm}). This type of reduction of ``friction'' and of the fluctuations of work in quantum engines has been studied theoretically in several recent works~\cite{Deng13,Paternostro14, Chotorlishvili16}. Another example of a quantum Otto engine/refrigerator is given in reference~\cite{Karimi16}.

A quantum heat engine does not necessary need any moving parts nor external drive. It was pointed out in reference~\cite{Scovil59} that a three-level maser can be interpreted as a thermal engine. This is possible because two of the transitions in the three-level system (section~\ref{sec:tls}) are coupled to heat baths of different temperatures and filtered from the others. Similar thermal engines and refrigerators can also be realized using electron transport. In this connection it can be noted that even a two state-system can be used as working medium if a periodic driving of the qubit is allowed. Reference~\cite{Gelbwaser-Klimovsky13} considers the Hamiltonian
\begin{equation}
\hat{H} (t) = \frac{\hbar}{2} \left[\omega_{0} + \xi (t)\right] \hat{\sigma}_{z} + \hat{\sigma}_{x} (\hat{B}_{\rm H} + \hat{B}_{\rm C}),
\end{equation}
where $\xi (t)$ is a periodic function and $\hat B_{\rm H}$ and $\hat B_{\rm C}$ are the operators of the hot and the cold bath respectively. Except for the two temperature baths, this system was extensively studied in section~\ref{sec:twostate}. Such a ``Floquet machine'' realizes a minimalistic two-level quantum heat machine, which can be operated either as a quantum heat engine (producing work) or as a quantum refrigerator (cooling a reservoir). This machine can be used to investigate fundamental thermodynamic limits, such as the attainability of zero temperature, i.e.\ Nernst's formulation of the third law of thermodynamics~\cite{Kolar12}. This resulted in unexpected connections between quantum physics and thermodynamics - for example in some engine models with a single-mode oscillator as the working medium the dynamical Casimir effect, a purely quantum phenomenon, prevents the machine from reaching absolute zero in a finite time, thus enforcing the third law~\cite{Benenti15}. Other fundamental connections have been investigated, for example the action of Maxwell's demon and the Landauer erasure principle formulated at the quantum level~\cite{Pekola16,Lebedev16}.

The thermal distribution and the dynamics of quantum states emerge also in fluctuation relations. Traditionally, fluctuation relations have been studied in classical systems~\cite{Bochkov77,Jarzynski97,Seifert12}, but recently there has been considerable activity in extending these to the quantum systems. Let us take the Jarzynski equality~\cite{Jarzynski97} as an example. We consider a process, starting from thermal equilibrium at temperature $T$, where the Hamiltonian $H(t)$ depends on time changing from $H_{\rm i}$ at the initial time to $H_{\rm f}$ at the final time. Jarzynski equality states that the work $W$ done in this process obeys
\begin{equation}
\langle e^{-\beta W} \rangle =e^{-\beta (F_{\rm f}-F_{\rm i})}. 
\label{e.Jarzynski}
\end{equation}
The average is over the initial thermal distribution and $\beta=1/k_{\rm B}T$ is a constant determined by the initial temperature. $F_{\rm i}$ and $F_{\rm f}$ are the free energies corresponding to $H_{\rm i}$ and $H_{\rm f}$, respectively, calculated at the initial temperature $T$, $e^{-\beta F}=\mathop{\rm Tr}e^{-\beta H}$. For a closed system in the classical limit (where the operators commute) the proof of equation~(\ref{e.Jarzynski}) is more or less trivial,
\begin{eqnarray}
\langle e^{-\beta W}\rangle
&=&\mathop{\rm Tr}[e^{\beta(F_{\rm i}-H_{\rm i})}e^{-\beta W}]
=e^{\beta F_{\rm i}}\mathop{\rm Tr}e^{-\beta (H_{\rm i}+W)}
=e^{\beta F_{\rm i}}\mathop{\rm Tr}e^{-\beta H_{\rm f}}\nonumber\\
&=&e^{-\beta (F_{\rm f}-F_{\rm i})}.
\label{e.Jarzynski1}\end{eqnarray}
Using proper definitions this result can be extended to the quantum case and to open systems~\cite{Campisi11,Jarzynski15}.

In order to get insight into the  Jarzynski equality~(\ref{e.Jarzynski}), consider a two-state system. At the initial time the system is degenerate. This gives $e^{-\beta F_{\rm i}}=2$ (choosing $E_{\rm i}=0$). At the final time, state 2 is lifted relative to state 1 to an energy much higher than $k_{\rm B}T$ so that its thermal occupation is negligible. Thus $e^{-\beta F_{\rm f}}=1$ and~(\ref{e.Jarzynski}) gives 
\begin{equation}
\langle e^{-\beta W}\rangle=\frac12.
\label{e.jzksc}\end{equation}
We can take a look at some special cases. If the states are uncoupled, their probabilities remain the same. The validity of~(\ref{e.jzksc}) in this case can easily be verified. Alternatively we can use the mathematical inequality $\langle e^{X}\rangle\ge e^{ \langle X\rangle}$, which gives
$\langle W\rangle\ge k_{\rm B}T\ln 2$. The process is known as {\em information erasure}, and we see that it costs an average work $k_{\rm B}T\ln 2$ at minimum~\cite{Landauer61}.
The minimum is achieved when the qubit is coupled to environment and the level spacing is varied quasistatically so that thermal occupation is preserved during the lift.

It is interesting to observe that the right hand side of the Jarzynski equality~(\ref{e.Jarzynski}) depends on the initial and final Hamiltonians, but there is no dependence on how the change proceeds in time. Therefore the left hand side, the  averaged exponentiated work $\langle e^{-\beta W}\rangle$, should also be independent of the time dependence in between. Thus, equation~(\ref{e.Jarzynski}) is valid for any driving from $H_{\rm i}$ to $H_{\rm f}$. However, if one is interested in other quantities besides equation~(\ref{e.Jarzynski}), or wants to verify it by numerical methods, one has to select some driving. Examples of  quantum systems studied thus far are two-state systems, coupled two-state systems, 1 dimensional potentials with linear, quadratic, quartic,  or hard wall potentials, 2 dimensional potentials, and simple many-particle systems, while the proposed drivings have been linear sweeps, sudden quenches, and periodic~\cite{Gong14,Jarzynski15}.


\section{Conclusions }\label{sec:conc}
In this review we discussed the topic of frequency modulation in several different quantum mechanical systems. We tried to keep our theoretical examples simple enough, so that the presentation of the conveyed physics would be clear. We have mostly presented results that have been subjected to an experimental test or demonstration, and as such their relevance is validated by laboratory practice. There are certainly other concepts, theoretical developments, and proposals that no doubt will find a connection to experiment in the forthcoming years.

The modulation of the frequency of a quantum system produces a variable quantum phase difference, and we emphasized its connection with the measured spectrum, both for coherent and incoherent modulations. Many of the experiments discussed are from the field of superconducting systems, where spectacular advances in reducing the decoherence and in developing the quantum control have been achieved recently, but the concepts have general applicability.

We started our analysis with an extensive discussion of the case of two-level systems, and showed that the coherent modulations lead to the formation of multi-photon sidebands, to the observations of coherent destruction of tunnelling, dynamic Stark effect and Landau–Zener–St\"uckelberg -interference. For the case of incoherent modulation, we reviewed the connection with classical theory of dephasing. We showed that the conventional exponential phase damping can be obtained with a Gaussian noise process and a short correlation time. Also, the phenomena of motional narrowing and motional averaging arise naturally when the phase experiences random telegraph noise or other kind of temporal fluctuations. Next, we discussed the paradigmatic case of the harmonic oscillator under frequency modulation, and  showed that this can lead to phenomena such as squeezing and dephasing.

As an example of coupled few-level systems, we analyzed the Jaynes-Cummings and the optomechanical Hamiltonians. The variation of the qubit frequency in a Jaynes-Cummings system can be harnessed to create fast two-qubit gates for quantum information processing. On the other hand, a modulation of the cavity frequency causes additional splitting of the vacuum Rabi doublet. In optomechanical systems, the modulations of the mechanical frequency shows enhancement of several quantum effects, such as entanglement between the mechanical and radiation degrees of freedom, squeezing, cooling, and quantum discord. Modulations in the cavity frequency are promising for the creation of strong coupling effects in the weak coupling regime.

Finally, we considered the effect of frequency modulation in a variety of many-body systems, from atomic gases to simulators of topological phase transitions in coupled qubits. We showed how the phenomenon of motional averaging and narrowing lead to the realization of quantum memories in hot atomic gases. With superconducting qubits, by manipulating adiabatically the drive strength and the detuning it is possible to emulate the Haldane model, which displays topological transitions. The same is achieved in ultracold gases, via a rather complex modulation of the trapping potential. In more simpler situations, when the lattice  modulation is sinusoidal, the result is that the tunneling is modified in the same way as the coupling for the qubit case: it becomes multiplied by a Bessel function. This leads to the phenomenon of dynamical localization and the subsequent transition  from the superfluid to Mott insulator state. As a final note, we gave an introduction of utilization of frequency modulations in quantum thermodynamics. 

Given the progress in the recent years, we can confidently predict that we will see more and more experiments where the modulation of frequency is used in clever ways. For almost any task in quantum information processing, this remains an invaluable tool. The topic of artificial gauge fields and possibly Floquet materials created by various forms of modulation will no doubt sustain the interest of the condensed-matter community. With a history as old as quantum mechanics itself, the problem of frequency modulation in quantum systems still remains an invaluable source of theoretical insights. The beautiful blend between theory and experimental techniques will likely continue to yield interesting results.

\ack
MPS acknowledges funding from the Finnish Academy of Science and Letters (Vilho, Yrj\"o and Kalle V\"ais\"al\"a Foundation), Alfred Kordelin Foundation, the National Graduate School of Materials Physics, Army Research Office W911NF-14-1-0011 and NSF DMR-1301798. JAT acknowledges support from the Academy of Finland (Centre of Excellence in Computational Nanoscience, projects 251748 and 284621). EVT acknowledges funding from  the Academy of Finland and Tauno T\"onning foundation. GSP acknowledges the Centre of Quantum Engineering (CQE) at Aalto University as well as support from the Academy of Finland (project 263457 and Center of Excellence in Low Temperature Quantum Phenomena and Devices - project 250280) and FQXi.


\section*{References}
\bibliographystyle{iopart-num}
\bibliography{FM_quantum_biblio_matti,FM_quantum_biblio_jani_v2,FM_quantum_biblio_erkki,FM_quantum_biblio_sorin}{}

\end{document}